\RequirePackage[2024-06-01]{latexrelease}
\documentclass[twocolumns]{aastex701}

\usepackage{amsmath}
\usepackage{bm}
\usepackage{graphicx}
\usepackage[caption=false]{subfig}
\usepackage{longtable}
\usepackage{booktabs}
\usepackage{array}  
\acceptjournal{ApJ}

\begin{document}

\title{The Impact of Galaxy Formation on Galaxy Biasing,\\ and Implications for Primordial non-Gaussianity Constraints}

\correspondingauthor{Lucia A. Perez, lucia.perez.phd@gmail.com}

\author[0000-0002-8449-1956]{Lucia A. Perez}
\affiliation{Center for Computational Astrophysics, Flatiron Institute, 162 5th Ave, New York, NY 10010, USA}
\affiliation{Department of Astrophysical Sciences,
Princeton University, 4 Ivy Lane, Princeton, NJ 08544, USA}
\email{lperez@flatironinstitute.org}

\author[0000-0002-3185-1540]{Shy Genel}
\affiliation{Center for Computational Astrophysics, Flatiron Institute, 162 5th Ave, New York, NY 10010, USA}
\affiliation{Columbia Astrophysics Laboratory, Columbia University, 550 West 120th Street, New York, NY 10027, USA}
\email{sgenel@flatironinstitute.org}

\author[0000-0001-8356-2014]{Elisabeth Krause}
\affiliation{Steward Observatory, University of Arizona, 922 N Cherry Ave, Tuscon, AZ 85719, USA}
\email{krausee@arizona.edu}

\author[0000-0003-2835-8533]{Rachel S. Somerville}
\affiliation{Center for Computational Astrophysics, Flatiron Institute, 162 5th Ave, New York, NY 10010, USA}
\email{rsomerville@flatironinstitute.org}

\begin{abstract}

The parameter $f_{\textrm{NL}}$ measures the local non-Gaussianity in the primordial energy fluctuations of the Universe, with any deviation from $f_{\textrm{NL}}=0$ providing key constraints on inflationary models.
Galaxy clustering is sensitive to $f_{\textrm{NL}}$ at large scale modes and the next generation of galaxy surveys will approach a statistical error of $\sigma_{f_{\textrm{NL}}}\sim1$. 
However, the systematic errors on these constraints are dominated by the degeneracy of $f_{\textrm{NL}}$ with the galaxy bias parameters $b_1$ (galaxy overdensities caused by mass perturbations) and $b_{\phi}$ (galaxy overdensities caused by primordial potential perturbations). It has been shown that the assumed scaling of $b_{\phi}(z)=2\delta_c (b_1(z)-1)$ is not accurate for realistically simulated galaxies, and depends both on the galaxy selection and the way that galaxies are modeled.  
To address this, we leverage the CAMELS-SAM pipeline to explore how varying parameters of galaxy formation affects $b_{\phi}$ and $b_1$ for various galaxy selections. We run separate-universe N-body simulations of $L=205 h^{-1}$ cMpc and $N=1280^3$ to measure $b_{\phi}$, and run 55 unique instances of the Santa Cruz semi-analytic model with varying parameters of stellar and AGN feedback.  
We find the behavior and evolution of a \texttt{SC-SAM} model's stellar-, SFR- and sSFR- to halo mass relationships track well with how $b_1$ and $b_{\phi}(b_1)$ change across redshift and selection for the SC-SAM. 
We find our variations of the SC-SAM encapsulate the $b_{\phi}$ behavior previously measured in IllustrisTNG, the Munich SAM, and Galacticus.
Finally, we identify sSFR selections as particularly robust to varied galaxy modeling.
\end{abstract}


\section{Introduction} \label{sec:intro}

A key outstanding topic in cosmology is the process that drove inflation. In particular, was inflation a single- or multifield process? These models can be broadly distinguished by the deviation from Gaussianity measured in the primordial perturbations of the Universe, with the parameter $f_{\text{NL}}^{\text{loc}}$ parameterizing how non-Gaussian these fluctuations are due to the local gravitational potential. Measurements of a non-zero $f_{\text{NL}}^{\text{loc}}$ rule out models of single-field slow roll inflation, and a precise measurement of $f_{\text{NL}}^{\text{loc}}$ can further clarify how inflation occurred (\citealt{Maldacena2003, CreminelliZaldarriaga2004, Cabass2022}). $f_{\text{NL}}^{\text{loc}}$ might also explain the S8 tension \citep{Stahl2024}.
In this work, we consider only primordial non-Gaussianity (PNG) of the local type, and shorten $f_{\text{NL}}^{\text{loc}}$ to $f_{\text{NL}}$; see \citet{Cabass2023, Planckfnl2020} for forecasts for other PNG templates.

The best current constraints on $f_{\text{NL}}$ come from analysis of the Cosmic Microwave Background (CMB) at $k \sim 0.05$ $h$ Mpc$^{-1}$ (intermediate angular scales), which has found $f_{\text{NL}}=-0.9 \pm 5.1$\footnote{We list 1$\sigma$ (68\%) errors throughout this work,
unless noted otherwise.} \citep{Planckfnl2020}, tightened slightly to $-0.1 \pm 5.0$ with reprocessing \citep{PlanckPR4}.
However, measurements with the CMB are nearing their cosmic-variance limit and are suppressed at small distance scales due to Silk damping, and cannot access many of the $k$ scales where non-Gaussianity could be further measured \citep{Ferraro2015}. The large scale structure of dark matter halos carries a distinctive signature of primordial local non-Gaussianity, as a scale-dependent modulation of galaxy clustering proportional to $k^{-2}$ in density (and therefore both $k^{-2}$ and $k^{-4}$ in the power spectrum; \citealt{Slosar2008, Dalal2008a}). 
Recent analyses are beginning to reach constraints competitive to the CMB's (e.g.\ \citealt{Chudaykin2026}) and future missions are expected to eventually surpass them, especially with tracers beyond two-point galaxy clustering: e.g.\ multitracer studies \citep{Hamaus2011, Gomes2020, BarreiraKrause2023}; different methods of analyzing large scale structure \citep[e.g.][]{Peron2024, Goldstein2024, Kvasiuk2025, Morawetz2025, Marinucci2025, Brown2025, Andrews2026Euclid, Andrews2026BORG}; Sunyaev-Zeldovich effects \citep{Adshead2024, Lague2025}; weak lensing surveys \citep{Hilbert2012}; line intensity mapping \citep{Camera2015, LiMa2017, MeerKLASS2026}; gravitational waves \citep{LiWang2024, ZhouLiWu2025}; and interesting cross-correlations with radio surveys \citep{Orlando2025, Kopana2025, Karagiannis2025} or CMB lensing \citep{Fabbian2025Quaia, Bermejo-Climent2025, Chiarenza2026}. In particular, the next generation of large scale galaxy surveys expect to reach errors of $1\sigma_{f_{\text{NL}}} \sim 1$ to confidently constrain single vs.\ multiple field inflation (\citealt{Giannantonio2012, SPHEREx2014, FerraroWilson2019, Mueller2019, Castorina2019, Heinrich2024, Euclid2026ICsprojected}).

However, all constraints involving galaxy surveys depend on how one models the connection between galaxies to the underlying dark matter field and halos (i.e.\ \textit{galaxy bias}). For analyses with the power spectrum in particular, $f_{\text{NL}}$ induces scale-dependent modulations $\propto b_1 b_{\phi} f_{\text{NL}} /k^2$ for $k \leq 0.01 h$ Mpc$^{-1}$.
The bias parameters $b_1$ and $b_{\phi}$ are defined as the response of the number density of galaxies to mass density $\delta_{\textrm{m}}$ and primordial potential perturbations $\phi$, respectively.
Though $b_1$ can be independently constrained for a galaxy population using scales where the $1/k^2$ PNG signal is negligible, the effects of $f_{\text{NL}}$ are completely degenerate with $b_{\phi}$. Most constraints with galaxy power spectra have so far relied on the so-called \textit{universality relation} built on the assumption that the halo mass function $n$ depends only on the height of density peaks, which itself only depends on the power spectrum \citep{Slosar2008,Ferraro2015}. 
This approximation is a good fit for dark matter halos, and a more general form of $\mathbf{b_{\phi} = 2 \delta_c (b_1 - p)}$ allows, in theory, other $p$ values to approximate specific galaxy populations. 
Table \ref{table:fNL_with_univ} lists recent constraints on $f_{\text{NL}}^{\text{loc}}$ with two-point clustering statistics of galaxies; nearly all rely the universality relation for their constraints.

\begin{longtable}{@{} p{1.5in} p{2in} p{3.25in} @{}}
\caption{Recent constraints of $f_{\mathrm{NL}}^{\mathrm{loc}}$ from two-point galaxy clustering, in chronological order. \label{table:fNL_with_univ}}\\
 \hline
 Citation & Constraint & Relevant details of study \\
 \hline
 \endfirsthead

 % \hline
 \multicolumn{3}{c}{\textit{Continuation of Table \ref{table:fNL_with_univ}}}\\
 \hline
 Citation & Constraint & Relevant details \\
 \hline
 \endhead

 \hline
 \endfoot

 % \hline
 % \multicolumn{2}{ c }{End of Table}\\
 \hline %\hline
 \endlastfoot

\cite{deBernardis2010}
& $f_{\mathrm{NL}} = 171_{-139}^{+140}$ (SDSS DR4);\; \newline $f_{\mathrm{NL}} = -93 \pm 128$ (SDSS DR7)
& WMAP7 CMB combined with SDSS DR4 red luminous galaxy power spectrum or DR7 LRG derived halo power spectrum; universality-adjacent relation with $p=1$. \\

\cite{Agarwal2014}
& $f_{\mathrm{NL}} = -17 \pm 68$ (LRGs);\; $f_{\mathrm{NL}} = 103^{+148}_{-146}$ (QSOs);\;  $f_{\mathrm{NL}} = 2_{-66}^{+65}$ (both)
& SDSS DR8 LRGs and quasars; universality-adjacent relation with $p=1$ for LRGs, $p=1.6$ for quasars. \\

\cite{Castorina2019}
& $-51<f_{\text{NL}}<21 (p=1,\ 95\%)$;\; \quad 
$-81<f_{\text{NL}}<26 (p=1.6,\ 95\%)$
& eBOSS DR14 NGC+SGC quasars Fourier space clustering; universality relation with both common $p$-values \\

\cite{Cabass2022}
& $f_{\mathrm{NL}} = -33 \pm 28 (p=0.55)$
& BOSS redshift-space galaxy power spectra and bispectra; universality with best fit from \citet{Barreira+2020_TNG} \\

\cite{Mueller2022}
& $f_{\mathrm{NL}} = -12 \pm 21$ ($p=1.6$);\; \newline
$-17 < f_{\mathrm{NL}}^{\mathrm{loc}} < 11$ ($p=1$)
& eBOSS DR16 quasars; universality function with both common $p$-values. \\

\citet{Barreira2022c}
& $f_{\mathrm{NL}} = 16 \pm 16 (p=-1)$ to \newline $f_{\mathrm{NL}} = 230 \pm 226 (p=2)$
& BOSS DR12 galaxy spectra re-analysis showing how changing $p$ and $b_1$, $b_{\phi}$ priors affects results from the same data \\

\cite{Krolewski2024}
& $f_{\mathrm{NL}} = -26^{+45}_{-40}\ (p=1.6);$ \newline $f_{\mathrm{NL}} = -18^{+29}_{-27}\ (p=1)$
& DESI EDR and DR1 QSOs cross-correlated with Planck 2018 CMB lensing; universality function with both common $p$-values. \\

\cite{Rezaie2024}
& $f_{\mathrm{NL}} = 34^{+25}_{-44}$
& DESI Legacy Survey DR9 LRGs; universality with $p=1$. \\

\cite{Cagliari2024}
& $-4 < f_{\mathrm{NL}} < 27$ ($p=1$);\; \newline 
$-23 < f_{\mathrm{NL}} < 21$ ($p=1.6$)
& eBOSS NGC+SGC DR16 quasars; universality function with both common $p$-values. \\

\cite{D'Amico2025}
& $f_{\mathrm{NL}} = 52 \pm 34$
& BOSS DR12 power spectrum and bispectrum; EFTofLSS at one-loop order for PNG, \textit{no} universality. \\ % VERIFY: arXiv v2 abstract of 2201.11518 quotes 7 +/- 31

\cite{Fabbian2025Quaia}
& $f_{\mathrm{NL}} = -20.5^{+19.0}_{-18.1}\ (p=1);$ \newline $f_{\mathrm{NL}} = -28.7^{+26.1}_{-24.6}\ (p=1.6)$
& Quaia quasars auto-correlation and cross-correlation with Planck PR4 CMB lensing; universality function with both common $p$-values. \\

\cite{Bermejo-Climent2025}
& $f_{\mathrm{NL}} = 39^{+40}_{-38}\ (p=1)$ (lensing cross-correlation only);\; \newline $f_{\mathrm{NL}} = 24^{+20}_{-21}\ (p=1$, with LRG auto-correlation)
& DESI Legacy Survey DR9 LRGs and Planck PR4 CMB two-point angular cross-correlation; universality function. \\

\cite{Chaussidon2025}
& $b_{\phi}f_{\mathrm{NL}} = 22^{+92}_{-63}$ (LRGs);\; \newline $b_{\phi}f_{\mathrm{NL}} = -13^{+56}_{-56}$ (QSOs);\; \newline $f_{\mathrm{NL}} = -3.6^{+9.0}_{-9.1}$ (both)%;\; 
% $f_{\mathrm{NL}} = 3.5^{+10.7}_{-7.4}$ (both, QSO $p=1$)
& DESI DR1; combined constraint assumes universality-like $b_{\phi}(b_1)$. \\ % ($p=1$ for LRGs, $p=1.6$ for QSOs). \\ %; the single-sample $b_{\phi}f_{\mathrm{NL}}$ values do \textit{not} assume universality. \\

\cite{Cagliari2025Bispectrum}
& $-6 < f_{\mathrm{NL}} < 20$ ($p=1$);\; \newline
$-23 < f_{\mathrm{NL}} < 14$ ($p=1.6$)
& eBOSS DR16 quasar optimal power spectrum and bispectrum; universality function with both common $p$-values. \\

\cite{Chudaykin2026}
& $f_{\mathrm{NL}} = -0.1 \pm 7.4$ (DESI galaxies + QSO-all);\; \newline $f_{\mathrm{NL}} = -0.0 \pm 4.1$ (with Planck PR4 CMB)
& DESI DR1 BGS, LRG1--3 and ELG2 galaxies (power spectrum and bispectrum multipoles, one-loop EFT) and full high-redshift quasar sample (large-scale tree-level $P_{\ell}$ and optimal $B_0$); universality with common $p$-values. \\

\cite{RosadoMarin2026}
& $f_{\mathrm{NL}} = 2.1^{+8.8}_{-8.3}$ ($p_{\mathrm{LRG}}=1$, $p_{\mathrm{QSO}}=1.6$);\; \quad $f_{\mathrm{NL}} = 1.9^{+7.6}_{-6.8}$ ($p_{\mathrm{LRG}}=0.55$, $p_{\mathrm{QSO}}=1.4$)
& DESI DR1 LRG and QSO auto power spectra plus the LRG$\times$QSO cross-power spectrum; universality function with common and new $p$-values, from \citep{Barreira+2020_TNG} and  \citep{Fondi2026}. \\

\cite{Chiarenza2026}
& $f_{\mathrm{NL}} = 2^{+28}_{-34}\ (p=1.6);$ \newline $f_{\mathrm{NL}} = 6^{+20}_{-24}\ (p=1)$
& DESI DR1 spectroscopic QSOs cross-correlated with Planck PR4 CMB lensing; universality function with both common $p$-values. \\

\cite{Fondi2026}
& $f_{\mathrm{NL}} = -3.3 \pm 9.2$ (optimal weights with $p_w=1.6$);\; \newline $f_{\mathrm{NL}} = 2.4 \pm 11.7$ (optimal weights with $p_w=1$)
& DESI DR1 QSO power spectrum (data and chains of \citealt{Chaussidon2025}); universality using a simulation-based prior on $p$ (from IllustrisTNG and CAMELS). \\

\cite{Brown2026}
& $f_{\mathrm{NL}} = -3^{+22}_{-21}$ (LRGs);\; \newline $f_{\mathrm{NL}} = 0^{+17}_{-16}$ (QSOs);\; \newline $f_{\mathrm{NL}} = -3 \pm 12$ (both)
& DESI DR1 configuration-space two-point correlation function (monopole) using simulation-based model; universality function with Gaussian priors $p_{\mathrm{LRG}}=1.0\pm0.1$ and $p_{\mathrm{QSO}}=1.6\pm0.1$. \\

 \end{longtable}

An important complication is the complexity that galaxy formation presents for modeling galaxy bias, particularly in the era of `precision cosmology' (\citealt{Narayanan2000, WechslerTinker2018, Wibking2019, SafiFarhang2021, Martinelli2021, Shiferaw2025}). \citet{Barreira+2020_TNG} showed that the universality relation is a poor descriptor of galaxy bias, and that different galaxy selections in the IllustrisTNG model showed different and nontrivial relations between $b_1$ and $b_{\phi}$. Other work has shown the risk of biased and inaccurate constraints when $b_{\phi}$ is mischaracterized or its prior too broad (\citealt{Barreira2020, M-D2021}, leading \citealt{Chudaykin2026} to fix their $p$ values), with \citet{Barreira2022c} showing the same data can give $f_{\text{NL}}= 16\pm 16$ or $f_{\text{NL}}= 230\pm 226$ under different $b_{\phi}(b_1)$ relations. 
Even efforts to constrain the degenerate $b_{\phi}f_{\textrm{NL}}$ (which are only useful for when $b_{\phi} \neq 0$ for the sample) are difficult to do in a prior-agnostic way, particularly in single redshift bins \citep{Euclid2026prep}.
There has been exciting in progress in utilizing the information from observed galaxy samples directly: \citep{YuNyugen2026} combining small-scale observed clustering with a simulation-based HOD approach, or using the redshift evolution of tracer densities  to directly measure $b_{\phi}$ \citep{Sullivan2025TimeEvolution, DalalPercival2025}. These methods come with caveats -- a parametric HOD galaxy-halo model in the former, and small-volume boxes and complex corrections in the observer plane  for the latter \citep{Nascimento2026}. 
Still, while effort is ongoing to theoretically improve the descriptions and constraints of bias (\citealt{Shiferaw2025} generally, \citealt{HadzhiyskaFerraro2025} for $b_{\phi}$ specifically), there is much to learn still by simulating galaxies more realistically. 

Though some work has been conducted using hydrodynamical simulations to study the effect of $f_{\text{NL}}$ on the resulting galaxy catalogs (\citealt{Barreira+2020_TNG, Stahl2023}), the simulations are far from the number and size of N-body simulations built for better understanding PNG (e.g.\ \citealt{QuijotePNG, Baldi2024, PNGunit, Hadzhiyska2024}). In particular, the complexity of hydrodynamical galaxy models makes it computationally expensive to run simulations large enough and with the required physics to study non-Gaussian galaxy bias. 
Clever approaches have enabled some encouraging results from the small CAMELS simulations \citep{CAMELS} to improve existing analyses \citep{Fondi2026} or confirm the veracity of the `time evolution' method \citep{Sullivan2025TimeEvolution, Nascimento2026}, with the caveat of large errors due to small volumes.

In this work, we leverage the flexibility and speed of the Santa Cruz semi-analytic model (\texttt{SC-SAM}; \citealt{Perez2023ConstrainingSuite, Somerville2015, Somerville2008}). Semi-analytic models (SAMs) of galaxy formation are a long-used technique to simulate galaxies in a cosmological context, using physically motivated recipes for astrophysical properties to populate galaxies onto dark matter halo merger trees (i.e.\ their formation and mass assembly history through accretion and mergers). SAMs are many times faster to run than hydrodynamical simulations, contain insights into astrophysical processed that halo occupation models lack, and also reproduce galaxy assembly bias \citep{Croton2007} in galaxy clustering measurements naturally. All together, SAMs offer a useful landscape for understanding how uncertainties and variations within a given model for the astrophysics of galaxies affect a galaxy clustering observable.
While other SAMs have been used for focused studies of non-Gaussian assembly bias (\citealt{Reid2010}, \citealt{Marinucci2023}, see \textsection \ref{subsec:TNGnSAMs}), this work innovates in both running a SAM on separate-universe N-body simulations to specifically study $b_{\phi}$ while also varying several parameters of galaxy feedback at once to probe how uncertainty in galaxy modeling affects $b_{\phi}$ for galaxy selections. 
We note the upcoming results in de Icaza-Lizaloa et al.\ (in prep), who run the fiducial \texttt{GALFORM} \citep{Gonzalez-Perez2020} and \texttt{SHARK} \citep{Lagos2018} SAMs on the PNGUnit simulations initialized with $f_{\textrm{NL}}^{\textrm{loc}}=100$ \citep{PNGunit}, as well as separate-universe simulations as we do, in order to approximate $b_{\phi}$ and $p$-value priors and their impact on $f_{\textrm{NL}}$ for DESI-like galaxy selections. We hope our work here further motivates the study of $b_{\phi}$ both across different models of galaxy formation as well as diverse parameterizations within them. 

In this work, we seek to better understand how variations in galaxy formation modeling affect measurements of the galaxy bias parameters $b_{\phi}$ and $b_1$, using the \texttt{SC-SAM}: 
\begin{itemize}
    \item \textsection \ref{sec:methods}: we describe our separate-universe experiment and simulations, and how we measure $b_{\phi}$ and $b_1$.
    \item \textsection \ref{sec:SCSAMbphi}: we discuss the relevant parts of the \texttt{SC-SAM}; the parameters we vary; the resulting stellar mass-, star formation rate- (SFR), and specific SFR- (sSFR) to-halo mass relationships; and the spread of 56 \texttt{SC-SAM} model variations we probe for the rest of the work.
    \item \textsection \ref{sec:bphiresults}: we study how $b_{\phi}$ changes for a given galaxy selection, for models where one parameter is varied at a time to its minimum and maximum (`one-at-a-time'), and for the 56 fully varied models. We primarily focus on $z=1$ in our main tezt, but present $z=0$ and $z=2$ for the `one-at-a-time' models in the text, and in Appendix \ref{app:bphi_zevol} \& \ref{app:b1bphi_zevol} for the varied models.
    \item \textsection \ref{sec:b1bphiresults}: we analyze the $b_{\phi}(b_1)$ relationships of our models under different galaxy selections, identifying strong deviations from universality but consistent behavior across \texttt{SC-SAM} parametrizations.
    \item \textsection \ref{sec:discussion}: we discuss the effect of first-order Eulerian galaxy assembly bias on our $b_1$ and $b_{\phi}$; compare to related works; and offer guidance for observational analyses.
    \item \textsection \ref{sec:conclusion}: we conclude and summarize our findings.
\end{itemize}

%---------------------------%
%---------------------------%
%---------------------------%
%---------------------------%
%---------------------------%
%---------------------------%

\section{Methodology and Simulations for $\lowercase{b}_{\phi}$} \label{sec:methods}

\subsection{Separate-Universe N-body Simulations for $b_{\phi}$} \label{subsec:SepUni}

In this section, we describe $b_{\phi}$ and how its effect on galaxy formation is modeled with the `separate-universe' technique.
The galaxy bias parameter $b_{\phi}$ is associated with PNG of the local type. This type of PNG is parametrized in terms of the primordial gravitational potential during matter domination $\phi(x)$, as local perturbations to a Gaussian distributed random field $\phi_{\mathrm{G}}(x)$:

\begin{equation}
    \phi(\bm{x}) = \phi_{\mathrm{G}}(\bm{x}) + f_{\textrm{NL}} [ \phi_{\mathrm{G}}(\bm{x})^2 - \langle \phi_{\mathrm{G}}(\bm{x})^2 \rangle ]\ .
    \label{eq:fNL_initialdef}
\end{equation}

Here, $\langle ... \rangle$ denotes an ensemble average. In simple single-field models of inflation, $f_{\textrm{NL}}$ vanishes. Therefore, any non-zero $f_{\textrm{NL}}$ detection would disrupt our understanding of the early Universe and how the seeds of structure formation were generated. The galaxy power spectrum probes local PNG through a scale-dependent bias feature, which appears so in the galaxy bias expansion:

\begin{equation}
\begin{split}
    \delta_g(\bm{x},z)^{\textrm{LO}} &= b_1(z) \delta_{\textrm{m}}(\bm{x},z) + b_{\phi}(z)f_{\mathrm{NL}}\phi(\bm{x})+ \epsilon(\bm{x})\ , \\
    &\mathrm{where} \quad \delta_{\textrm{m}}(\bm{k},z)=\mathcal{M}(k,z) \phi(\bm{k}) \quad \mathrm{and} \quad
    \mathcal{M}(k,z)= \frac{2}{3} \frac{k^2 T_{\textrm{m}}(k,z)}{\Omega_{\textrm{M}}H_0^2}.
    \end{split}
    \label{eq:bphi_initialdef}
\end{equation}

Here, $\delta_{\textrm{m}}(\bm{x},z)$ is the dark matter density perturbations, and $\epsilon(\bm{x})$ encapsulates the stochasticity describing random modulations in the galaxy density due to statistically uncorrelated short-scale modes \citep{Assassi2015}. Additionally, $T_{\textrm{M}}$ is the matter transfer function, $\Omega_{\textrm{M}}=0.3$ is the fractional matter density measured today, and $H_0$ is the Hubble expansion rate measured today. Finally, note that by definition $\delta_g(\bm{x},z) \equiv n_g(\bm{x},z)/ \bar{n}_g(z) -1$, where $n_g(\bm{x},z)$ is the local galaxy density of galaxies, and $\bar{n}_g(z)$ is the mean comoving number density of galaxies, and $\delta_g(\bm{x},z)$ measures the density contrast of galaxies at some redshift.

In this notation, $b_1$ is the linear Eulerian bias, and $b_{\phi}$ describes the response of the galaxy distribution to non-zero $f_{\textrm{NL}}$ induced by PNG (see \citealt{Desjacques2018} for a detailed review). The bias parameters describe the response of galaxy number counts to long-wavelength perturbations of $\delta_{\textrm{m}}(\bm{x},z)$ for $b_1$, and $\phi(\bm{x})$ perturbations with local PNG for $b_{\phi}$.
With the leading-order galaxy bias model of Eq.\ \ref{eq:bphi_initialdef}, the galaxy power spectrum can be written as $(2\pi)^3 P_{gg}(k,z) \delta_{\text{D}} (\bm{k} - \bm{k}') = \langle \delta(\bm{k},z) \delta(\bm{k}',z) \rangle$, where $\delta_{\text{D}}$ is the Dirac delta function, and with:

\begin{equation}
    \begin{split}
    P_{gg}(k,z)^{\textrm{LO}} &= b_1^2P_{\textrm{mm}}(k,z) + 2b_1b_{\phi}f_{\mathrm{NL}}
P_{\textrm{m}\phi}(k,z)+ b_{\phi}^2 f_{\mathrm{NL}}^2 P_{\phi \phi}(k) + P_{\epsilon \epsilon}(k) \\
    &= \Bigg[ b_1^2 + \frac{2b_1 b_{\phi}f_{\mathrm{NL}}}{\mathcal{M}(k,z)} + \frac{b_{\phi}^2f_{\mathrm{NL}}^2}{\mathcal{M}(k,z)^2} \Bigg] P_{\textrm{mm}}(k,z) + P_{\epsilon \epsilon}.
    \end{split}
\end{equation}

Here, $P_{ab}$ is the cross-power spectrum of fields $a,b$ (with m 
'$\equiv \delta_{\textrm{m}}$)
% and $g\equiv \delta_g$)
; $P_{\epsilon \epsilon}$ is the $k$-independent power spectrum of the noise. Note that we begin to omit the redshift dependence of the bias parameters. On scales $k < 0.1\ h$ Mpc$^{-1}$, the matter transfer function is scale-independent, meaning $f_{\mathrm{NL}}$ induces scale-dependent corrections $\propto b_1 b_{\phi}f_{\mathrm{NL}} k^{-2}$ and $b_{\phi}^2f_{\mathrm{NL}}^2 k^{-4}$ relative to $P_{\textrm{mm}}$. This can put bounds on $f_{\mathrm{NL}}$, but its effect is completely degenerate with $b_{\phi}$ and partially degenerate with $b_{1}$. Therefore, constraints on $f_{\mathrm{NL}}$ from the galaxy distribution depend critically on our knowledge of $b_{\phi}$.

The `separate-universe' technique \citep{Lemaitre1933, BarrowSaich1993, Cole1997} takes advantage of the equivalence between the response of galaxy formation to long-wavelength perturbations and its response to changes in the background cosmology, in order to efficiently predict galaxy bias.  If one assumes the physics of galaxy formation operates on sufficiently small scales relative to the size of the perturbations, then the perturbations effectively act as a modified background to the galaxies forming on those small scales (or, the peak background split argument; \citealt{Kaiser1984, Bardeen1986, ColeKaiser1989, Slosar2008}). One can then invoke the separate-universe argument: local structure formation inside long-wavelength perturbations in a fiducial cosmology is equivalent to global structure formation at cosmic mean in an appropriately modified cosmology. For $b_{\phi}$, the modified cosmology uses a different amplitude of the primordial scalar power spectrum $\mathcal{A}_s$, or its low-redshift parameterization $\sigma_8$.

The use of separate-universe simulations for local PNG is recent (e.g. \citealt{Dai2015a}), and our application for galaxy bias analyses was first explained in \citet{Barreira+2020_TNG}. In summary, local PNG induces a non-vanishing bispectrum in the primordial graviational potential $\phi(\bm{x})$, which peaks when long-wavelength perturbations couple with the power spectrum of two small-scale modes. In practice, the long-wavelength perturbations of $\phi(\bm{x})$ cause a modulation in the amplitude of the small-scale primordial scalar power spectrum. These modulations affect subsequent formation of structure inside the perturbations. Specifically, the primordial potential power spectrum evaluated locally around some $\bm{x}$ can be written as:

\begin{equation}
    P_{\phi \phi}(k_{\mathrm{short}}, z|\bm{x}) = P_{\phi \phi}(k_{\mathrm{short}}, z)) [1+4f_{\mathrm{NL}}\phi(\bm{x})]
\end{equation}
where $k_{\mathrm{short}}$ reminds us that these are small-scale modes compared to the wavelength of the $\phi(\bm{x})$ perturbation. A sufficiently long-wavelength perturbation in sensed by galaxies as a spatially uniform change to the variance of the small scale fluctuations in which they form. In other words, galaxies form as if in a separate cosmology with perfectly Gaussian initial conditions, but a modified amplitude of the primordial power spectrum. If $\phi_L$ is the amplitude of the long-wavelength perturbation of the potential, the modified amplitude is:

\begin{equation}
    \Tilde{\mathcal{A}}_s = \mathcal{A}_s[1+\delta \mathcal{A}_s], \quad \mathrm{with} \quad \delta \mathcal{A}_s=4f_{\mathrm{NL}}\phi_{L}
\end{equation}

One can assume the `universality' of the halo mass function, where the mass of halos depends only on the height of the density peak as $\nu=\delta_c / \sigma(M)$, with $\delta_c$ as the critical overdensity and $\sigma^2(M)$ as the variance of the density field smoothed at mass $M$. 
Explicitly, $b_{\phi}$ is a non-Gaussian bias parameter that can be calculated as the derivative of the halo mass function with respect to the power spectrum amplitude. Expressing the density field in terms of $\mathcal{A}_s$, one can calculate $b_{\phi}$ as: 

\begin{equation}
    b_{\phi}(z) = 4 \frac{\textrm{d\ ln}n_g (z)}{\textrm{d}\delta \mathcal{A}_s} \Bigg|_{\delta \mathcal{A}_s=0}.
    \label{eq:bphi_exact}
\end{equation}

We can evaluate this by first-order finite-differencing between our fiducial, high $\mathcal{A}_s$, and low $\mathcal{A}_s$ simulations. For a given galaxy sample \textbf{S}, $b_{\phi}$ and its error\footnote{Here we follow the convention \citet{Barreira+2020_TNG} does for their bias measurements. They found that simply using Poisson errors in each bin is approximately a factor of 2 overestimate of the error given the separate universe simulations share the same initial conditions; see their \textsection 2.2 for their complete justification.} is defined so:

\begin{equation}
    b_{\phi} (z, \textbf{S}) = \frac{b_{\phi}^{high} (z, \textbf{S}) + b_{\phi}^{low} (z, \textbf{S})}{2},
    \quad \textrm{with error} \quad
    \sigma_{b_\phi} (z, \textbf{S}) = \frac{|b_{\phi}^{high} (z, \textbf{S}) - b_{\phi}^{low} (z, \textbf{S})|}{2},
\label{eq:bphi_def_error}
\end{equation}

In terms of $\mathcal{A}_s$, $b_{\phi}^{high}$ and $b_{\phi}^{low}$ are, for a given sample  \textbf{S}:

\begin{equation}
    b_{\phi}^{\text{high}} (z, \textbf{S}) = \frac{4}{\delta_{\mathcal{A}_s}^{\text{high}}} \Biggl[ \frac{N_g^{\text{high}\ \mathcal{A}_s}(z, \textbf{S})}{N_g^{\text{fid}\ \mathcal{A}_s}(z, \textbf{S})} -1 \Biggr] , \quad
    b_{\phi}^{\text{low}} (z, \textbf{S}) = \frac{4}{\delta_{\mathcal{A}_s}^{\text{low}}} \Biggl[ \frac{N_g^{\text{low}\ \mathcal{A}_s}(z, \textbf{S})}{N_g^{\text{fid}\ \mathcal{A}_s}(z, \textbf{S})} -1 \Biggr]
\label{eq:bphi_high_low_As}
\end{equation}

Here, $N_g(z, \textbf{S})$ denotes the number of galaxies at some redshift $z$ within the selection \textbf{S}, and is calculated at each $\mathcal{A}_s$ simulation. We emphasize that the selections \textbf{S} we address in this work are bins galaxy properties, not thresholds.  Additionally, 

\begin{equation}
 \delta_{\mathcal{A}_s}=\frac{\mathcal{A}_{s,\text{larger}}}{\mathcal{A}_{s,\text{smaller}}}-1, \quad \text{so that, e.g.} \quad \delta_{\mathcal{A}_s}^{\text{low}} = \frac{\mathcal{A}_{s,\text{fid}}}{\mathcal{A}_{s,\text{low}}}-1
\label{eq:bphi_high_low_As2}
\end{equation}

However, given that the initial conditions we use for our \texttt{AREPO} N-body simulations are initialized for a given $\sigma_8$, it is straightforward to calculate $b_{\phi}^{\text{high/low}} (z, \textbf{S})$ directly in terms of $\sigma_8$. By definition, $\sigma_8 = \int dk\ P_{\mathrm{lin}}(k)\ W^2(k= 8\ \mathrm{Mpc} h^{-1})$, where $W$ is the Fourier Transform of a top-hat, therefore giving the linear $P_k$ on 8\ Mpc $h^{-1}$ scales. Since $P_k \sim \mathcal{A}_s$, then $\sigma_8 \sim \sqrt{\mathcal{A}_s}$ and $\delta\sigma_8/\sigma_8 = \delta \mathcal{A}_s/ 2\mathcal{A}_s$, giving:

\begin{equation}
    b_{\phi}^{\text{high}} (z, \textbf{S}) = \frac{2}{\delta_{\sigma_8}^{\text{high}}} \Biggl[ \frac{N_g^{\text{high}}(z, \textbf{S})}{N_g^{\text{fid}}(z, \textbf{S})} -1 \Biggr] , \quad
    b_{\phi}^{\text{low}} (z, \textbf{S}) = \frac{2}{\delta_{\sigma_8}^{\text{low}}} \Biggl[ \frac{N_g^{\text{low}}(z, \textbf{S})}{N_g^{\text{fid}}(z, \textbf{S})} -1 \Biggr] \quad \mathrm{with} \quad \delta_{\sigma_8}=\frac{\sigma_{8,\text{larger}}}{\sigma_{8,\text{smaller}}}-1
\label{eq:bphi_high_low_sig8}
\end{equation}

For our separate-universe simulations, $\delta_{\sigma_8}$ is calculated by comparing either the ``high" cosmology ($\sigma_8=0.88$) against the fiducial cosmology ($\sigma_8=0.8$), or the fiducial against the ``low" cosmology ($\sigma_8=0.72$), for a relative difference $\delta_{\sigma_8}\sim0.1$. We run three N-body \texttt{AREPO} simulations, sharing the same Gaussian random initial conditions\footnote{Computed with \texttt{CAMB} from \citet{CAMB} and evolved using second-order Lagrangian Perturbation code \texttt{2LPTic} documented by R. Scoccimarro here: \url{https://cosmo.nyu.edu/roman/2LPT/}.}, of $L=205\ h^{-1}$ Mpc, $N_{\mathrm{DM}}=1280^3$, and $\Omega_{\textrm{M}}=0.3$, differing only in their $\sigma_8$ values. All were run through the \texttt{ROCKSTAR} halo finder with the demand that each halo include at least 100 particles, and then \texttt{ConsistentTrees} for the merger histories to feed the \texttt{SC-SAM}. The \texttt{SC-SAM} is also run with the same internal random seed for all SAMs, guaranteeing the separate-universe setup.
For figures that do not involve $b_{\phi}$, we use the `fiducial' $\sigma_8=0.8$ simulation trees for all SAMs, meaning the halos and their histories are identical, and any differences that appear come from the \texttt{SC-SAM}'s generation and characterization of galaxies under the given parameters.

We note that the error measurement for $b_{\phi}$ in Equation \ref{eq:bphi_def_error} depends on the number of galaxies in each selection \textbf{S} \citep{Barreira+2020_TNG}. However, \citet{PNGunit} found these same errors in their separate-universe simulations had some resolution dependence, and did not encompass the dispersion in the universality relation the much larger and precise PNG-UNIT simulations measure. Though the separate-universe simulations we test here are both larger and higher resolution than theirs, and therefore may not reflect the same inconsistencies, we still suggest readers take our errors as optimistic. 

\begin{figure*}
 \centering
 \includegraphics[width=\textwidth]{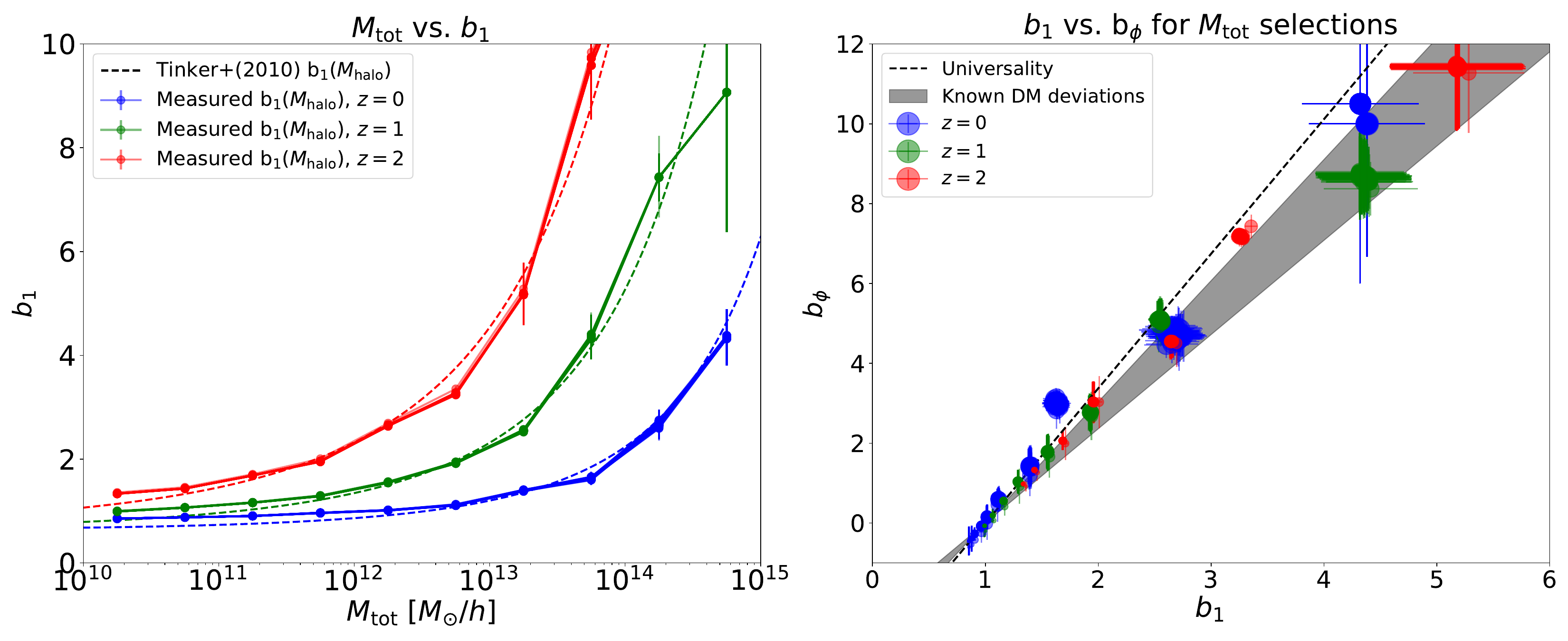}
 \caption{Comparison of halo bias relations  $b_1$ vs.\ $M_{\textrm{tot}}$ (left) and $b_{\phi}$ vs.\ $b_1$ (right) measured in \texttt{SC-SAM} galaxies and results from the literature, validating out separate-universe simulations. Galaxies are selected by \textit{total (halo) mass} across the 56 models at $z=0, 1, 2$, and we measure $b_1$ and $b_{\phi}$ for all bins in all models to create these plots.}
 As expected for N-body simulations, $b_1$ vs.\ $M_{\textrm{tot}}$ is consistent with what the \citet{Tinker2010} model (dashed lines, left), and $b_{\phi}$ vs.\ $b_1$ closely follows the universality relationship for dark matter $b_{\phi}=2 \delta_c (b_1 -1)$ (dashed line, right) or  known deviations from this relationship given halo definitions $b_{\phi}=q \times 2 \delta_c (b_1 -1)$, $q \in [0.7, 0.9]$ (dark grey shading). The size of points in the right plot corresponds to the total mass selection (following those in the left plot), with higher mass selections plotted as larger and proceeding in dex-bins of 0.5.
 \label{fig:totDMmass_bphib1}
\end{figure*}

%---------------------------%
%---------------------------%
%---------------------------%

\subsection{Computation of $b_1$ and the universality relation for $b_{\phi}(b_1)$} \label{subsec:b1calcmethod}

We calculate $b_1$ directly with the galaxy-matter power spectra for each sample selection. We emphasize that we have \textit{not} carried out separate-universe simulations to measure $b_1$ (as e.g.\ \citealt{Barreira+2020_TNG} did by varying $\Omega_{\textrm{M}}$). This choice makes our $b_{\phi}$ calculations and $b_{\phi}$ vs.\ $b_1$ comparisons more robust within each sample. For a given selection \textbf{S}, and the matter power spectrum $P_{mm}$ and galaxy-matter cross power spectrum $P_{gm}$, we solve for:

\begin{equation}
    \frac{P_{gm}}{P_{mm}} = b_1 + A_1 k^2.
\label{eq:b1calc}
\end{equation}

Linear theory gives that, to leading order, $P_{gm}/P_{mm} \approx b_1$, and higher order corrections beyond that scale roughly by $k^2$. The approximation in Eq. \ref{eq:b1calc} is accurate as $k \rightarrow 0$, and we confirm that our choice of $k_{\textrm{max}} = 0.15\ \mathrm{Mpc}\ h^{-1}$ does not affect $b_1$ beyond our listed errors for our simulations. We take $A_1$ as a free parameter for these higher order corrections and set it aside after fitting for $b_1$. Both $P_{gm}$ and $P_{mm}$ have their shot noise subtracted away before fitting, and we fit\footnote{We use \texttt{Pylians} \citep{Pylians} to calculate the power spectra from 3D density fields created with an $N=64$ mesh and TSC mass-assignment scheme.} \textcolor{red}{Eq.
\ref{eq:b1calc}} where $P_{gm}>0$ and $k \leq 0.15\ \mathrm{Mpc}\ h^{-1}$. We use non-linear least squares\footnote{\textit{scipy.optimize.curve\_fit}} to fit Eq. \ref{eq:b1calc}, and take the $1\sigma$ error on $b_1$ from the diagonal of the covariance reported. 

When examining the relationship between $b_1$ and $b_{\phi}$, it is easiest to start with the theoretical relationship for dark matter halos, called the \textit{universality relation}. 
\citet{Slosar2008} derived this relationship under the peak background split argument and the extended Press-Schechter formalism, finding it will describe objects whose halo occupation depends only on halo mass (i.e.\ objects with no assembly bias).
The exact description of $b_{\phi}$ in Eq \ref{eq:bphi_exact} simplifies under a barrier crossing model (with barrier height $\delta_c$, the threshold overdensity for spherical collapse; \citealt{McDonald2008}, \citealt{Ferraro2015}) so:

\begin{equation}
    b_{\phi} = 2\delta_c (b_1 - p), \quad p=1\ \mathrm{for\  objects\ with\ no\ assembly\ bias,\ 1.6\ for\ recent\ mergers}
\label{eq:universality}
\end{equation}

Though this describes N-body simulations to within 10$\%$, several works have shown that this universality relation can overestimate the $b_{\phi}(b_1)$ slope for simulated DM halos (e.g.\ \citealt{WagnerVerde2012, Biagetti2017, Lazeyras2023, PNGunit, Hadzhiyska2024, Lucie-Smith2023, DalalPercival2025, SullivanChen2025, Shiveshwarkar2026}).
\citet{Barreira+2020_TNG} summarized these known variations to universality by halos so:

\begin{equation}
    b_{\phi} = q \times 2\delta_c (b_1 - 1), \quad q \in [0.7, 0.9].
\label{eq:universalityDMvars}
\end{equation}

In Figure \ref{fig:totDMmass_bphib1}, we confirm that the behavior of $b_1$ and $b_{\phi}(b_1)$ for our simulations and \texttt{SC-SAM} catalogs is as expected. In the left panel, we show our measured $b_1$ for total (halo) mass selections is consistent with what the \citet{Tinker2010} model predicts, across all redshifts we use.
In the right panel, we show the $b_{\phi}(b_1)$ for the same total (halo) mass-selected samples across all our \texttt{SC-SAM} catalogs, and the true universality relationship and known deviations from it between $0<z<2$. 
Larger points in $b_{\phi}(b_1)$ plots through out this work indicate larger mass selections, and we do not connect points from the same models to keep the plots easier to interpret.
Fig.\ \ref{fig:totDMmass_bphib1} confirms that all our \texttt{SC-SAM} variations are consistent with the universality relationship, and measured deviations (Eq. \ref{eq:universalityDMvars}). As the \texttt{SC-SAM} builds galaxies starting from the information in the dark matter merger trees, varying the astrophysical parameters does not affect the distribution of the underlying dark matter halos. 

%---------------------------%
%---------------------------%
%---------------------------%
%---------------------------%
%---------------------------%
%---------------------------%

\section{The \texttt{SC-SAM} for $\lowercase{b}_{\phi}$} \label{sec:SCSAMbphi}

\subsection{The Santa Cruz Semi-Analytic Model of Galaxy Formation} \label{subsec:SCSAMintro}

Semi-analytic models (SAMs) of galaxy formation are embedded within a dark matter halo merger tree that represents how halos grow and merge in the context of a chosen $\Lambda$CDM cosmology. 
Within the merger tree framework, SAMs compute how mass and metals move between different `reservoirs' using a set of coupled ordinary differential equations. These `reservoirs' correspond to e.g.\ intergalactic medium (IGM), circumgalactic medium (CGM), interstellar medium (ISM), stars, etc. For example, nearly all SAMs assume that gas accretes into the CGM at a rate proportional to the growth rate of the dark matter halo. Most SAMs also adopt a cooling model for how rapidly gas in the CGM cools and accretes into the ISM, often following \citet{WhiteFrenk1991}. 

The \texttt{SC-SAM} used here is the one used to create the CAMELS-SAM suite \citep{CAMELS-SAM}, but run atop the separate-universe simulations described at the end of \textsection \ref{subsec:SepUni}. This version of the \texttt{SC-SAM} is similar to that in \citet{Somerville2008} and \citet{Somerville2015}, with small updates described in \citet{Gabrielpillai2022}. We also direct readers to other descriptions of the \texttt{SC-SAM} and results based on it, in \citet{Porter2014a}, as well as recent mock observations made in \citet{Somerville2021} and \citet{Yung_i}.

The SAM approach offers flexibility and speed over hydrodynamic simulations, while incorporating more physical understanding than empirical models (e.g.\ halo occupation distribution, most subhalo-abundance matching, models like \texttt{UNIVERSEMACHINE}, \citealt{UNIVERSEMACHINE}).
The main differences between SAMs lie in the specific equations and how they are computed, and parameters and functions that are adopted to describe star formation, stellar feedback, and black hole growth and AGN feedback (e.g.\  \citealt{Lu2014,SomervilleRomeel2015, Knebe2018}). The overall takeaway from these studies comparing the implementations and predictions of SAMs is that, in spite of the many differences in implementation and choice of physical recipes, most SAMs produce similar predictions for key quantities such as the stellar mass function, suggesting a similar stellar to halo mass relationship \citep{SomervilleRomeel2015}. 

A more complete description of the variations of the \texttt{SC-SAM} parameters can be found in \citet{CAMELS-SAM}. We briefly summarize the relevant parts here. The \texttt{SC-SAM} implementation of star formation is similar to those implemented in many other SAMs, differing in details. 
The version of the \texttt{SC-SAM} used here partitions gas into different phases (ionized, molecular, and atomic), and its adopted star formation relation uses only the molecular gas density (rather than total gas density). However, \citet{Somerville2015} showed that this has a rather minor effect on most predictions of the \texttt{SC-SAM} compared to others. 
The \texttt{SC-SAM} prescription gives the mass outflow rate from stellar driven winds and supernovae according to the galaxy's SFR and the halo's circular velocity:

\begin{equation}
  \dot{m}_{\mathrm{out}}=\epsilon_{\mathrm{SN}} \Bigg( \frac{V_0}{V_c} \Bigg)^{\alpha_{\mathrm{rh}}} \dot{m}_*
\label{eq:SCSAM_mdotout}
\end{equation}

In this prescription, $\dot{m}_*$ is the SFR; $V_0$ is a normalization constant set to 200 km s$^{-1}$; and $V_c$ is the maximum circular velocity of the galaxy's disc. $V_c$ is assumed to be the host dark matter halo's maximum circular velocity, which is controlled by the depth of the halo's potential well. The fiducial values of these parameters when the \texttt{SC-SAM} is calibrated to fit several $z=0$ galaxy observables are $\epsilon_{\text{SN}}=1.7$ and $\alpha_{rh}=3.0$ \citep{Gabrielpillai2022}.

The heating of the hot halo gas is assumed to be caused by energetic radio jets from radiatively inefficient accretion onto black holes. The rate of accretion onto the black hole from the hot halo is given by:

\begin{equation}
  \dot{m}_{\mathrm{radio}}=\kappa_{\mathrm{radio}} \Bigg[ \frac{kT}{\Lambda [T,Z_h]} \Bigg] \Bigg( \frac{M_{\mathrm{BH}}}{10^8 M_{\odot}} \Bigg)
\label{eq:SCSAM_mdotradio}
\end{equation}
 
Here, $kT$ is the temperature of gas within the Bondi accretion radius (approximated as $r_A \equiv 2GM_{\text{BH}}/c_s^2$, and with $T$ as the virial temperature of the gas in the halo); and $\Lambda[T,Z_h]$ is the temperature- and metallicity-dependent atomic cooling function \citep{S-D1993}. This radio mode accretion heats the hot halo gas at a rate that is proportional to $\dot{m}_{\mathrm{radio}}$, and can partially or completely offset cooling and accretion into the ISM. This allows the SAM to control the strength of the feedback from the jet mode by varying the parameter $\kappa_{\mathrm{radio}}$. The AGN feedback mainly affects the most massive galaxies (\citealp{Somerville2008} and Appendix A in \citealt{CAMELS-SAM}). The treatment of black hole growth and AGN feedback in the \texttt{SC-SAM} is also somewhat different from the implementation in other SAMs, but this also turns out to have a rather minor effect on most galaxy properties. The fiducial value is  $\kappa_{\mathrm{radio}}=0.002$ \citep{Somerville2008, CAMELS-SAM}, and controls the rate of accretion onto the black hole and resulting radio mode BH feedback.

The values of these and other parameters in the fiducial \texttt{SC-SAM} model were selected by tuning the model``by hand'' to reproduce a set of key $z=0$ relationships (\citealt{Somerville2008, Somerville2015, Somerville2021, Yung_i}), such as the stellar mass function, cold gas fraction, mass-metallicity relation for stars, and black hole mass vs.\ bulge mass relation (\citealt{Bernardi2013, Moustakas2013, Baldry2012, Rodriguez-Puebla2017, Catinella2018, Calette2018, Gallazzi2005, Kirby2011, McConnell2013, Kormendy2013}). See \citet{Somerville2008} and \citet{Somerville2015} for more details on the ``by hand'' calibrations.

In summary, the \texttt{SC-SAM} is implemented for 55\footnote{The number of models is a somewhat arbitrary, and chosen to balance the cost of $b_1$($b_{\phi}$) computations while still mostly uniformly sampling the CAMELS-SAM latin hypercube. 55 was the nearest round number of CAMELS-LH simulations in our parameter range, and all are contained within the first 75 simulations of the original latin hypercube.} varied models so:

\begin{equation}
    \begin{split}
    \dot{m}_{\mathrm{out}} = (\epsilon_{\mathrm{SN}} \times \mathsf{SN1})\ \dot{m}_*\ \Bigg( \frac{V_0}{V_c} \Bigg)^{(\alpha_{\mathrm{rh}}+\mathsf{SN2})}
    % \mathrm{A}_{\mathrm{SN2}})} 
    \quad \mathrm{for}\ \mathsf{SN1}=[\frac{1}{4}, 4]\ \mathrm{and}\ \mathsf{SN2}=[-2, 2], \\
    % &  
    \dot{m}_{\mathrm{radio}} = (\kappa_{\mathrm{radio}} \times \mathsf{AGN}) \Bigg[ \frac{kT}{\Lambda [T,Z_h]} \Bigg] \Bigg( \frac{M_{\mathrm{BH}}}{10^8 M_{\odot}} \Bigg) \quad \mathrm{for}\ \mathsf{AGN}=[\frac{1}{4}, 1.8].
    \end{split}
    \label{eq:summary_params}
\end{equation}

%---------------------------%
%---------------------------%
%---------------------------%

\subsection{One-parameter-at-a-time \texttt{SC-SAM} models} \label{subsec:SCSAM1P}

\begin{figure}
    \centering
    \includegraphics[width=\linewidth]{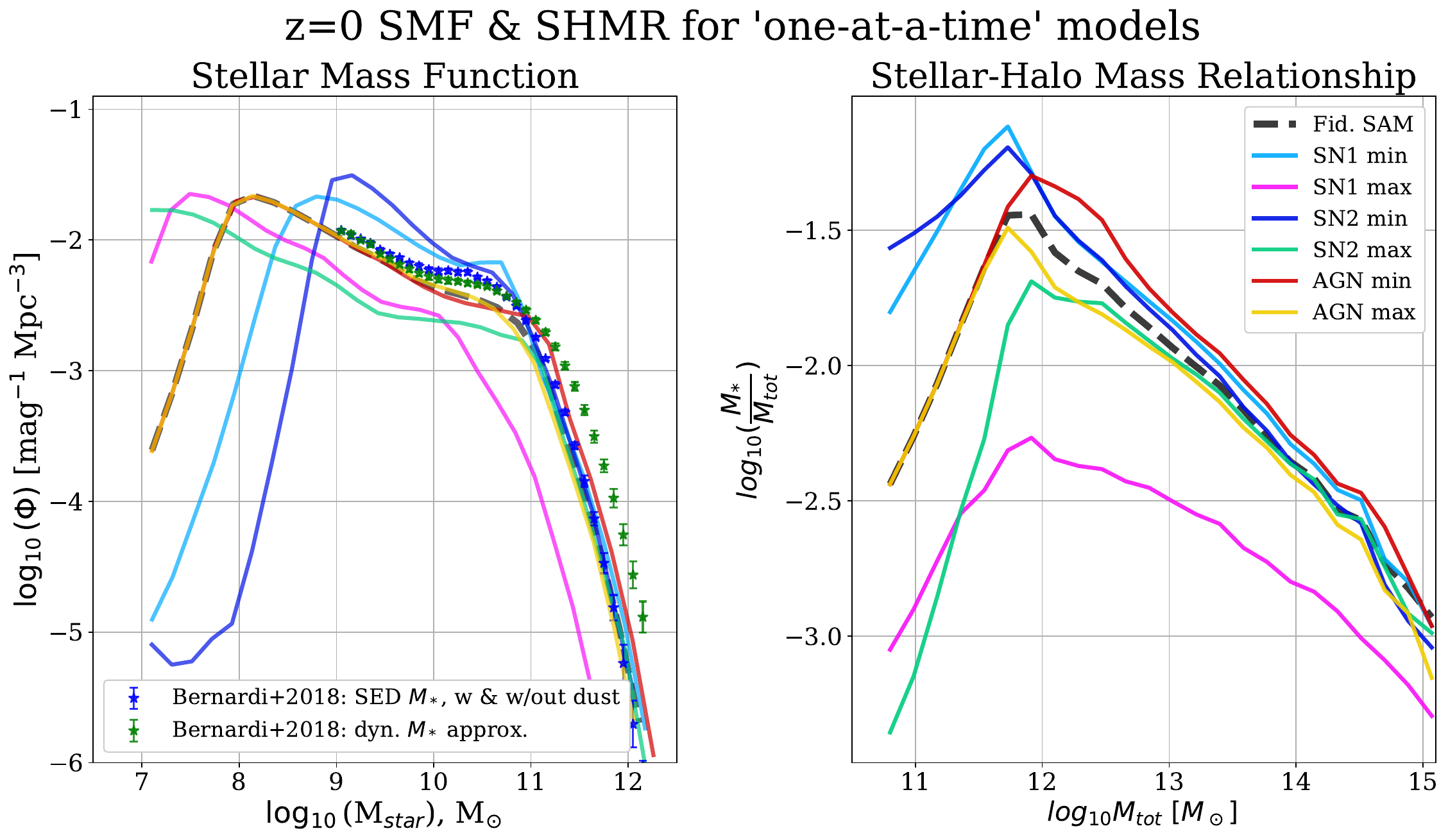}
    \caption{The $z=0$ stellar mass function (left) and stellar-halo mass relationship (right) for the `one-at-a-time' \texttt{SC-SAM} models, showing the effect of the minimum and maximum of each parameter with all others held constant (including cosmology). The fiducial model is plotted thick, dotted grey line. The $z=0$ SMF observations used for assessing `realism' are shown in blue and green stars. \textsf{AGN} affects the just the high mass end of the SMF, but raises/lowers the SHMR at its minimum/maximum, respectively. The two A$_{\text{SN}}$ parameters strongly warp the SMF and SHMR in unique ways. }
    \label{fig:1P_SMFs}
\end{figure}

To help build intuition around the galaxy-halo connection and galaxy properties the \texttt{SC-SAM} predicts, we compare \texttt{SC-SAM} models where the parameters have been varied one-at-a-time to their minimum and maximum values. Figure \ref{fig:1P_SMFs} shows the $z=0$ SMFs (left) and SHMRs (right) for these models run on the same merger trees of our central fiducial separate-universe simulation with $\Omega_{\textrm{M}}=0.3$ and $\sigma_8 = 0.8$. The minimum/maximum of \textsf{SN1}, the amplitude of stellar-driven mass outflow in Eq. \ref{eq:SCSAM_mdotout}, is colored in cyan/magenta respectively. The minimum/maximum of \textsf{SN2}, the exponent in the same Eq. \ref{eq:SCSAM_mdotout}, is colored in blue/teal respectively. Finally, the minimum/maximum of \textsf{AGN}, the amplitude of the rate of accretion leading to the radio jet mode of the supermassive black hole in Eq. \ref{eq:SCSAM_mdotradio}, is colored in red/gold respectively. For comparison, the fiducial \texttt{SC-SAM} model (featured in \citealt{CAMELS-SAM, Gabrielpillai2022}) is shown in black. 

Each parameter changes the properties of the galaxies placed into the halos given the modeling of feedback in the \texttt{SC-SAM}, changing the stellar-halo mass relationship. The observed SMFs, the left subplot in Fig.\ \ref{fig:1P_SMFs}, are an observable consequence of this\footnote{The left-hand drop in the SMF results from the limiting resolution of the smallest halos in the merger tree, and then the stellar mass those halos achieve given a model's SHMR.  Our implementation of \texttt{ROCKSTAR} begins to track halos with at least 100 dark matter particles, for an approximate mass of $3.4\times 10^{10}$ M$_{\odot}$. However, we restrict future plots and analyses to only use galaxies in halos of at least 150 particles (roughly log$_{10}$M$_{H}>10.7$). }. 
For example, the effect of \textsf{AGN} closely tracks the fiducial model until the high-mass turnover near M$_{\textrm{halo}}> 10^{11.9}$ M$_{\odot}$. The maximum \textsf{AGN} increases the accretion rate of hot gas onto the hot halo, and further powers the radio jets: the resulting galaxies are lower mass, driving the right side of the SHMR down sharply, and slightly suppressing the high-mass end of the SMF. Conversely, the minimum \textsf{AGN} allows massive galaxies to better hold onto their hot gas, and the massive edge of the SMF to reach higher masses. The SN feedback parameters strongly affect the full range of the SHMR and the SMF in unique ways. The minimum of either parameter decreases the mass outflow from a galaxy caused by SNe and stellar winds, meaning galaxies of all masses are able to grow more massive (yielding a higher SHMR, especially at lower halo masses, and an SMF notably higher and shifted to higher masses). Inversely, their maximum suppresses the stellar masses across small halo masses and lowers the SMF. \textsf{SN1} and \textsf{SN2} differ in the magnitude and shape of their effect of the SHMR.
However, the maximum \textsf{SN1} sharply suppresses the entire SHMR (rather than just the low-mass left edge) and therefore SMF. 

\begin{figure}
    \centering
    \includegraphics[width=\linewidth]{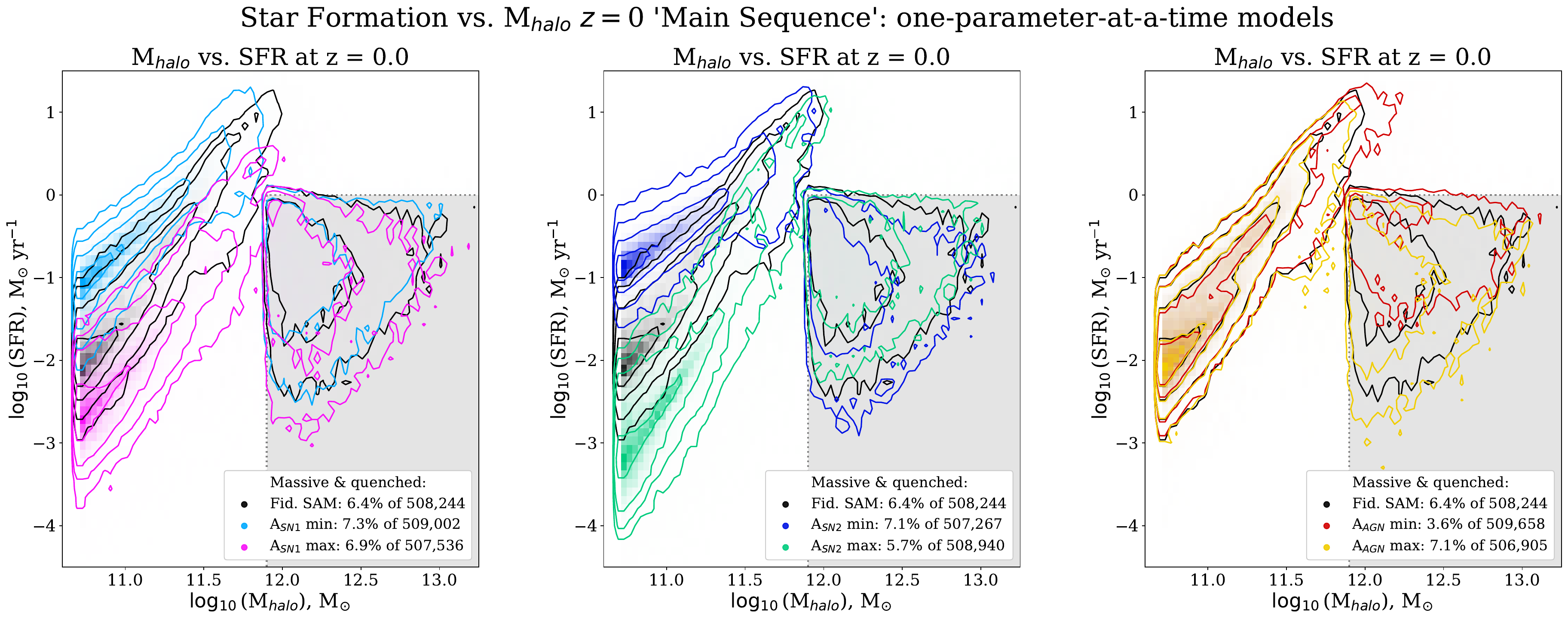}
    \includegraphics[width=\linewidth]{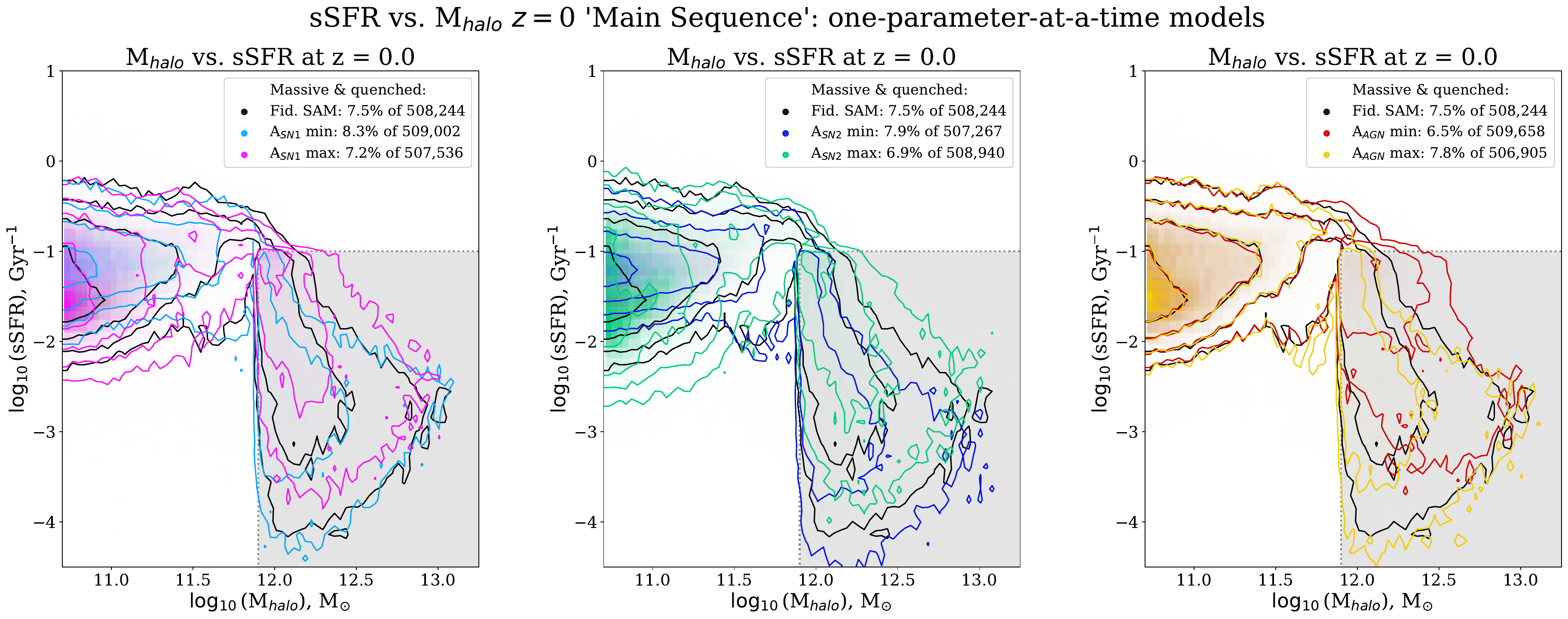}
    \caption{The $z=0$ instantaneous SFR vs.\ M$_{\textrm{halo}}$ (upper) and specific SFR vs.\ M$_{\textrm{halo}}$ (lower) `main sequences' for central galaxies in the `one-at-a-time' \texttt{SC-SAM} models. (See $z=1$ and $z=2$ in Appendix \ref{app:bphi_zevol}.) The fiducial \texttt{SC-SAM} is plotted in black alongside the other parameter pairs: \textsf{SN1} min/max in cyan/magenta (left), \textsf{SN2} min/max in blue/teal (center), and \textsf{AGN} min/max in red/gold (right). For emphasis, we highlight a region of massive and quenched galaxies with M$_{\textrm{halo}}> 10^{11.9}$ M$_{\odot}$ and SFR $< 0$ M$_{\odot}$ yr$^{-1}$ or sSFR $< -1$ Gyr$^{-1}$. 
    The contours show the 2D galaxy number density enclosing approximately [12, 39, 68, 87]\% of the sample (2D Gaussian equivalent of 0.5, 1, 1.5, and 2$\sigma$); the contours include all galaxies in the approximate `massive and quenched' corner and the first 70,000 galaxies of each model (for figure legibility). 
    The legends indicate the total number of central galaxies in each model, and what percentage of galaxies are in the grey `massive and quenched' regions. 
    All models share the same fundamental shape in their sSFR vs.\ M$_{*}$ distribution by design in the \texttt{SC-SAM}, but differences in their stellar-halo mass relationships change their SFR vs.\ M$_{\textrm{halo}}$ relationship. }
    \label{fig:1P_sSFHMRs}
\end{figure}

Next, we can also examine how the galaxy-halo relationship affects star formation for these `one-at-a-time' parameter extremes. We first note that all \texttt{SC-SAM} models have very similar instantaneous SFR vs.\ stellar mass relationships (not pictured here due to their similarity), as echoed in their SFR vs.\ halo mass relationships in Figure \ref{fig:1P_sSFHMRs} -- all parametrizations follow roughly the same slope for their positively-correlated diagonal `main sequence' and producing massive quenched galaxies in similar stellar masses and SFRs (and therefore similar distributions of sSFRs). This result is quite general, and arises from the self-regulated nature of star formation in the \texttt{SC-SAM}. Figure \ref{fig:1P_sSFHMRs} shows the differing $z=0$ SFR vs.\ halo mass relationship for central galaxies (top), and $z=0$ sSFR vs.\ halo mass relationship for central galaxies (bottom). (See Appendix \ref{app:bphi_zevol} for $z=1$ and $z=2$ version of this Figure.) For legibility, we emphasize the region occupied by massive quenched galaxies (i.e.\ the right-hand edge of the SHMR in Fig. \ref{fig:1P_SMFs}): all galaxies in that region are plotted, and then the first 70,000 galaxies in each model. The left/center/right figures show the minimum and maximum \textsf{SN1}/\textsf{SN2}/\textsf{AGN} alongside the fiducial SAM, respectively. These figures allow one to visualize what halo masses correspond to some (s)SFR selection, and fundamentally explain the $b_{\phi}$ values we measure across our \texttt{SC-SAM} models. Roughly, we categorize a massive quenched galaxy to have log$_{10}$M$_{\text{halo}} > 11.9$ M$_{\odot}$ and log$_{10}$SFR$<0$ M$_{\odot}$ yr$^{-1}$ or log$_{10}$sSFR$<-1$ Gyr$^{-1}$, at any of our target redshifts, and list in the plot legends the rough percentage of galaxies that are massive and quenched in each \texttt{SC-SAM} model.

\textsf{SN1} shifts the (instantaneous) SFR vs.\ M$_{\textrm{halo}}$ up and down, but maintains mostly the same number of massive quenched galaxies. A strong amplitude in Eq. \ref{eq:SCSAM_mdotout} reduces the stellar mass of galaxies, and reduces the material available to power star formation; meaning halos of the same mass will show lower SFRs for higher \textsf{SN1} values. \textsf{SN2} does much of the same, but the maximum \textsf{SN2} reaches even smaller SFRs at the smallest halo masses. This is due to how the maximum \textsf{SN2} model sees a strong positive slope at the low-mass end of the SHMR, meaning the smallest halos have the smallest masses, and consequently the smallest SFRs. The two extremes of \textsf{SN2}, given their strong effect on the SHMR, also show a notable difference in how many massive quenched galaxies they make, where the maximum model makes fewer but stronger star-forming massive galaxies (due to how the maximum model pops slightly above the minimum at the highest halo masses of the SHMR). For \textsf{AGN}, the SFR vs.\ M$_{\textrm{halo}}$ `main sequence' is very similar for both the minimum and maximum, but like the SHMR, strongly diverge past M$_{\textrm{halo}}> 10^{11.9}$ M$_{\odot}$. The minimum \textsf{AGN} parameter yields the highest SHMR and SMFs at the high mass ends, also resulting in more strongly star forming objects, and therefore fewer massive quenched objects than its opposite.

In the specific SFR vs.\ M$_{\textrm{halo}}$ relationships, these behaviors collapse somewhat: the new `main sequence' becomes roughly flat, and the primary differences in the models appear in the massive and quenched corners. The extremes of \textsf{SN1} and \textsf{SN2} slightly shift the flat sSFR vs.\ M$_{\textrm{halo}}<10^{11.9}$ M$_{\odot}$ relationship up and down. However, all models show similar halo mass distributions for a given sSFR `slice'. We now proceed to vary all these \texttt{SC-SAM} parameters together.

%---------------------------%
%---------------------------%
%---------------------------%

\subsection{Variations in \texttt{SC-SAM} models and their `Realism'} \label{subsec:SCSAMmodels}

\begin{figure}
    \centering
    \includegraphics[width=\linewidth]{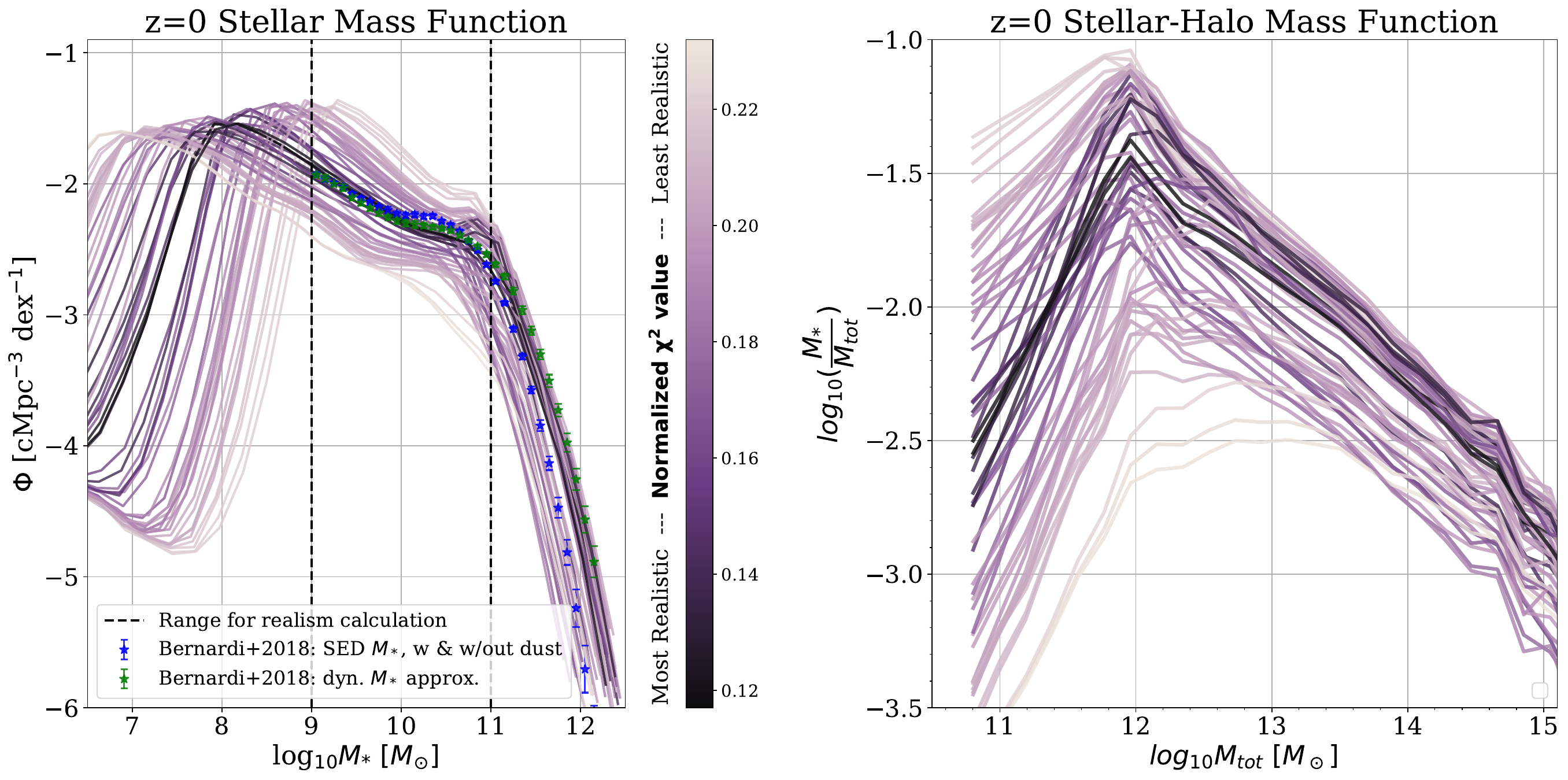}
    \caption{The $z=0$ stellar mass function (SMF, left) and stellar-halo mass relationship (SHMR, right) for the 56 \texttt{SC-SAM} models in this work. Each model is colored by its relative `realism' when compared to the \citet{Bernardi2018} $z=0$ SMF measurements (green and blue) between 9 $< \log M_{*} (M_{\odot}) <$ 11, with least-to-most realistic as beige-to-black. In this work, this `realism' measure is used qualitatively to built intuition; see our partner work, \citet{MoorePerez2026}, for a more precise handling of model realism compared to a target observation when creating priors on $b_{\phi}$.
    }
     \label{fig:SMFnSHMRallmodels}
\end{figure}

Figure \ref{fig:SMFnSHMRallmodels} examines the SMF (left) and predicted SHMR (right) of the 55 unique variations of the \texttt{SC-SAM} and the fiducial model. For this figure, all 56 models were run on the same merger trees from the central fiducial separate-universe simulation with \{$\Omega_{\text{M}}$, $\sigma_8$\}=\{0.3, 0.8\} and $L=205\ h^{-1}$ Mpc. To help contextualize these models in the coming analyses, we assign each model a `realism' $\chi^2$ score according to its agreement with the $z=0$ observed SMF \citep{Bernardi2018}, and color them from most to least `realistic' (black to beige). 

We limit our `realism' comparisons with the SMFs between $9<\textrm{log}_{10}M_{*} <11$ to account for where most \texttt{SC-SAM} models are complete and exclude where inferred SMFs from observations disagree due to modeling choices (dotted vertical lines in Fig. \ref{fig:SMFnSHMRallmodels}). 
In Figure \ref{fig:SMFnSHMRallmodels}, the two $z=0$ SMF calculations of \citet{Bernardi2018} are for the same base observations, but with each using a different method to estimate galaxy mass: the common  spectral energy distribution fitting method (blue, both with and without dust), or a dynamical mass estimation method (green).  Additionally, choices in galaxy profiles and the stellar initial mass function also affect the SMF, particularly beyond the `knee' of $\sim 10^{11}$ M$_{\odot}$ (see \citealt{Bernardi2018}). Though we plot the full SMFs in the left panel of Figure \ref{fig:SMFnSHMRallmodels}, we focus on the initial plateau of $9<\textrm{log}_{10}M_{*} <11$ to assess the realism of an \texttt{SC-SAM} model given these uncertainties in measuring stellar masses from observations.

We compute each model's `realism' according to their $\chi^2$ values compared to the $z=0$ SMFs from \citet[green and blue lines]{Bernardi2018}. For N$_{M_*}=20$ steps of mass between $9<\textrm{log}_{10}M_{*} <11$ in an observed SMF $\Phi_Y$, the $\chi^2$ `realism' factor is calculated between the observed SMF and each SAM model's SMF measurement so:

\begin{equation}
    \chi^2_{Y} = \sum_{\mathrm{N}_{M_*}} \frac{\chi^2 (M_{i})}{\mathrm{N}_{M_*}}, \quad \mathrm{with} \quad \chi^2 (M_{i}) = \frac{\Phi_{Y,i} - \Phi_{\mathrm{model}}(M_{i})}{\sigma_{\mathrm{\Phi_{Y,i}}}^2}.
    \label{eq:realismchi2}
\end{equation}

The data are taken directly from Table B2 of \citet{Bernardi2018}, and a central $\Phi_{Y}$ and $\sigma_{\mathrm{\Phi_Y}}$ is calculated from the upper and lower bounds listed for both $\Phi^{\mathrm{M14}_d}_{\mathrm{SerExp}}$ (blue) and $\Phi^{ \alpha_{\mathrm{JAM}} - \mathrm{M14}_d}_{\mathrm{SerExp}}$ (green). We average the $\chi^2$ measured for each $\Phi_{\mathrm{SerExp}}$ for a final $\chi^2$ value for each \texttt{SC-SAM} model. The `realism' $\chi^2$ values range from 0.1 (e.g.\ for the fiducial SAM calibrated to earlier SMFs in \citealt{Bernardi2013}) to 0.24 (highly warped SMFs). We use these values qualitatively in this work to identify where more `realistic' models live in the $b_{\phi}(b_1)$ plane, but take more robust and detailed approach in our partner work, \citet{MoorePerez2026}. There, we present a framework for building quantitative $b_{\phi}(b_1)$ priors while conditioning on observables like the SMF and SHMR. Briefly, \citet{MoorePerez2026} first learn to emulate a target observable given the parameters varied in the base CAMELS-SAM \citep{CAMELS-SAM} suite of 1000 simulations; then trains an interpolator to predict $b_{\phi}$ for a DESI-ELG-like galaxy sample using the separate-universe simulations we present in this work; and then combines these tools into a galaxy-formation marginalized prior on $b_{\phi}$ for the target sample. For a direct comparison of their approach to our qualitative $\chi^2$ statistic, see their Figure 5: conditioning on just the relative realism of a given model from this work (the thin dotted lines) yields much broader priors than their more advanced method to marginalize over galaxy formation (solid lines). For the purposes of our work here that preceded \citet{MoorePerez2026}, we maintain the relative $\chi^2$ measure to qualitatively guide the interpretation of our $b_{\phi}$ and $b_1(b_{\phi})$ across galaxy formation models.
% but upcoming work with these data will present a framework for building quantitative $b_{\phi}(b_1)$ priors while weighing how closely a galaxy model aligns with a chosen observable (Moore, Perez, \& Krause in prep).

%---------------------------%
%---------------------------%
%---------------------------%
%---------------------------%
%---------------------------%
%---------------------------%

\section{\lowercase{b}$_{\phi}$ across \texttt{SC-SAM} models} \label{sec:bphiresults}

We now will examine how $b_{\phi}$ changes for different variations of the \texttt{SC-SAM} model under different galaxy selections. 
We will primarily present and discuss $z=1$ here, and leave the $z=0$ and $z=2$ plots and the discussion of redshift evolution to Appendix \ref{app:bphi_zevol}. We choose to focus on $z=1$ in this primary text as it best illustrates the pattern that evolves over redshift, helps prepare for the higher redshift samples to be targeted by future surveys, and can be compared directly to other SAM results in \textsection \ref{subsec:TNGnSAMs} and App.\ \ref{app:PNGwSAMsdetails}.

As described in \textsection \ref{subsec:SepUni}, $b_{\phi}$ is calculated for a given galaxy selection. Though we confirm the behavior does not change with binning after tests, throughout this work we plot only one choice for binning galaxies. These are the edges of the bins, with the logscale center plotted when relevant:
\begin{itemize}
    \item Stellar Mass: [8, 8.5, 9, 9.5, 10, 10.5, 11, 11.5, 12, 12.5, 13] log$_{10}$M$_{\odot}$
    \item SFR: [-3, -2.5, -2, -1.5, -1, -0.5, 0, 0.5, 1,  1.5, 2] log$_{10}$M$_{\odot}$yr$^{-1}$
    \item sSFR: [-3, -2.5, -2, -1.5, -1, -0.5, 0, 0.5, 1,  1.5, 2] log$_{10}$Gyr$^{-1}$
\end{itemize}
We plot the $1\sigma$ error bars calculated in Eq. \ref{eq:bphi_def_error}, and exclude bins with fewer than 500 galaxies (where these errors would dominate). Finally, we remind readers that all SAMs are run atop the same three merger trees behind the separate-universes method discussed in \textsection \ref{subsec:SepUni}, and that differences in $b_{\phi}$ result from differences in number counts for a given selection across the different \texttt{SC-SAM} models. We encourage readers to refer to \textsection \ref{subsec:OneataTime} and Figures \ref{fig:1P_SMFs}, \ref{fig:1P_sSFHMRs}, \ref{fig:1P_z0n2_SFHMRs}, and \ref{fig:1P_z0n2_sSFHMRs} for the $z=0,1,2$ stellar mass functions, stellar-halo-mass relatiosnhips, and (specific-)star forming main sequences for the one-at-a-time models, to better understand which halos a particular stellar mass or (s)SFR selection yields and the expected $b_1(b_{\phi})$ (Fig.\ \ref{fig:totDMmass_bphib1}) that selection might yield.

%---------------------------%
%---------------------------%
%---------------------------%

\subsection{Feedback Parameters `one at a time'} \label{subsec:OneataTime}

\begin{figure*}
 \centering
 \includegraphics[width=\textwidth]{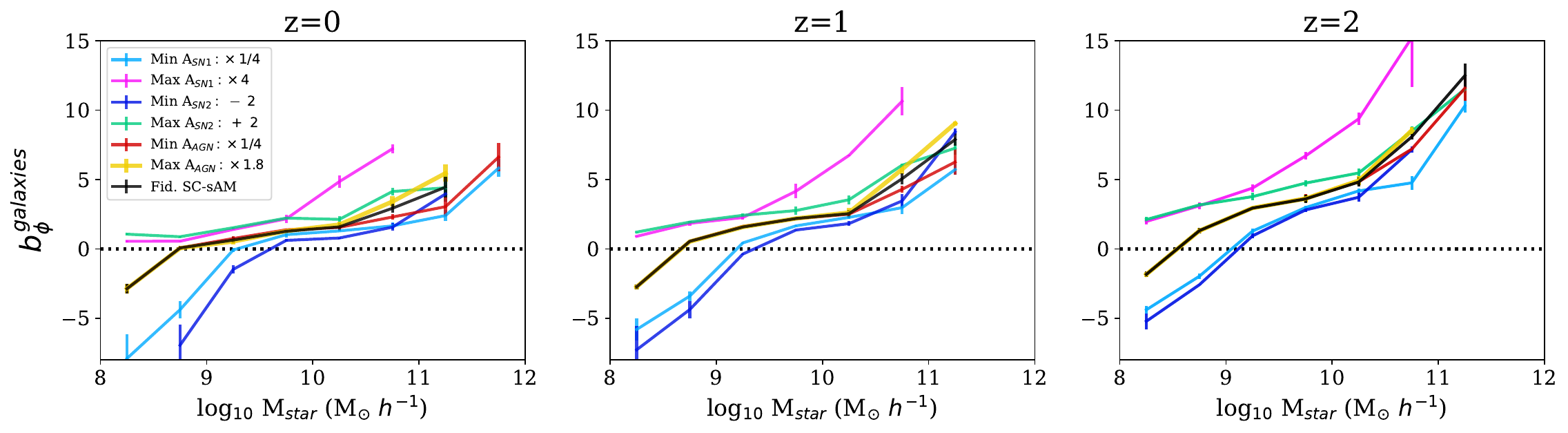}
 \includegraphics[width=\textwidth]{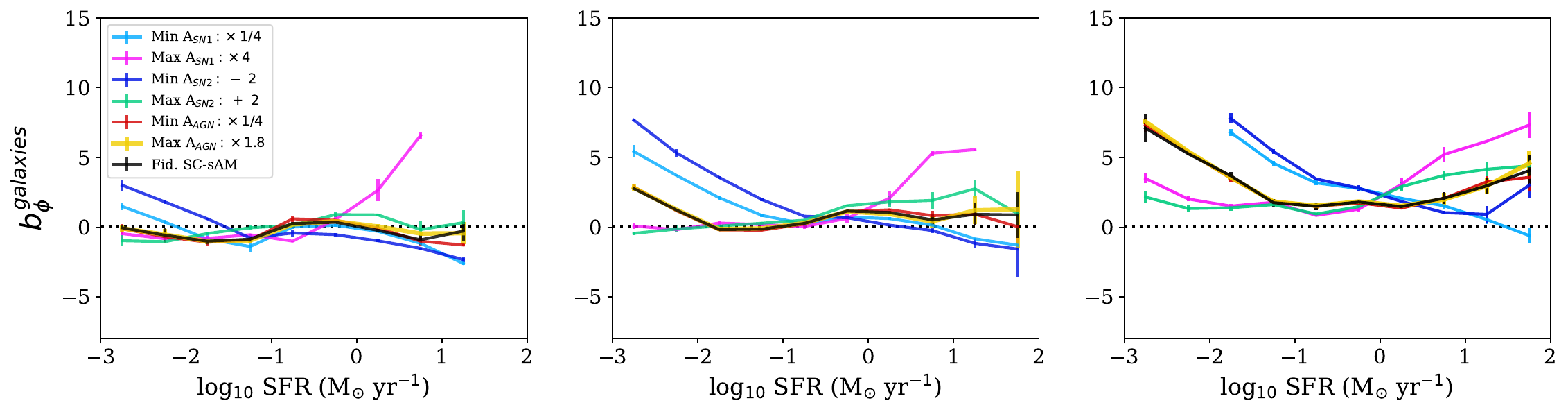}
 \includegraphics[width=\textwidth]{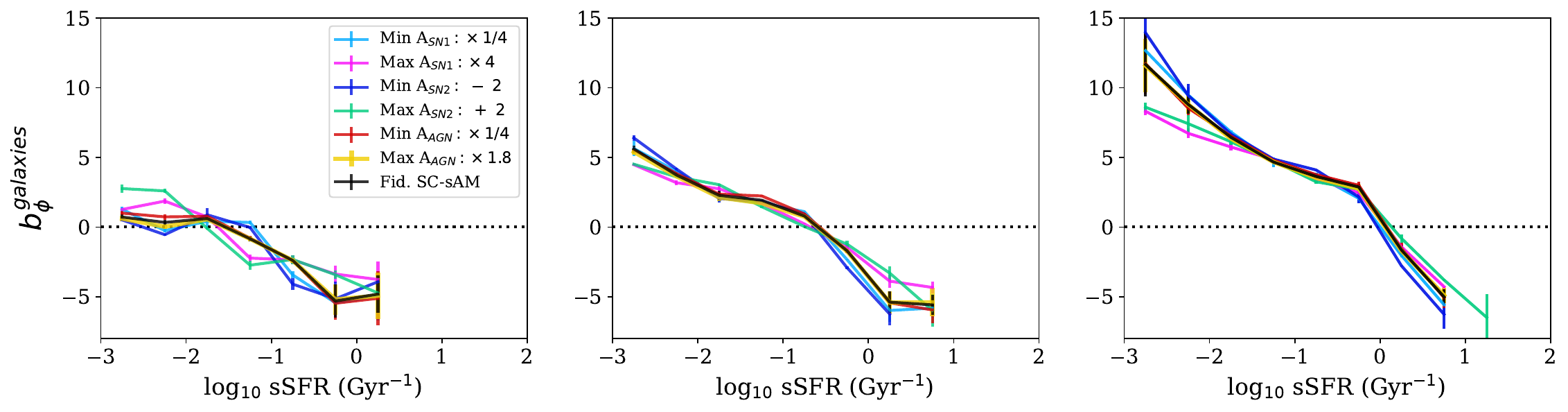}
 \caption{The behavior of $b_{\phi}$ as a function of stellar mass (top row), SFR (middle row), and specific SFR (bottom row) when changing the selected \texttt{SC-SAM} astrophysical parameters one at a time, for $z=0, 1, 2$ (left, middle, right columns respectively). The black curve is the fiducial model, fit to several $z=0$ scaling relations. The colored lines correspond to \texttt{SC-SAM} runs where each parameter was varied to its minimum or maximum while all others were held to the fiducial values. We exclude bins whose selections led to fewer than 500 galaxies, and errors are calculated as described in Eq. \ref{eq:bphi_def_error}. 
 }
 \label{fig:bphi_1Psets}
\end{figure*}

We first examine the dependence of $b_{\phi}$ on our \texttt{SC-SAM} one-(parameter)-at-a-time models under a stellar mass, star formation rate (SFR), or specific SFR selection. 
Figure \ref{fig:bphi_1Psets} shows $b_{\phi}$ vs.\ stellar mass (top), $b_{\phi}$ vs.\ SFR (center), and their $b_{\phi}$  vs.\ specific SFR (bottom) for all parameters' minimum and maximum \texttt{SC-SAM} models. We show here $z=0$ (left), 1 (center), and 2 (right), but in further sections will focus on $z=1$ and discuss redshift evolution in Appendix \ref{app:bphi_zevol}. The fiducial \texttt{SC-SAM} model is plotted in a solid black for comparison, with $b_{\phi}=0$ emphasized as a dotted line. 

The behavior of $b_{\phi}$ with stellar mass is reminiscent of what was found with the IllustrisTNG model in \citet{Barreira+2020_TNG}. All models show $b_{\phi}$ increasing with stellar mass selection, a logical consequence of the tight connection between stellar mass and halo mass, and the similar behavior of increasing $b_{\phi}$ for increasing halo mass (e.g.\ Figure 2 in \citealt{Barreira+2020_TNG}). Each parameter can be thought of as shifting the fiducial $b_{\phi}$ vs.\ M$_{*}$ relationship left or right, corresponding to its effect on the SHMR. When a parameter lifts the right-hand side of the SHMR (corresponding to larger stellar masses for the same halo masses), the $b_{\phi}$ vs.\ M$_{*}$ relation shifts right, towards higher stellar mass selections. 

Most stellar mass selections, for example, correspond to higher mass halos when \textsf{SN1} is higher, leading to higher $b_{\phi}$ values. Models with SHMR `higher' than fiducial on the left edge (i.e.\ which tend to form higher stellar mass galaxies within their smallest halos), such as the minimum \textsf{SN1} and \textsf{SN2}, result in strongly negative $b_{\phi}$ values at the lowest halo massos. 
Finally, \textsf{AGN} aligns with the fiducial model in both the SMF and $b_{\phi}$ values, diverging past log$_{10}$M$_{*}>10.5$, where their SHMRs notably split.

Similarly, the behavior of $b_{\phi}$ with (instantaneous) star formation rate (SFR) aligns with the SFR vs.\ M$_{\textrm{halo}}$ relationships in the upper panel of Figure \ref{fig:1P_sSFHMRs}. Most notably, all models show higher $b_{\phi}$ in general at higher redshifts, corresponding to the $b_{\phi}$ of halos increasing more sharply at higher redshift (i.e.\ though the same SFR slice gets slightly smaller halos overall, all halos have higher $b_{\phi}$ at higher redshift in a given mass bin). For any given model, the $b_{\phi}$ for some range of star formation rate corresponds to the mass distribution for the halos in that range, and the $b_{\phi}$ that implies. 
For example, note which \texttt{SC-SAM} models show the highest $b_{\phi}$ for the $z=0$ and $z=1$ $b_{\phi}$ vs.\ SFR relationships in the center row Fig. \ref{fig:bphi_1Psets}. The minimum \textsf{SN2} model has the highest $b_{\phi}$ at the smallest SFRs (log$_{10}$SFR $<-2$ M$_{\odot}$ yr$^{-1}$), and the maximum \textsf{SN1} model has the highest $b_{\phi}$ for high SFRs (log$_{10}$ SFR$>0.5$ M$_{\odot}$ yr$^{-1}$). Examining the top row of Figure \ref{fig:1P_sSFHMRs} SFR vs.\ M$_{\text{halo}}$ relationships, one sees that the minimum \textsf{SN2} model (dark blue) shows a `main sequence' that begins near log$_{10}$SFR$>-1$ M$_{\odot}$ yr$^{-1}$. For this relationship with higher SFRs across all halo masses, selections under log$_{10}$SFR$<-2$ M$_{\odot}$ yr$^{-1}$ strongly favor the massive and quenched halos, whose masses imply a large value of $b_{\phi}$. Similarly, though less apparent in Figure \ref{fig:1P_sSFHMRs}, the halo mass distribution of the maximum \textsf{SN1} model (magenta) for large SFRs (log$_{10}$SFR$>0.5$) leans notably more toward higher halo masses than all other models, leading to a higher $b_{\phi}$ for that SFR selection. 

For sSFR selections, the spread in $b_{\phi}$ by parameter collapses into a narrow relationship, where $b_{\phi}$ begins positive and crosses to strong negative values. Here again, comprehending which halos inhabit a given sSFR range in Figure \ref{fig:1P_sSFHMRs} clarifies why all models align in their $b_{\phi}$. The sSFR vs.\ M$_{\textrm{halo}}$ `main sequences' across all models show very similar flat behavior (corresponding to the near-identical slope in SFR vs.\ M$_{\textrm{*}}$ that the \texttt{SC-SAM} feedback regulates to). Additionally, the massive and quenched regions of all models are also quite alike in their spread. Still, the comparatively small differences between different models in a given sSFR arise from the mass distribution and implied $b_{\phi}$ for the halos of galaxies with that sSFR. 

Altogether, we find that the $b_{\phi}$ for a given model for stellar mass, SFR, or specific SFR selections is well-explained by understanding: 1) the relationship of halo mass with stellar mass, SFR, and specific SFR for that model, with 2) how $b_{\phi}$ is understood to behave for different halo masses at different redshifts. 

%---------------------------%
%---------------------------%
%---------------------------%

\subsection{All parameters varied}\label{subsec:AllParams}

We now consider the behavior of $b_{\phi}$ for various stellar mass, SFR, or sSFR selections when allowing the three \texttt{SC-SAM} feedback parameters to all vary simultaneously within their full range. 
As a reminder: for amplitudes $\epsilon_{\textrm{SN}}$ and $\kappa_{\textrm{radio}}$, we multiply the fiducial value by prefactors $0.25<$A$_{\textrm{SN1}}<4.0$ or $0.25<$A$_{\textrm{AGN}}<1.8$ respectively; and for exponential $\alpha_{\textrm{rh}}$, we add to the fiducial value the prefactor $-2<$A$_{\textrm{SN2}}<2$. 

\begin{figure*}
 \centering
 \includegraphics[width=0.98\textwidth]{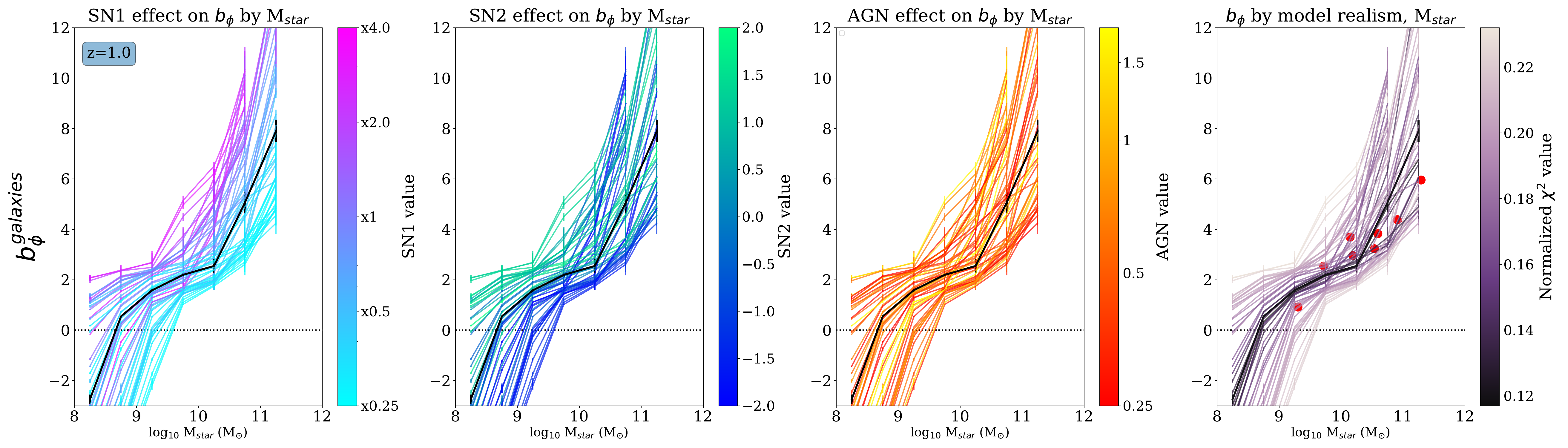}
 \caption{$b_{\phi}$ vs.\ stellar mass at  $z=1$  across all 56 \texttt{SC-SAM} models (see Appendix \ref{app:bphi_zevol} for other redshifts). Plots in each row are identical, but are colored according to each model's value of SN1, SN2, and AGN (left to right) or relative `realism' (right-most plot, most-to-least realistic in black-to-white; see Figure \ref{fig:SMFnSHMRallmodels} and the end of \textsection 2.5). We mark $b_{\phi}=0$ with a black dotted line to guide the eye, and plot the fiducial \texttt{SC-SAM} model in black for reference. Finally, the $z=1\ b_{\phi}$ measurements from \citet[Fig. 4]{Barreira+2020_TNG} are shown as red circles in the far right plot.}
 \label{fig:bphi_stemass}
\end{figure*}

\begin{figure*}
 \centering
 \includegraphics[width=0.98\textwidth]{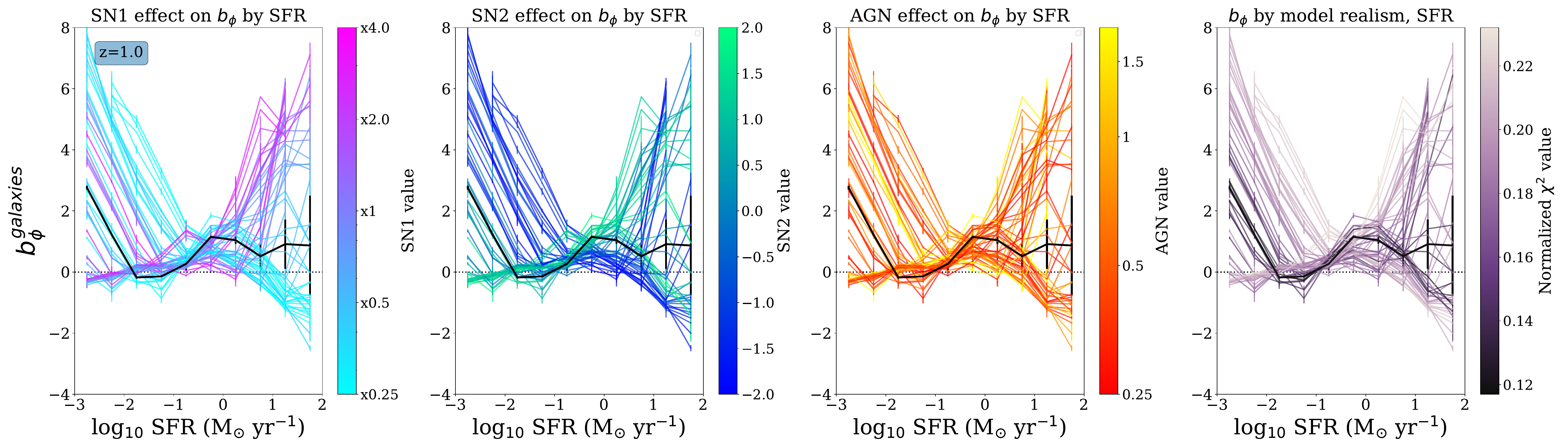}
 \caption{Like Figure \ref{fig:bphi_stemass}, but for SFR (M$_{\odot}$ yr$^{-1}$) selections.}
 \label{fig:bphi_sfr}
\end{figure*}

\begin{figure*}
 \centering
 \includegraphics[width=0.98\textwidth]{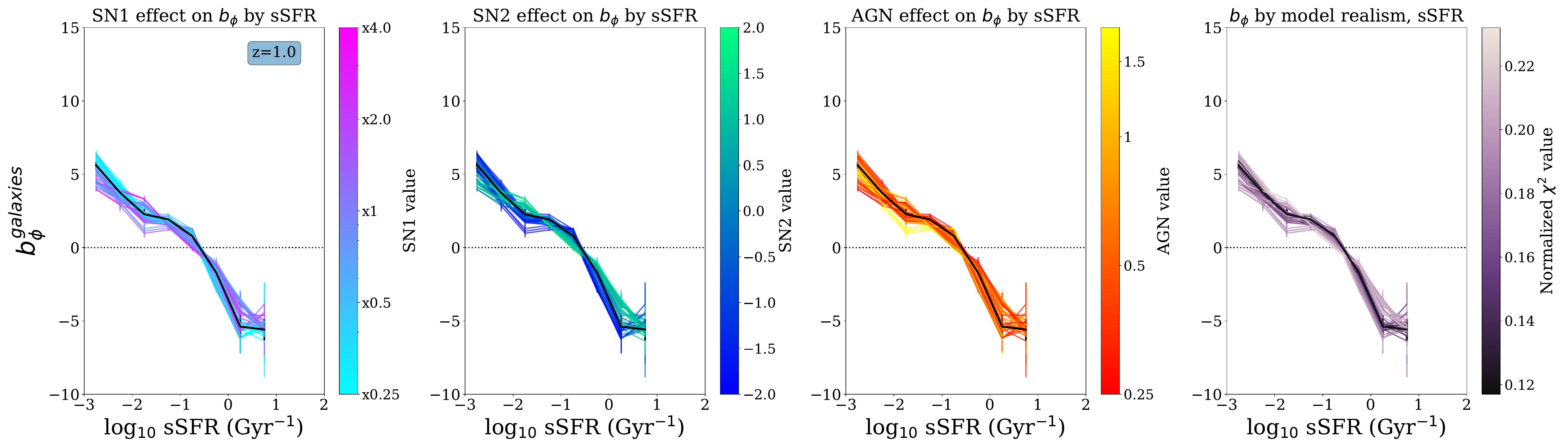}
 \caption{Like Figures \ref{fig:bphi_stemass} and \ref{fig:bphi_sfr}, but for specific SFR selections (where sSFR = SFR$\div$ M$_{\mathrm{star}}$, Gyr$^{-1}$).}
 \label{fig:bphi_ssfr}
\end{figure*}

Figures \ref{fig:bphi_stemass}, \ref{fig:bphi_sfr}, and \ref{fig:bphi_ssfr} are much like Figure \ref{fig:bphi_1Psets}, but for the 56 \texttt{SC-SAM} models. 
All plots show the same information (e.g.\ $b_{\phi}$ for stellar mass selections), but each with a colorbar showing one parameter's values in the particular model (three left columns), or the models' `realism' (right column). The far left column of plots colors each \texttt{SC-SAM} model by its unique \textsf{SN1} value, ranging from  cyan (0.25) to magenta (4.0). The center-left column of plots instead color each model by its unique \textsf{SN2} value, ranging from royal blue (-2) to teal (2). The center-right column of plots colors each \texttt{SC-SAM} model by its unique \textsf{AGN} value, ranging from red (0.25) to gold (1.8). This helps decipher the effects of varying the \texttt{SC-SAM} galaxy model on these galaxy bias parameters. Finally, the far right plot colors each plot instead by its relative `realism', as described in \textsection \ref{subsec:SCSAMmodels} and illustrated in Figure \ref{fig:SMFnSHMRallmodels}. 

Figure \ref{fig:bphi_stemass} explores how $b_{\phi}$ changes for \textbf{stellar mass selections} across the \texttt{SC-SAM} models at $z=1$. This still aligns closely with the intuition built in Figure \ref{fig:bphi_1Psets}: selections past the `knee' of a model's SMF (often  near $10^{11}$ M$_{\odot}/h$) show the highest $b_{\phi}$ values. Additionally, we see the most overt influence on the $b_{\phi}$ vs.\ M$_{\textrm{*}}$ comes from the A$_{\textrm{SN}}$ prefactors, aligning with their strong effect on the SMF and SHMR.
Finally, we plot as red circles the measured $b_{\phi}$ from \citet[Fig. 4]{Barreira+2020_TNG}; for all but a few $z=0$ stellar mass selections (see Appendix \ref{app:bphi_zevol}), our \texttt{SC-SAM} models encompass the IllustrisTNG results. 

Next, Figure \ref{fig:bphi_sfr} examines how $b_{\phi}$ changes for (instantaneous) \textbf{SFR selections} across the \texttt{SC-SAM} models at $z=1$. The patterns from varying each parameter `one-at-a-time' in Figure \ref{fig:bphi_1Psets} begin to blur when all three parameters' effects combine. Still apparent is the polarizing effect of the A$_{\textrm{SN}}$ parameters: at the smallest SFRs, the models with very weak A$_{\textrm{SN1/SN2}}$ have much higher $+b_{\phi}$ than other models, and at high SFRs, strong (weak) A$_{\textrm{SN}}$ models see notably higher $+b_{\phi}$ (lower $-b_{\phi}$) than fiducial. 
Finally, the closeness to $b_{\phi}=0$ for $z=0,\ 1$ suggests that selections that closely track the instantaneous SFR might not be optimal targets for $f_{\text{NL}}$ studies.

Next, Figure \ref{fig:bphi_ssfr} explores how $b_{\phi}$ changes for \textbf{specific SFR selections} across the \texttt{SC-SAM} models at $z=1$.
For our 56 \texttt{SC-SAM} models, $b_{\phi}$ continues to display the encouraging narrow pattern from Figure \ref{fig:bphi_1Psets}: low sSFR samples result in $b_{\phi}>0$, and higher sSFR samples in $b_{\phi}<0$, with a similar turnover across all the models. As before, this results from the similarity of the sSFR vs.\ M$_{\textrm{halo}}$ distributions across parameter variations. All models go from clearly positive to negative $b_{\phi}$ at sSFR=0.1 Gyr$^{-1}$, the approximate ceiling for the massive and quenched objects at $z=1$. Indeed, the sSFR where the relationships cross from positive to negative $b_{\phi}$ is the central sSFR `plateau' value for the `main sequence' of the sSFR vs.\ M$_{\textrm{halo}}$ relationships at a given redshift.
An individual parameters' influence is most different at low sSFR bins, where the massive quenched objects are most prevalent, but strongly standardizes until becoming somewhat uncertain again at the highest sSFR bins (corresponding to rare galaxies at the very top edges of a parameter's sSFR vs.\ M$_{\textrm{halo}}$ `main sequence'). 

%---------------------------%
%---------------------------%
%---------------------------%
%---------------------------%
%---------------------------%
%---------------------------%

\section{\lowercase{b}$_{\phi}$ vs.\ \lowercase{b}$_1$ across \texttt{SC-SAM} models} \label{sec:b1bphiresults}

Next, we examine the $b_{\phi}$ versus $b_1$ relationship for different $z=1$ galaxy selections across our 56 \texttt{SC-SAM} models. Previous work with IllustrisTNG showed galaxies diverged from the universality relation, and differently so, depending on how the galaxies were selected \citep{Barreira+2020_TNG}, even within the single IllustrisTNG model. How does this present within the \texttt{SC-SAM}, and when varying the strength of stellar and AGN feedback simultaneously?

\begin{figure*}
 \centering
 \includegraphics[width=0.95\textwidth]{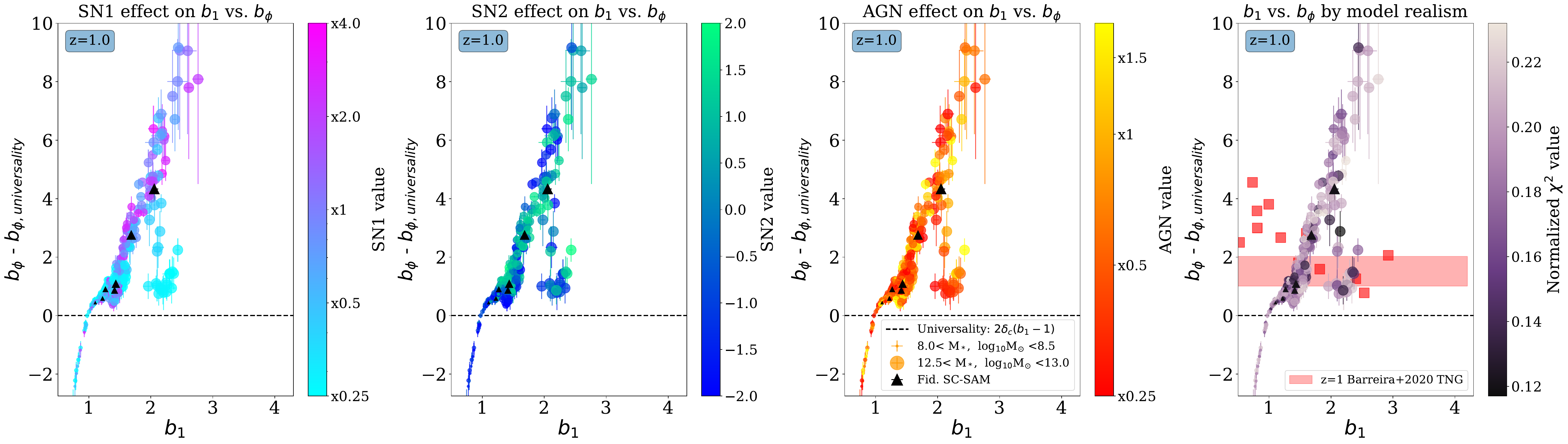}
 \caption{ $b_{\phi}-b_{\phi, \textrm{univ.}}$ vs.\ $b_1$ relationship at $z=1$  across all \texttt{SC-SAM} models for \textbf{stellar mass selections}. The $y$-axis is normalized to the predicted $b_{\phi}$ from the universality relation (Eq. \ref{eq:universality} with $p=1$), so that deviations from zero indicate $b_{\phi}$ far from predicted. Colored as earlier figures, with the fiducial \texttt{SC-SAM} model as black triangles.
 The size of markers corresponds qualitatively to the selection, where the largest points indicate the largest stellar mass bins. The rightmost figure, colored by model `realism', also replots the data from \citet[Fig. 3]{Barreira+2020_TNG} in red squares, and their proposed modification to the universality relation for stellar mass selections (Eq. \ref{eq:AB_mod_universality}).
 % \texttt{SC-SAM} models show a large positive deviation from the universality relation throughout, where models with weak \textsf{SN1} drop close to universality. % at $z>1$.
 }
 \label{fig:b1Vbphi_stemass}
\end{figure*}

\begin{figure}
 \centering
 \includegraphics[width=0.95\textwidth]{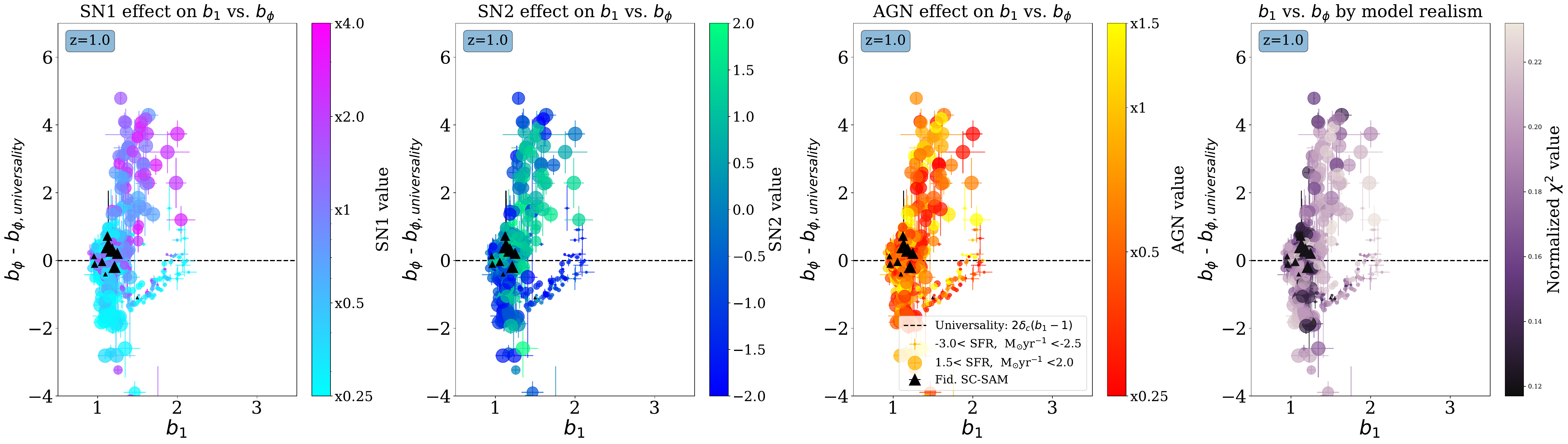} 
 \caption{Like Figure \ref{fig:b1Vbphi_stemass}, but for \textbf{SFR selections}. The \texttt{SC-SAM} models span a wide and mostly featureless range of $b_{\phi}$ in a narrow range of $b_1$. As Figure \ref{fig:1P_sSFHMRs} shows, all models show a similarly wide spread of halo masses across most of their SFR selections, leading to a $b_1$ measurement roughly corresponding to the mean mass. 
 The lowest value SFR selections show uniquely diverging behavior to mostly negative $b_{\phi}$ and slightly higher $b_1$, apparently driven by the efficiency of small \textsf{SN1} and/or \textsf{SN2} in creating more massive and quenched galaxies compared to `main sequence' objects.}
 \label{fig:b1Vbphi_sfr}
\end{figure}

\begin{figure}
 \centering
 \includegraphics[width=0.95\textwidth]{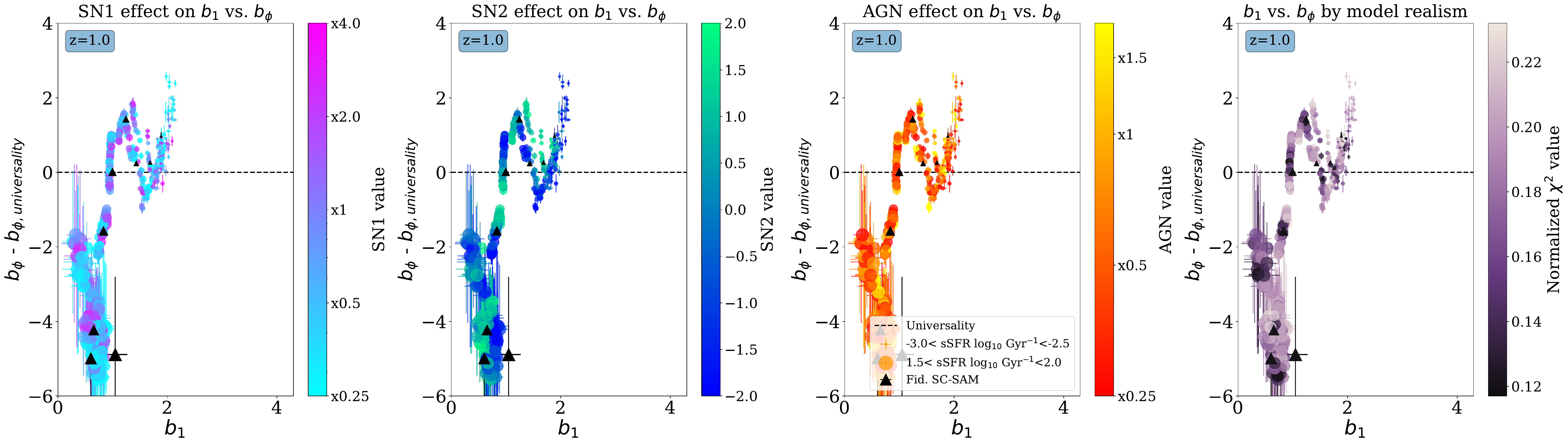}
 \caption{Like Figs.\ \ref{fig:b1Vbphi_stemass} and \ref{fig:b1Vbphi_sfr}, but for specific SFR selections. As in Figs.\ \ref{fig:bphi_1Psets} and \ref{fig:bphi_ssfr}, sSFR selections follow a tight $b_{\phi}-b_{\phi, \textrm{univ.}}$ vs.\ $b_1$ relationship whose $b_{\phi}$ spread is narrow appears agnostic to \texttt{SC-SAM} model variations, thanks to the mostly model-agnostic sSFR vs.\ M$_{\textrm{halo}}$ relationship found in Figure \ref{fig:1P_sSFHMRs}.}
 \label{fig:b1Vbphi_ssfr}
\end{figure}

Figures \ref{fig:b1Vbphi_stemass}, \ref{fig:b1Vbphi_sfr}, and \ref{fig:b1Vbphi_ssfr} present $b_{\phi, \mathrm{measured}} - b_{\phi, \mathrm{universality}}$ versus $b_1$ for a given galaxy selection at $z=1$.  
Rather than plotting simply $b_{\phi}$ vs.\ $b_1$ (as in e.g.\ \citealt{Barreira+2020_TNG} Fig.\ 3), we instead normalize all $b_{\phi}(b_1)$ values against what the universality relation predicts with $p=1$ for the measured $b_1$ of the selection (Eq. \ref{eq:universality}). This way, values below/above the black-dashed line at zero indicate a lower/higher $b_{\phi}$ than the universality relation implies, respectively, and values close to zero indicate the galaxy sample follows the universality relation. 
As in the previous section, the panels in each of these three figures show the same measurement, but with the color of each point corresponding to that \texttt{SC-SAM} model's value for a given parameter (cyan-to-pink for \textsf{SN1}, blue-to-teal for \textsf{SN2}, and red-to-gold for \textsf{AGN}). The rightmost panel colors each model's points based on its `realism' when compared to the $z=0$ SMFs of \citet[see Figure \ref{fig:SMFnSHMRallmodels}]{Bernardi2018}.
Additionally, the size of each point corresponds to how small or large the corresponding selection is compared to others; and the shared legend shows the minimum and maximum selection bin that is shown in each row of plots. As before, we only plot results for bins with at least 500 galaxies, and we do not connect $b_{\phi}-b_{\phi, \textrm{univ.}}$ vs.\ $b_1$ measurements within the same \texttt{SC-SAM} for legibility. Given that some \texttt{SC-SAM} models yield many more galaxies across all mass ranges(e.g.\ Fig. \ref{fig:1P_SMFs}), the largest selections/points within Figures \ref{fig:b1Vbphi_stemass}-\ref{fig:b1Vbphi_ssfr} will preferentially come from those more extreme models. Our partner work in \citet{MoorePerez2026} accounts for this bias in the generation of $b_{\phi}$ priors by conditioning on e.g.\ the observed stellar mass function in \citet{Bernardi2018}.

Figure \ref{fig:b1Vbphi_stemass} plots $b_{\phi}-b_{\phi, \textrm{univ.}}$ vs.\ $b_1$ for \textbf{stellar mass selections}, here limited to bins within $8<$ log$_{10}$M$_{*}$ (M$_{\odot})<13$ and $z=1$. Here, most apparent is the clear departure from the universality relation for most stellar mass selections and \texttt{SC-SAM} models. Taken all together, the \texttt{SC-SAM} models appear to favor a steeper $b_{\phi}(b_1)$ than seen in \citet{Barreira2020} for TNG (horizontal red band in the right panel). 
All models appear similar in the smallest stellar mass bins, with the smallest points creating a tight tail of $b_1\approx1$ and $-3< b_{\phi}-b_{\phi, \textrm{univ.}}<0$ (or approximately $-3<b_{\phi}<0$). All models also show a `fanning-out' of $b_{\phi}$ values for the selections with highest $b_1$ (often the highest mass and most clustered galaxies, aligning with larger error bars for smaller sample sizes). There is a selection of models with very small \textsf{SN1} that find most of their high mass selections bunched close to the universality relation and the \citet{Barreira+2020_TNG} TNG measurements, near $b_1 \sim 2.5$ and $1< b_{\phi}-b_{\phi, \textrm{univ.}}<3$ (or roughly $4<b_{\phi}<7$). Nearly all other models and selections follow a linear relationship far above the universality relation, with a slope of approximately $+6$. The other astrophysical parameters do not show a similar influence, and we note that models across the SMF-`realism' spectrum inhabit both the small-\textsf{SN1} island and the dominant linear relationship. This trend holds across redshift, with the location of the small-\textsf{SN1} island and the rough slope of all other models shifting; see Appendix \ref{app:b1bphi_zevol} for figures.

For stellar mass selections, \citet{Barreira+2020_TNG} derived their own best fit within IllustrisTNG separate-universe simulations. We display their stellar mass selected $b_{\phi}(b_1)$ rescaled to our axes as red dots, as well as their derived best fit in a red region:
\begin{equation}
    b_{\phi,\mathrm{TNG\ M*}}=2\delta_c (b_1-p), \quad p \in [0.4, 0.7].
\label{eq:AB_mod_universality}
\end{equation}

We find some disagreement with the IllustrisTNG results, particularly for the smaller stellar mass bins. Given the agreement in Figure \ref{fig:bphi_stemass}, there seems to be a mild tension with the $b_1$ measurements at some stellar mass selections, which is then visually exaggerated by our rescaled $b_{\phi}$ $y$-axis. We discuss this further in \textsection \ref{subsec:TNGnSAMs}.

Figure \ref{fig:b1Vbphi_sfr} plots $b_{\phi}-b_{\phi, \textrm{univ.}}$ vs.\ $b_1$ for \textbf{SFR selections}. These differ substantially from stellar mass selections in shape, given the different relationship between halo mass and SFR for a given \texttt{SC-SAM} model. Nearly all selections are situated within $1 < b_1 <2$ and $-2 <b_{\phi}-b_{\phi, \textrm{univ.}} <4$, and seem most affected by the value of \textsf{SN1/2}. 
The threshold of the SFR selection is important here, with a binary path appearing: the lowest SFR bins (the smallest points) 
align much more closely with the expected slope, but dip lower in $b_{\phi}$; and all other (higher) SFR bins cluster near $b_1 \approx 1$ and $-2.5<b_{\phi}<0$ (or, $\pm 2.5$ about what universality predicts). This becomes clearer with the context of the SFR vs.\ M$_{\textrm{halo}}$ relation (upper panel of Figure \ref{fig:1P_sSFHMRs}): first, the shape of the SFR vs.\ M$_{\textrm{halo}}$ distribution means that most SFR selections include a very wide range of halo masses, from the dominant lower-mass `main sequence' to the massive quenched region of rarer objects. This leads to most selections clumping about the mean $b_1$ of all the halos in the selection. In Appendix \ref{app:b1bphi_zevol}, we see this clumping in $b_1$ is strongest at $z=0$, and relaxes with higher redshift as models and selections broaden over $b_1$ values.

However, we see models like the minimum \textsf{SN1/2} push the `main sequence' to higher average SFRs while creating extra-quenched massive galaxies, leading to small-SFR selections with slightly larger  $b_1 > 1$. Inversely, the highest value \textsf{SN1/2} yields high SFR samples with highly positive $b_{\phi}$ above universality, consistent with their effect in Figure \ref{fig:bphi_sfr}. For \textsf{AGN}, its possible effect on $b_{\phi}-b_{\phi, \textrm{univ.}}$ vs.\ $b_1$ is not coherent, agreeing with the Figure \ref{fig:1P_sSFHMRs} \textsf{AGN} panel. 
When judging by SFR `realism', there is some evidence that less realistic models show higher $b_{\phi}$, but this becomes less apparent at higher redshift (see Appendix \ref{app:b1bphi_zevol}).

Figure \ref{fig:b1Vbphi_ssfr} plots $b_{\phi}-b_{\phi, \textrm{univ.}}$ vs.\ $b_1$ for \textbf{specific SFR selections}. Echoing the behavior in Figure \ref{fig:bphi_ssfr}, we see nearly all models displaying similar behavior with no distinguishing influence of any particular parameter or the `realism' of a model. All models show a rough `zig-zag' pattern, where the largest sSFR selections appear at the smallest $b_1<1$ and negative $b_{\phi}-b_{\phi, \textrm{univ.}}$, follow a sharp upwards trajectory until reaching $b_{\phi}-b_{\phi, \textrm{univ.}} \sim +2$ at a slightly larger $b_1$, dipping again just below universality, and finally reaching a peak  positive $b_{\phi}-b_{\phi, \textrm{univ.}}$ and peak $b_1$. As Appendix \ref{app:b1bphi_zevol} discusses, increasing redshift sees the pattern from the SFR selections, broadening the same `zig-zag' shape to cover more $b_1$ and $b_{\phi}-b_{\phi, \textrm{univ.}}$ values. 

As before, the sSFR vs.\ M$_{\textrm{halo}}$ relationships in the lower panel of Figure \ref{fig:1P_sSFHMRs} help clarify why this `zig-zag' shape arises. High-value sSFR selections catch the top of the flat `main-sequence', covering a broad and mostly low halo mass range. These samples are therefore expected to have small measured $b_1$, and also strongly negative $b_{\phi}$ values from Figure \ref{fig:bphi_ssfr}. The rise and drop `zig-zag' for moderate sSFR bins are at higher $b_1$ as more of the massive quenched galaxies make it into the selection, and also likely come from the crossover points from below to above $b_{\phi}=b_{\phi, \textrm{univ.}}$ seen in Figure \ref{fig:bphi_ssfr}. All together, this might indicate that, though more complex, sSFR selections are agnostic to the particular galaxy formation modeling. Appendix \ref{app:b1bphi_zevol} examines sSFR selections at $z=0$ and $z=2$.

%---------------------------%
%---------------------------%
%---------------------------%
%---------------------------%
%---------------------------%
%---------------------------%

\section{Discussion}
\label{sec:discussion}

\subsection{The Contribution of Galaxy Assembly Bias} \label{subsec:assemblyBias}

Next, we assess the influence of \textit{galaxy assembly bias} to these measurements. We define galaxy assembly bias as in \citet{Croton2007}, where the clustering of a galaxy sample depends on more than simply the mass of the host halo\footnote{This study was preceded by observational and theoretical studies in \citet{Yang2006, AbbaSheth2006, Reed2007}. Galaxy assembly bias has also been described and studied in the lens of halo occupation, and whether halo properties beyond halo mass cause galaxies of some properties to inhabit a halo; see \citet{WechslerTinker2018} and \citet{Desjacques2018} for relevant reviews.}. 
Appendix \ref{app:AssemblyBias} details our brief assessment of galaxy assembly bias, and its effect on $b_1$ and $b_{\phi}(b_1)$, for the one-parameter-at-a-time \texttt{SC-SAM} models. We summarize the relevant results here.

First, each one-at-a-time model leads to a different amount of assembly bias, and in a unique way for our three galaxy properties.
Our one-at-time \texttt{SC-SAM} models show a signal of 5-20$\%$ assembly bias when selecting to a stellar mass threshold across \texttt{SC-SAM} one-at-a-time models (depending on the model and selection). This agrees with \citet{Hadzhiyska2021}, who found the fiducial \texttt{SC-SAM} model recreated a $\sim 10\%$ level of assembly bias for stellar mass selections, consistent with what is observed in IllustrisTNG. They found the galaxy assembly bias was due to the halo's formation histories. For SFR and sSFR, we find a noisier but inverted assembly bias signal, such that galaxies shuffled in their halo mass bin are \textit{more} correlated than their original distribution. This qualitatively agrees with the assembly bias measured for simulated blue galaxies in \citet{Croton2007}.
More deeply understanding this behavior, and why parameter variations change the \texttt{SC-SAM} galaxy assembly bias, is intriguing but beyond the scope of our work.

Next, we compare $b_1$ measured on samples with and without their halo's formation history to identify how galaxy assembly bias will affect our measured $b_1$ and $b_{\phi}(b_1)$. We shuffle galaxies to halos of a similar halo mass,  and then place them into stellar mass, SFR, or sSFR bins before measuring $b_1$ as in the rest of our work. 
For bins of stellar mass and SFR, we see good 1:1 alignment for all but the largest stellar mass bins or smallest SFR bins. This means that, if we were to shuffle galaxies into halos of a similar mass and then measure $b_{\phi}(b_1)$ for stellar mass or SFR selections like in Figure \ref{fig:b1Vbphi_stemass}, the $b_{\phi}(b_1)$ relationship would \textit{not} significantly change. 
The only notable effect is for small sSFR selections: because they include both very low and high mass objects, the shuffled $b_1$ blurs the the signals from all halos to $b_1~\sim 1$. After shuffling, the low-value sSFR $b_{\phi}(b_1)$ measurements stack vertically at $b_1~\sim 1$. 
It is difficult to separate the effect of the halo mass differences from galaxy assembly signals, and more detailed analyses are beyond the scope of this work. However, we note that if we exclude what are obvious massive quenched galaxies, our measured $b_1$ are minimally affected by the effects of galaxy assembly bias. Finally, we remind readers that because $b_{\phi}$ is defined as the change in galaxy numbers induced by primordial long-wavelength perturbations, there is no change in our measured $b_{\phi}$ if we separate galaxies from their host halo's formation history.

%---------------------------%
%---------------------------%
%---------------------------%

\subsection{Comparison to IllustrisTNG and other SAMs} \label{subsec:TNGnSAMs}

Our choice to cover a very broad stellar and AGN feedback parameter space in the \texttt{SC-SAM}, and to focus on fundamental galaxy properties like stellar mass and SFR, aims to encompass the behavior of other galaxy formation models with just the \texttt{SC-SAM}. To our knowledge, few other studies have used galaxy formation models to probe the bias parameters involved in constraints of $f_{\text{NL}}$. 
Here, we compare to three relevant studies: \citet{Barreira+2020_TNG}, who used hydrodynamically simulated IllustrisTNG galaxies to measure $b_{\phi}$ and $b_{\phi}(b_1)$; and \citet{Reid2010} and \citet{Marinucci2023}, who used SAMs to study non-Gaussian assembly bias corrections beyond the response of the halo mass function. We contextualize and compare against their measurements here. Finally, though we do not compare to their results here, we note that \citet{Nascimento2026} found consistent $b_{\phi}$ measured in both SIMBA and TNG when comparing the separate universe method and the evolution of tracer counts method \citep{DalalPercival2025}.

\textbf{IllustrisTNG: }
We find decent consistency when comparing the $b_{\phi}$(M$_*$) for IllustrisTNG to our \texttt{SC-SAM} models (rightmost panels of Figures \ref{fig:bphi_stemass} and \ref{fig:bphi_stemass_0n2}). However, our combined $b_{\phi}-b_{\phi, \textrm{univ.}}$ vs.\ $b_1$ disagrees with theirs at all redshifts (Figs. \ref{fig:b1Vbphi_stemass}, \ref{fig:b1Vbphi_stemass_0n2}), especially at small masses/low $b_1$ value samples. We note that the rescaling of $b_{\phi}$ accentuates the disagreement in the $b_1$ measured for a particular stellar mass bin; plotting $b_{\phi}$ vs.\ $b_1$ directly conceals the differences given the strong positive slope of the relationship. Though we note that \citet{Barreira+2020_TNG} used a $\delta_{\textrm{m}}$ separate universe approach to evaluate $b_1$ (compared to how our $b_1$ are calculated with the $P_{gm}$ of an \texttt{SC-SAM} model run atop our fiducial central cosmology ($\Omega_{\text{M}}=0.3,\ \sigma_8=0.8$) and that simulation's $P_{mm}$), the differences likely result from the differences between IllustrisTNG and the \texttt{SC-SAM} as galaxy formation models. 
Additionally, IllustrisTNG includes baryonic effects that the \texttt{SC-SAM}, by construction, does not include. Given our $k_{\text{max}}$, baryonic effects might lead to a few percent overestimate on our measured $P_{mm}$ and $P_{gm}$, and a small underestimation of $b_1$ \citep{Debackere2020}.
% } 
Further work extending the IllustrisTNG analysis to SFR and sSFR selections could help better clarify how well our span of \texttt{SC-SAM} models covers the TNG model in $b_{\phi}$. 

To compare with the results of \citet{Reid2010} and \citet{Marinucci2023}, it is helpful to separate $b_{\phi}$ -- the total non-Gaussian bias measured for a sample of tracers (given selection \textbf{S} at some $z$, as we describe in \textsection \ref{subsec:SepUni}) -- into two components: the non-Gaussian halo bias (i.e.\ the response of the halo mass function alone to changing $\sigma_8$) and the non-Gaussian halo assembly bias (i.e.\ the response of however the tracers inhabit halos to changing $\sigma_8$). In \citet{Reid2010} and \citet{Marinucci2023}, they are written so:

\begin{equation}
\begin{split}
    \textrm{Marinucci et al.\ (2023, Eq.\ 6) : } b_{\phi,\ \textrm{galaxies}} = \bar{b}_{\phi}\ +\ \Delta b_{\phi,\ \textrm{HOD}} \\
    \textrm{Reid et al.\ (2010, Eq.\ 18) : } A_{\textrm{NG, galaxies}} = A_{\textrm{NG}}^{\textrm{all}}\ +\ \Delta A_{\textrm{NG}}^{\textrm{l,h }z_{form}}\, \\
    \textrm{where }  b_{\phi}= \frac{2 A_{\textrm{NG}}}{D(z_0)}, \textrm{ with } D(z_0)=\frac{g(z)}{(1+z)} \textrm{ and } D(z=1) \approx 0.44,
\end{split}
\label{eq:allNGassemblybiasdefs}
\end{equation}

Here, $D(z_0)$ is the linear growth function normalized to $(1+z)^{-1}$ in the matter dominated era, and $g(z)$ is the growth suppression due to non-zero $\Lambda$ (with $g(z=1)\approx 0.88$)
For both these works, the first term in the sum (either $\bar{b}_{\phi}$ or $A^{\textrm{all}}_{\textrm{NG}}$) comes from the non-Gaussian halo bias, which is to first order well approximated by the universality relation (see Appendix \ref{app:PNGwSAMsdetails} for more details.) Here, we focus on their measurements for the non-Gaussian assembly bias for galaxy samples.

\textbf{Galacticus: } \citet{Marinucci2023} use the \texttt{Galacticus} SAM to probe non-Gaussian assembly bias through the response of a Halo Occupation Distribution (HOD) model to some long-wavelength perturbation. First, they populate galaxies onto halos using the \texttt{Galacticus} SAM of galaxy formation, and then use the HOD framework as the lens of analysis for those galaxies, following the exploration of non-Gaussian assembly bias of \citet{Voivodic2021}. \citet{Marinucci2023} define and measure $\Delta b_\phi^{c,s}$, which is the response of the (average per halo) number of selected centrals/satellites to long-wavelength perturbations of $\phi(\textit{\textbf{x}})$ and compute $\Delta b_\phi^{c,s}$ with the peak background split for simulated halos at some target $M_h$ at $z$ thusly: 

\begin{equation}
    \Delta b_\phi^{c,s} (\textbf{S}|M_h, z)\equiv 2 \frac{\partial \textrm{ln}N_{c,s}}{\partial \textrm{ln} \sigma_8} (\textbf{S}| M_h,z)\ \Rightarrow\
    \frac{1}{|\delta_{\sigma_8}|} \left[ \frac{N_{c,s}^{\text{high}}(\textbf{S}|M_h, z) - N_{c,s}^{\text{low}}(\textbf{S}|M_h, z)}{N_{c,s}^{\text{fid}}(\textbf{S}|M_h, z)} \right]
\label{eq:Deltabphi_cs}
\end{equation}

Here, $N^{\text{fid/high/low}}_{c,s}$ is the total number of central/satellite galaxies over $N_h\sim 10^5$ halos with $M_h$ at some $z$, generated at some fiducial/high/low $\sigma_8$ values (like what we do in \textsection \ref{subsec:SepUni}). 
\citet{Marinucci2023} used extended Press-Schechter (EPS; \citealt{LaceyCole1993}) merger trees to build up their halos and histories for their separate-universe measurements. % with \texttt{Galacticus}. 
EPS merger trees build up the formation history of the target `root' halo backwards in time with a Monte Carlo algorithm, using a splitting algorithm to determine if a merger happened in the previous step, and then a conditional probabilistic mass function for the progenitor halos.
Crucial to note here, EPS merger trees do not have any halo assembly bias -- 
\citet{Furlanetto2006} show that Gaussian halo bias in the EPS formalism is independent of its formation history,
and furthermore, there is no environmental information for which to measure the clustering against the background density as we do in App.\ \ref{app:b1AssemblyBias}.

\citet{Marinucci2023} run \texttt{Galacticus} over $N_h\sim10^{4-5}$ EPS merger trees with $z=1$ root halos with masses between $[3\times 10^{10}, 10^{12}]$ M$_{\odot}$.
Because they only sample three $z=1$ root masses, we cannot fully integrate over halo mass and directly compare to our $b_{\phi}(\textbf{S}, z=1)$.
However, we are able to mimic their experiment by similarly selecting central\footnote{We do not compare to their $\Delta b_{\phi}^{s}$ measurements given the mass cuts we have applied to maintain completeness across \texttt{SC-SAM} models.} galaxies across our \texttt{SC-SAM} models and comparing to the $\Delta b_{\phi}^{c}(M_{*}|$M$_h, z=1)$ in their Figure A1 -- see App.\ \ref{app:PNGwSAMsdetails} for full details.
The \texttt{SC-SAM}'s $\Delta b_{\phi}^{c}($M$_*|$M$_h, z=1)$ sit higher than nearly all in \texttt{GALACTICUS} across the halo mass groups and stellar mass bins. The best consistency is with the \texttt{GALACTICUS} $\Delta b_{\phi}^{c}($M$_*|$M$_h=10^{11}, z=1)$ values and the two one-at-a-time minimum A$_{\textrm{SN}}$ \texttt{SC-SAM} models. 
The best agreement overall is at the highest $z=1$ stellar mass bins for all probed halo masses.

Given that our merger trees include the effects of the environment and galaxy assembly bias\footnote{Where galaxy assembly bias is defined compared to halo clustering, as in \textsection \ref{subsec:assemblyBias} and Appendix \ref{app:AssemblyBias}.} that EPS merger trees do not, this may explain our higher $\Delta b_{\phi}^{c}($M$_*|$M$_H, z=1)$.
Halo assembly bias appears to be highest for low mass halos \citep{GaoWhite2007}, perhaps explaining why the lowest mass halos selection show the best agreement with \citet{Marinucci2023} for some \texttt{SC-SAM} parameters.
Still, this experiment highlights the diversity of measurements that arise from galaxy model variations, and the challenge of comparing the many ways non-Gaussianity is parametrized.

\textbf{Munich SAM}:  \citet{Reid2010} describe and detect the dependence of the non-Gaussian halo bias signal induced by $f_{\textrm{NL}}$ on halo formation history.
They define the non-Gaussian halo assembly bias as a correction factor $\Delta A_{\text{NG}}$, as we described in Eq.\ \ref{eq:allNGassemblybiasdefs}, which is added to the $A^{\text{all}}_{\text{NG}}$ (measured on all halos in the same mass range, not simply those that fall into the galaxy selection). 
They assume universality describes $A^{\text{all}}_{\text{NG}}$ well, with $A^{\text{all}}_{\text{NG}} \approx \delta_c (b_G -1)$, where $b_G$ is the Eulerian scale-independent contribution to the halo bias (functionally equivalent to our $b_1$). By comparing their Eq.\ 17 to our Eq.\ \ref{eq:bphi_initialdef}, one finds our $b_{\phi}= 2A_{\textrm{NG}}/D(z_0)$ in Eq.\ \ref{eq:allNGassemblybiasdefs}. We take their reported values of $A^{\text{all}}_{\text{NG}}$ and $\Delta A^{\text{gal}}_{\text{NG}}$ and convert them into $b_{\phi}-b_{\phi, \textrm{ univ.}}$ to directly compare to our results with the \texttt{SC-SAM}.

For their $z=1$ Munich SAM galaxies with M$_* \geq 8\times 10^{10}\ h^{-1}$ M$_{\odot}$, we find $b_{\phi}-b_{\phi, \textrm{ univ.}}=7.93$ for their reported $b_G$=2.3; $\Delta A^{\text{gal}}_{\text{NG}} =0.51$; and $A_{\textrm{NG}}^{\textrm{all}} =2.2$ (estimated via universality given their $b_G$). 
Their galaxies with SFR $>24$ M$_{\odot}$ yr$^{-1}$, we find $b_{\phi}-b_{\phi, \textrm{ univ.}}=2.4$ for their reported  $b_G=1.3$, $\Delta A^{\text{gal}}_{\text{NG}}=0.24$; and estimated $A_{\textrm{NG}}^{\textrm{all}} =0.51$.  For both, we assume $p=1$ in the Eq.\ \ref{eq:universality} universality function. 
These are consistent with the $b_{\phi}-b_{\phi, \textrm{ univ.}}$ vs.~$b_1$ values for the highest stellar mass and SFR selections in our Figs.\ \ref{fig:b1Vbphi_stemass} and \ref{fig:b1Vbphi_sfr}, and also $b_{\phi}-b_{\phi, \textrm{ univ.}}$ that we measure for the same threshold selections in App.\ \ref{app:PNGwSAMsdetails} Figure \ref{fig:Reid2010}. This indicates our varied parametrizations of the \texttt{SC-SAM} likely encapsulate the $b_{\phi}$ behavior of the Munich SAM run in \citet{Bertone2007} and studied in \citet{Reid2010}.

The behavior of $b_{\phi}$ and $b_1$ for other SAMs and their modeling variations remains understudied. Though our strong variations across the \texttt{SC-SAM} model create a broad range of measurements that encompass much of these other works' findings, other SAMs and galaxy models might show behavior the \texttt{SC-SAM} cannot produce. We see general agreement (with some caveats) with the IllustrisTNG model \citep{Barreira+2020_TNG} and the SAMs in \citet{Reid2010} and \citet{Marinucci2023}.
Given that SAMs are often tuned to reproduce the same nearby-Universe observables, like the stellar mass function, and also work off a merger tree-based framework, they might be expected to perform similarly for these observables (as they have for others, e.g.\ \citealt{Somerville2015}, particularly if they are shown to produce galaxy assembly bias). We also note the upcoming work in de Icaza-Lizaloa et al.\ (in prep) probing the behavior of $b_{\phi}$ in the 
% Initial comparisons with the 
fiducial \texttt{GALFORM} (\citealt{Gonzalez-Perez2020, Benson2010}) and \texttt{SHARK} \citep{Lagos2018} SAMs, and which will further
inform our understanding of how galaxy physics affects $b_{\phi}f_{\textrm{NL}}$ constraints.

%---------------------------%
%---------------------------%
%---------------------------%

\subsection{Implications for Survey Design} \label{subsec:obs}
 
Surveys to constrain $f_{\textrm{NL}}$ have not yet considered $b_{\phi}$ as an angle for optimization, and we hope this work gives actionable guidance toward that.
A crucial first question: even with limited certainty on the galaxy formation that describes our universe, how likely are we to have galaxy selections with strongly non-zero $b_{\phi}$? 
We find there are \textit{no} \texttt{SC-SAM} models for which \textit{all} galaxy selections have $b_{\phi}$ consistent with zero. Phrased differently, there always exists some selection in stellar mass, SFR, and specific SFR for which we measure $b_{\phi}$ to be more than $3\sigma$ from zero ($b_{\phi} \geq 3\sigma_{b \phi}$) for \textit{any} model. Therefore, even if our Universe is best represented by the \texttt{SC-SAM} model variation whose $b_{\phi}$ tends to be zero most often, a quarter of all galaxy selections made on the data can still be expected to have $\sigma_{b \phi} \geq 3$, and therefore a more confidently detectable $f_{\textrm{NL}}$ signal. 
In Table \ref{tab:bf_sigma3} 
we compare the smallest and largest percentages of selections with $\sigma_{b \phi} \geq 3$ across the 56 \texttt{SC-SAM} models.

\begin{table}[t]
\centering
\caption{The percentages of selections with $b_\phi \ge 3\sigma_{b_\phi}$ across the 56 \texttt{SC-SAM} models; note that every model we test finds, even in the worst case, 10$\%$ of selections that measure $b_{\phi}$ at least 3$\sigma$ away from 0.}
\label{tab:bf_sigma3}
\begin{tabular}{lccc}
\toprule
Selection & $z=0$ & $z=1$ & $z=2$ \\
\midrule
M$_{\textrm{star}}$ & 30--60\% & 45--70\% & 45--70\% \\
SFR          & 15--50\% & 10--55\% & 25--80\% \\
sSFR         & 30--85\% & 45--70\% & 47--65\% \\
\bottomrule
\end{tabular}
\end{table}

Next, we can consider which galaxy selections might be robust to uncertainties in galaxy modeling with the \texttt{SC-SAM}, particularly to disentangle the degeneracies of $b_1 b_{\phi} f_{\textrm{NL}}$. Phrased differently: how vulnerable are our assumptions of $b_{\phi}$ or $b_{\phi}(b_1)$, at a fixed selection, to galaxy formation uncertainties?

For \textit{stellar mass} selections (Figure \ref{fig:b1Vbphi_stemass}, \ref{fig:b1Vbphi_stemass_0n2}), $z=0$ finds nearly all models and selections following a relatively narrow positive linear relationship. This relationship broadens across probable $b_{\phi}$ significantly at higher redshift, though nearly all $b_{\phi}$ values remain confidently positive and non-zero. This is consistent with the results of \citet{Barreira+2020_TNG}, and tracks with the $b_{\phi}$ of halos and how a model's SHRM might change (Fig. \ref{fig:1P_SMFs}). 
For example, \citet{sdss3BOSSclustering} find the linear biases of their samples to hover around $b_{1,\textrm{BOSS}}(r) \sim 2$ for $0.43 < z < 0.7$; our Figures \ref{fig:b1Vbphi_stemass} and \ref{fig:bphi_stemass_0n2} might predict that that sample's $b_{\phi}$ will average to somewhere between $+2 < b_{\phi} < +8$, if any of our \texttt{SC-SAM} model variations approximate the real Universe.  

For \textit{SFR} selections (Figure \ref{fig:b1Vbphi_sfr}, \ref{fig:b1Vbphi_sfr_0n2}), the $b_{\phi}(b_1)$ relationships are more unpredictable. This appears to result from how the different parameters shift and warp the SFR vs.\ M$_{\text{halo}}$ relationship (Fig. \ref{fig:1P_sSFHMRs}). However, if one is able to observe enough very low or high value SFR galaxies in a sample, both groups occupy distinct loci in $b_{\phi}(b_1)$, particularly at higher redshifts. This might be challenging given uncertainties in observational selections that tie to instantaneous SFR, particularly in the challenging multiscale modeling required for SF-based emission lines \citep{Olsen2018}. 

\textit{sSFR} (Figure \ref{fig:b1Vbphi_ssfr}, \ref{fig:b1Vbphi_ssfr_0n2}), however, appears agnostic to our parameter variations: all $b_{\phi}$(sSFR) and $b_{\phi}(b_1)$ relationships follow a singular path (i.e.\ one $b_1$ value ties to a narrow range of $b_{\phi}$). sSFR may be an easier target observationally, as it has been found to be a good predictor of galaxy color in IllustrisTNG (Fig. 6 in \citealt{Nelson2018}) and  for the \texttt{SC-SAM} (R.\  Somerville, private communication, and Perez et al.\ in prep). Generally, a smaller (bluer) \textit{g-r} color corresponds to higher sSFR at all halo masses. This robustness to parameter variations results from the parametrization-agnostic shape of the sSFR(M$_{\text{halo}}$) distribution, which in turn comes from the self-regulating processes in the \texttt{SC-SAM}. 

Finally, the partner work of \citet{MoorePerez2026} uses our $b_{\phi}$ data to develop a framework for building informative priors given a set of comparison observations. Given that the largest deviations in $b_{\phi}$ often align with `less realistic' models, and with the knowledge of which parameter combinations yield the most realistic models, one can calculate a galaxy-formation marginalized prior for $b_{\phi}$. Using the notation from this work for the SMF, their Eq.\ 1.4 $b_{\phi}$ prior is:

\begin{equation}
    p(b_{\phi}|\Phi_{Y}) =\int d \boldsymbol{\theta}_{\rm{gf}}\ p\left(b_{\phi}|\boldsymbol{\theta}_{\rm{gf}} \right) p\left(\boldsymbol{\theta}_{\rm{gf}}|\Phi_{Y} \right).
\end{equation}

where $\theta_{\textrm{gf}}\equiv$ \textsf{SN1, SN2, AGN}.
From this, \citet{MoorePerez2026} generate initial priors for $f_{\textrm{NL}}$ analyses in a DESI-like ELG sample (a stellar mass and sSFR combined selection), finding their method reduces the uncertainties in $b_{\phi}$ by at least 88\% using the $z=0$ SMF. In combination with work like that in \citet{Shiferaw2025} building galaxy formation-motivated priors for other forms of galaxy bias, or the methods in \citep{YuNyugen2026} to leverage observed small scale clustering for a $b_{\phi}$ prior, we move closer to extracting precise cosmological information hand-in-hand with improved galaxy formation modeling.

\section{Conclusion} \label{sec:conclusion}
In this work, we seek to more deeply understand how variations in a galaxy formation model influence the $b_{\phi}$, a scale-dependent bias parameter induced by primordial non-Gaussianity of the local type (parametrized as $f_{\textrm{NL}}^{\textrm{loc}}$). 
$f_{\text{NL}}$ is a key target of cosmological galaxy surveys, and its non-zero detection will clarify the process of inflation. The large-scale structure of galaxies has scale-dependent signals induced by $f_{\text{NL}}$, and upcoming missions hope to measure $f_{\text{NL}}$ to within $\sigma_{f\text{NL}}=1$. However, galaxy power spectra can only measure the degenerate product $b_{\phi}f_{\text{NL}}$ -- therefore, any and all uncertainty in $b_{\phi}$ propagates to constraints on $f_{\text{NL}}$.

We use the Santa Cruz Semi-Analytic Model (\texttt{SC-SAM}) for galaxy formation, and vary three parameters of stellar and AGN feedback within it to study how the $b_{\phi}$ and $b_{\phi}(b_1)$ measured on galaxies varies across galaxy formation modeling. We run the \texttt{SC-SAM} models atop the merger trees of three separate-universe simulations over the amplitude of the primordial potential perturbation. With a handful of `one-at-a-time' models where each parameter is set to its minimum and maximum at a time, we contextualize the changes that result in the stellar-, SFR-, and sSFR- to halo mass relationships as the parameters change. We run 56 \texttt{SC-SAM} models where all three parameters are allowed to vary, and assess each one's `realism' when compared to the $z=0$ stellar mass functions \citep{Bernardi2018}. We present $b_{\phi}$ for galaxy samples selected by stellar mass, SFR, and sSFR for the central and representative $z=1$ ($z=0,\ 2$ in the Appendices). Additionally, we calculate $b_{\phi}(b_1)$ across all models for these three galaxy selections. The so-called universality relation for $b_{\phi}(b_1)$ has already been shown to not apply to galaxies due to galaxy assembly bias, and our work further emphasizes its failure for galaxy selections. 

Our work is novel in:
\begin{itemize}
    \item the breadth of \texttt{SC-SAM} model parameter variations we test, covering a large span of possible stellar mass functions and stellar-halo mass relationships;
    \item the connection between $b_{\phi}$, $b_{\phi}(b_1)$, and the galaxy-halo connection across the diversity of a galaxy formation model;
    \item the assessment of SFR and specific SFR selections alongside stellar mass;
    \item and the centralized comparison with other measurements of $b_{\phi}$ and related parameterizations with different galaxy formation models.
\end{itemize}

These are our results:
\begin{itemize}
    \item We find good agreement in the $b_{\phi}$ vs.\ stellar mass measurements of \citet{Barreira+2020_TNG}. We find a steeper $b_{\phi}(b_1)$ than \citet{Barreira+2020_TNG} see for stellar mass selections, with the strongest deviations for low-mass selections with small $b_1$. 
    \item SFR selections yield chaotic $b_{\phi}$ values mostly close to zero, and are strongly affected by our supernova feedback parameters and roughly increasing in $b_{\phi}$ with redshift. SFR selections also show a complex $b_{\phi}(b_1)$ centered about the expectations of universality, due to the many varied halo masses that can have a particular SFR. However, small SFR selections tend to dip below universality.
    \item We find sSFR is robust to even extreme parameter variations in both $b_{\phi}$ and $b_{\phi}(b_1)$. This results from the \texttt{SC-SAM} self-regulating the star forming main sequence to one slope, and the resulting common sSFR vs.\ M$_{\text{halo}}$ relationship that appears mostly insensitive to our parameter selections. This is a novel result, and confirming if it is true for other types of galaxy models and their variations would solidify sSFR as a model-agnostic galaxy selection for $f_{\textrm{NL}}$ constraints.
    \item We probe the first-order galaxy assembly bias across \texttt{SC-SAM} models, finding the expected assembly bias signal for stellar mass selections (where detaching galaxies from their halos' formation histories decreases their clustering).
    \item However, we find \textit{negative} assembly bias signals for SFR and sSFR selections (where detaching galaxies from their halos' formation histories \textit{increased} their clustering).
    \item We confirm that these galaxy assembly bias signals have little effect on our measured $b_{\phi}(b_1)$ relationships.
\end{itemize}

We also compare our results to previous work studying non-Gaussianity assembly bias with other semi-analytic models of galaxy formation: the Munich SAM in \citet{Reid2010}, and \texttt{Galacticus} in \citet{Marinucci2023}:
\begin{itemize}
    \item \citet{Reid2010} derive the non-Gaussian assembly bias correction factor $\Delta A_{\textrm{NG}}$ in the context of halo formation time. They measure $\Delta A_{\textrm{NG}}$ and $b_1$ for a run of the Munich SAM using the halo formation histories and clustering respectively, for stellar mass and SFR \textit{threshold} selections. We transform their reported measurements into the $b_{\phi}-b_{\phi \textrm{, univ.}}$ format we use, and measure $b_{\phi}$ for the same threshold selections across our \texttt{SC-SAM} models.
    We find their measurements in agreement with ours, indicating our \texttt{SC-SAM} variations encapsulate the $b_{\phi}$ behavior of the Munich SAM.
    \item \citet{Marinucci2023} measure the non-Gaussian assembly bias correction in the context of how a halo occupation distribution model responds to perturbations from non-Gaussianity. We repeat their experiment for stellar mass selections in central galaxies, finding agreement at the highest stellar masses and for \texttt{SC-SAM} models with very high SMFs. The often higher $\Delta b_{\phi}$ of our \texttt{SC-SAM} models is likely due to (standard) galaxy assembly bias that our simulation-measured merger trees show but extended Press-Schechter formalism merger trees do not include.
\end{itemize}

Finally, we summarize our results into guidance for observational analyses, emphasizing how there are many possible selections with confidently non-zero $b_{\phi}$ values even across our dramatic galaxy modeling. Partner work in \citet{MoorePerez2026} builds off our results to offer a framework for building informative priors given a set of reference observations, forging a path to physically-informed and observationally-grounded constraints on $f_{\textrm{NL}}$.

\begin{acknowledgments}
This project was built on an idea and a lot of code from Alex Barreira, for whom we are very deeply thankful. 
This work was carried out with the support and supercomputing resources of the Flatiron Institute, a part of the Simons Foundation. 
We thank those whose comments improved and influenced this work: Oliver Philcox, Sebastian Wagner-Careña, William Coulton, James Sullivan, Marco Marinucci, and the anonymous referee. 
\end{acknowledgments}

\vspace{5mm}

\software{
    \texttt{2LPTic} (\url{https://cosmo.nyu.edu/roman/2LPT/}), 
    \texttt{CAMB} \citep{CAMB}, 
    \texttt{ConsistentTrees} \citep{ConsistentTrees},
    \texttt{Cython} \citep{cython:2011}
    \texttt{matplotlib} \citep{Hunter:2007}, 
    \texttt{nbodykit} \citep{NbodyKit}, 
    \texttt{numpy} \citep{numpy}, 
    \texttt{pandas} \citep{mckinney-proc-scipy-2010, pandas_13819579}, 
    \texttt{Pylians} \citep{Pylians}, 
    \texttt{python} \citep{python}, 
    \texttt{Rockstar} \citep{Rockstar},
    and
    \texttt{scipy} \citep{2020SciPy-NMeth, scipy_14880408}.
This research has made use of NASA's Astrophysics Data System. A description of the numerical algorithms employed by the \texttt{AREPO} code is given in the original code papers \citep{2010MNRAS.401..791S,2011MNRAS.418.1392P,2013MNRAS.432..176P,2016MNRAS.455.1134P} and the release paper of the public version \citep{2020ApJS..248...32W}. Beige-to-purple colorbar from J. Borrow's \textit{swiftascmaps} project (DOI 10.5281/zenodo.5649258).
Software citation information aggregated using \texttt{\href{https://www.tomwagg.com/software-citation-station/}{The Software Citation Station}} \citep{software-citation-station-paper, software-citation-station-zenodo}.
}

%%%%%%%%%%%%%%%%%%%%%%%%%%%%%%%%%%%%%%%%
%%%%%%%%%%%%%%%%%%%%%%%%%%%%%%%%%%%%%%%%
%%%%%%%%%%%%%%%%%%%%%%%%%%%%%%%%%%%%%%%%
%%%%%%%%%%%%%%%%%%%%%%%%%%%%%%%%%%%%%%%%
%%%%%%%%%%%%%%%%%%%%%%%%%%%%%%%%%%%%%%%%

\newpage

\appendix

\section{Redshift evolution of $\lowercase{b}_{\phi}$ across galaxy selections} \label{app:bphi_zevol}

\begin{figure}
    \centering
    \includegraphics[width=\linewidth]{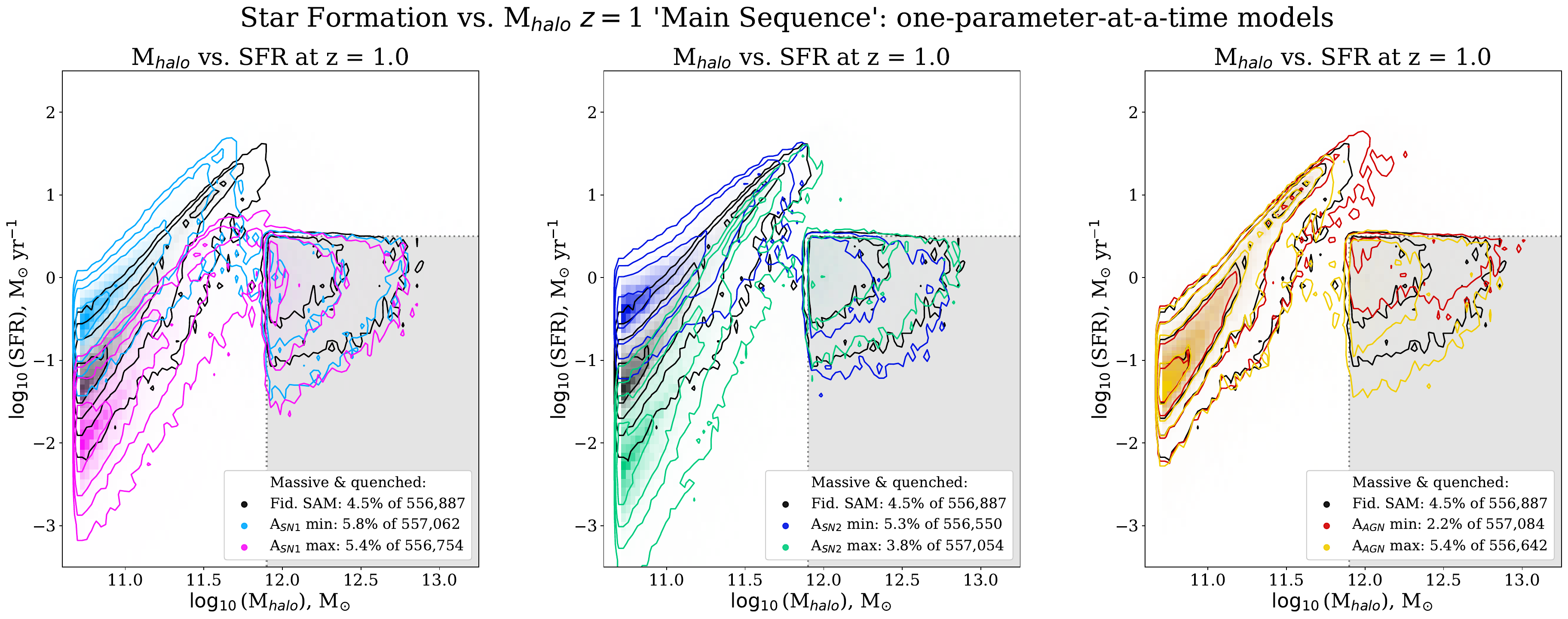}
    \includegraphics[width=\linewidth]{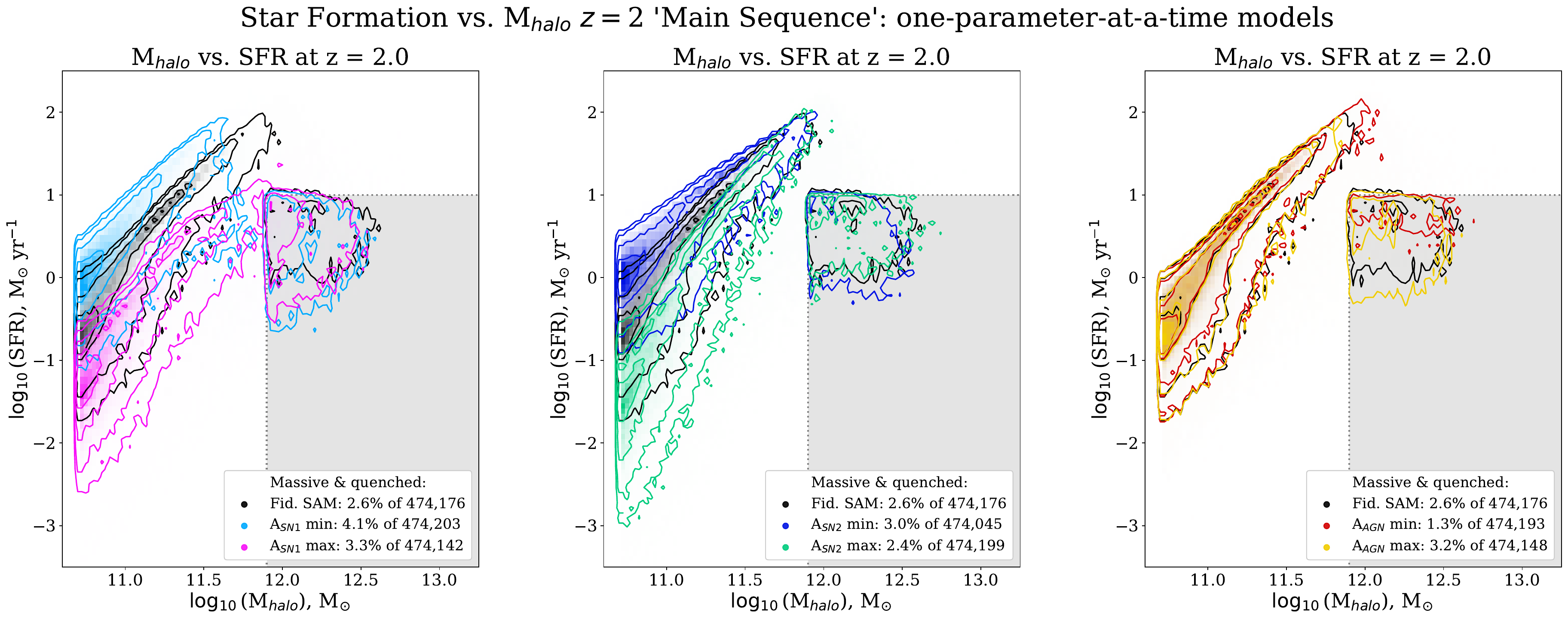}
    \caption{The $z=1$ (upper) and $z=2$ (lower) instantaneous SFR vs.\ M$_{\textrm{halo}}$ (upper) relationship, like Figure \ref{fig:1P_sSFHMRs}, for our one-at-a-time parameters. }
    \label{fig:1P_z0n2_SFHMRs}
\end{figure}

\begin{figure}
    \centering
    \includegraphics[width=\linewidth]{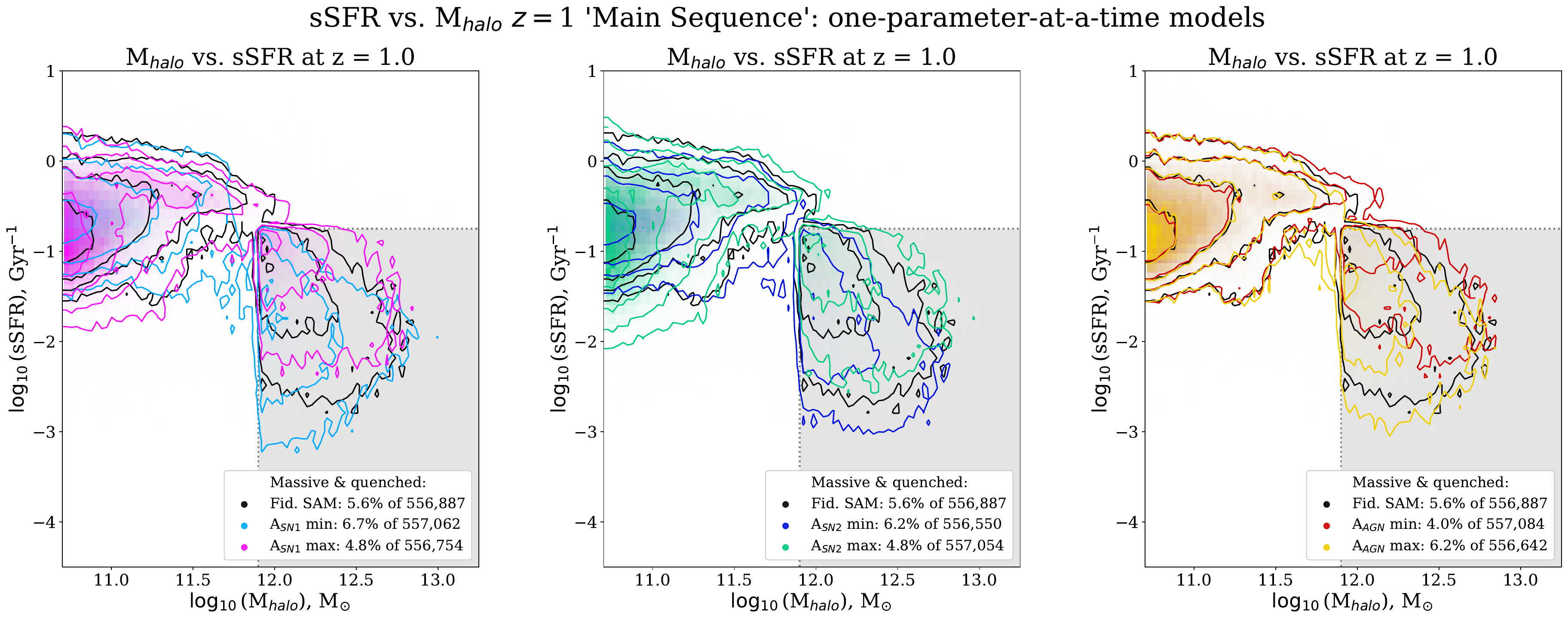}
    \includegraphics[width=\linewidth]{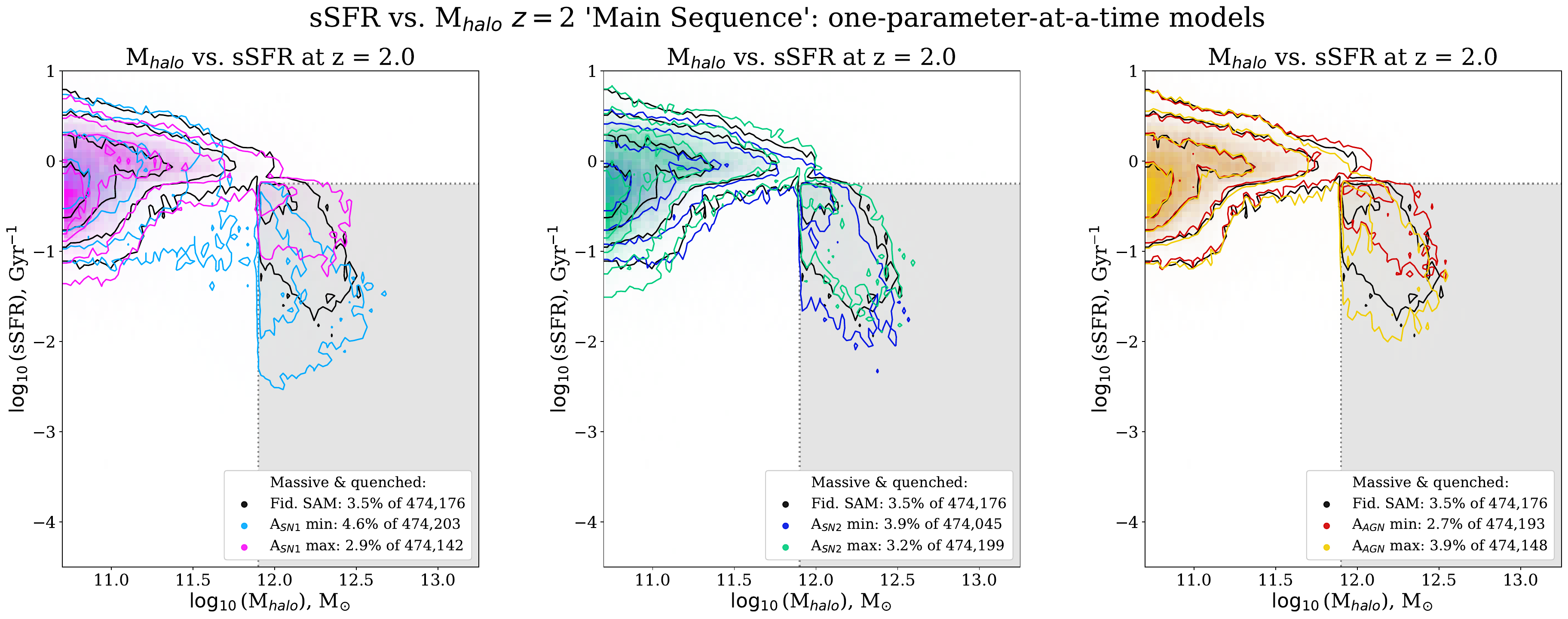}
    \caption{The $z=1$ (upper) and $z=2$ (lower) specific SFR vs.\ M$_{\textrm{halo}}$ (upper) relationship, like Figure \ref{fig:1P_sSFHMRs}, for our one-at-a-time parameters. }
    \label{fig:1P_z0n2_sSFHMRs}
\end{figure}

In this section, we discuss the redshift evolution of $b_{\phi}$ for our different galaxy selections. 
For context and completeness, we recreate Figure \ref{fig:1P_sSFHMRs}'s M$_{\textrm{halo}}$ vs.\ SFR or sSFR relationships for $z=1, 2$ in Figures \ref{fig:1P_z0n2_SFHMRs} and \ref{fig:1P_z0n2_sSFHMRs}, respectively. The effect of each \texttt{SC-SAM} parameter is quite consistent across redshifts. We see both the M$_{\textrm{halo}}$ vs.\ SFR and M$_{\textrm{halo}}$ vs.\ sSFR relationships shift upwards, to higher (s)SFR at higher redshifts.

First, we recreate Figure \ref{fig:1P_sSFHMRs} for $z=1$ and $z=2$ to help contextualize results. Figure \ref{fig:1P_z0n2_SFHMRs} shows the SFR vs.\ M$_{\textrm{halo}}$ relationship for the one-at-a-time minimum/maximum models, and Figure \ref{fig:1P_z0n2_sSFHMRs} shows the sSFR vs.\ M$_{\textrm{halo}}$ relationship. Both these relationships shift up to higher SFRs across all models as star formation increases to its peak at cosmic noon. We again create qualitative regions of massive and quenched galaxies, noting how their numbers decrease at higher redshifts.

\begin{figure*}
 \centering
 \includegraphics[width=\textwidth]{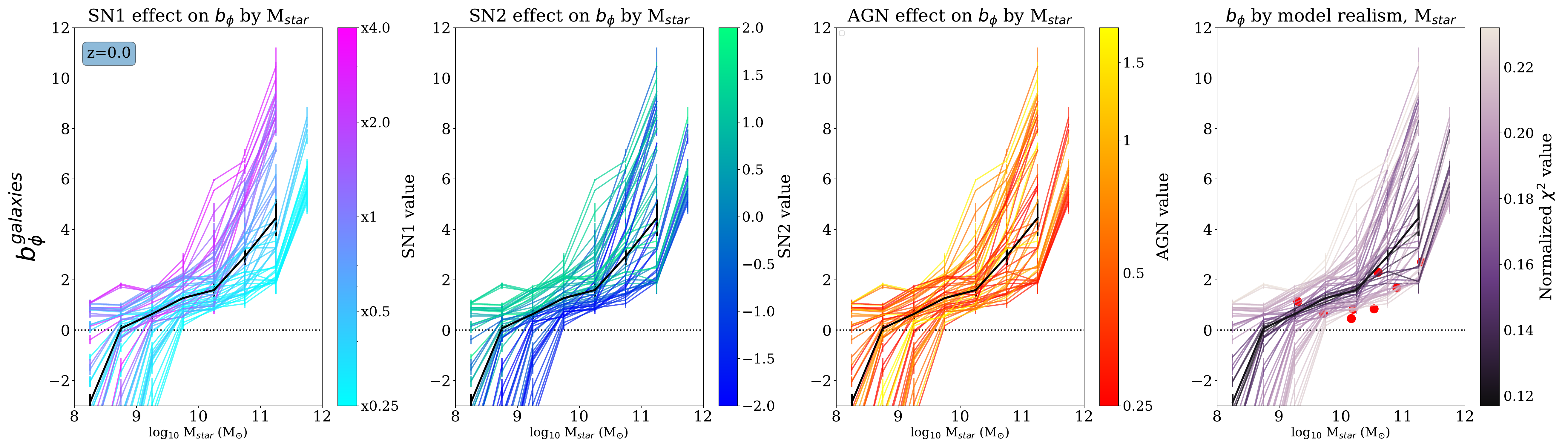}
 \includegraphics[width=\textwidth]{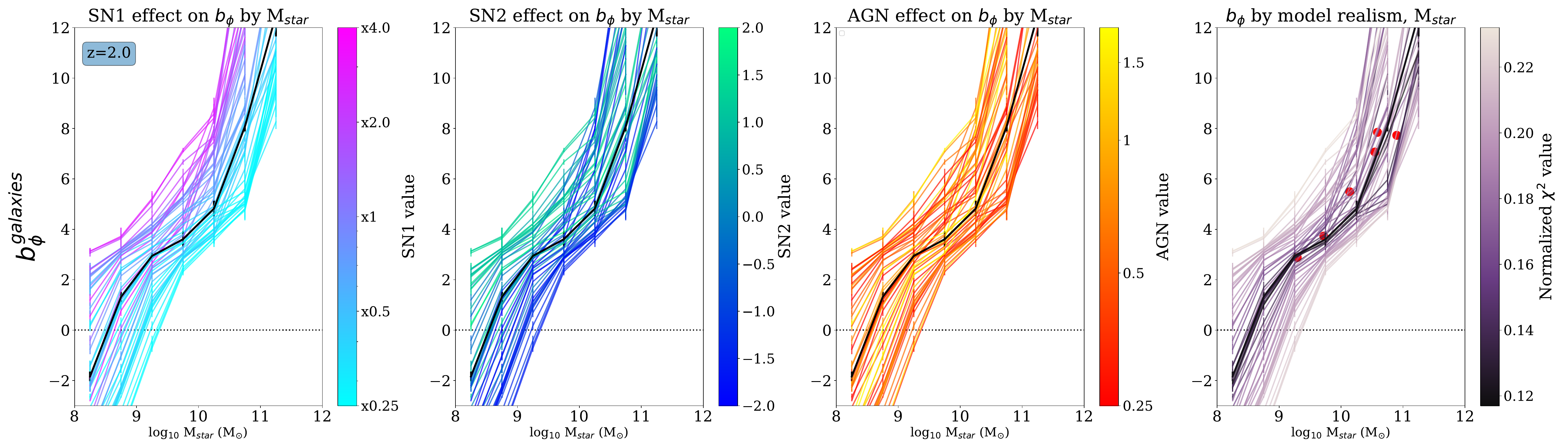}
 \caption{$b_{\phi}$ vs.\ stellar mass at $z=0$ (top) and $z=2$ (bottom) across all 56 \texttt{SC-SAM} models. Plots in each row are identical, but are colored according to each model's value of \textsf{SN1}, \textsf{SN2}, and \textsf{AGN} (left to right) or relative `realism' (right-most plot, most-to-least realistic in black-to-beige; see Figure \ref{fig:SMFnSHMRallmodels} and the end of \textsection 2.5). We mark $b_{\phi}=0$ with a black dotted line to guide the eye, plot the fiducial \texttt{SC-SAM} model in black for reference, and show the relevant data from \citet{Barreira+2020_TNG} Figure 4 as red circles.}
 \label{fig:bphi_stemass_0n2}
\end{figure*}

\textbf{Stellar mass}, Figure \ref{fig:bphi_stemass_0n2}: 

All models, including the fiducial \texttt{SC-SAM}, show $b_{\phi}$ increasing with stellar mass selection, and that higher redshifts have larger $b_{\phi}$ overall, as expected given $b_{\phi}$(M$_{\textrm{halo}}$). The tangent-like shape also shifts somewhat left to lower masses at higher redshifts; logical given that galaxies are less massive at higher redshift and form first at the most dense peaks. The influence of \textsf{SN1/2} is present across redshift, with strong \textsf{SN1/2} values pushing the relationships up to higher $b_{\phi}$ and left to lower masses comparatively. We find general agreement with the measurements from \citet{Barreira+2020_TNG} across redshifts, barring a few stellar mass selections.

\begin{figure*}
 \centering
 \includegraphics[width=\textwidth]{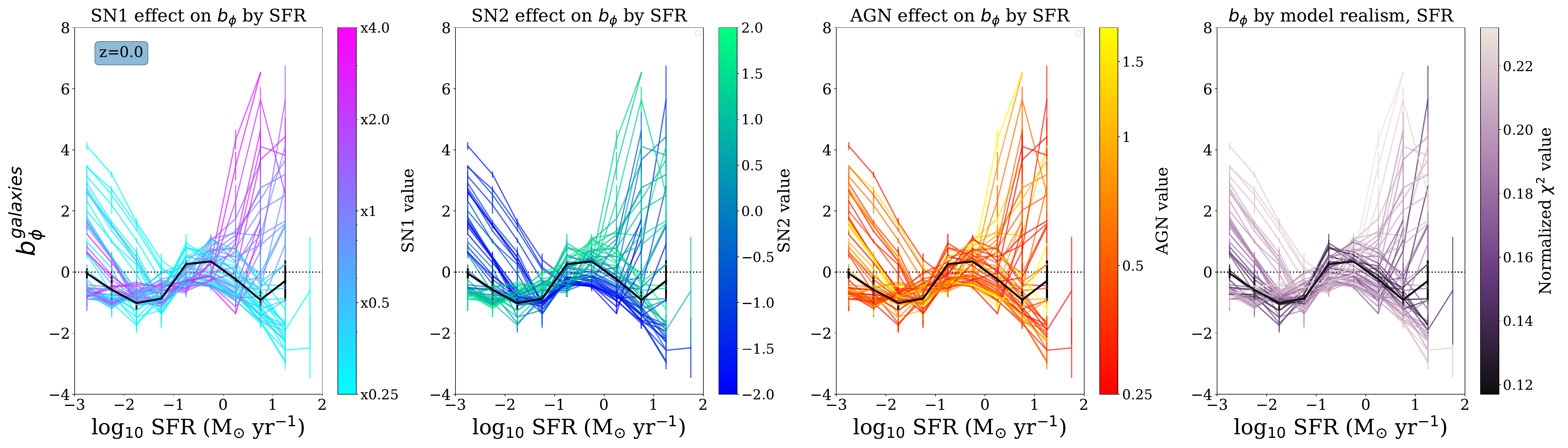}
 \includegraphics[width=\textwidth]{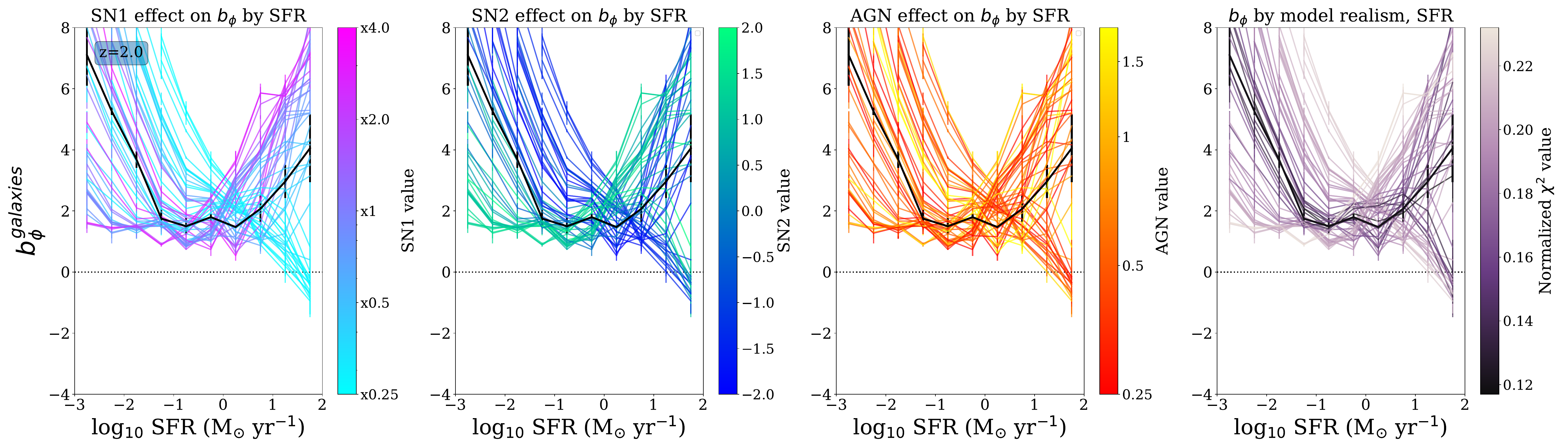}
 \caption{Like Figure \ref{fig:bphi_stemass}, but for SFR (M$_{\odot}$ yr$^{-1}$) selections at $z=0$ (top) and $z=2$ (bottom).}
 \label{fig:bphi_sfr_0n2}
\end{figure*}

\textbf{Star Formation Rate}, Figure \ref{fig:bphi_sfr_0n2}: 

At $z=0$, nearly all models dance around $b_{\phi} \sim 0$, particularly the more realistic models. As with stellar mass, large \textsf{SN1/2} values yield to strongly positive $b_{\phi}$ at high SFR selections but models large unrealistic when compared to the $z=0$ SMFs. 

We find that $z=2$ (cosmic noon, the peak of star formation in the universe) sees the same pattern but with the fiducial SAM's $b_{\phi}\approx 2$ at most SFRs. All together, the $b_{\phi}$ vs.\ SFR relations all shift upwards with higher redshift, with cosmic noon ($z=2$) seeing nearly all models and SFR$>0.01$ M$_{\odot}$ yr$^{-1}$ selections clearing $b_{\phi}>0$. This is consistent with the $b_{\phi}$ of all halos increasing more sharply at higher redshift.

\begin{figure*}
 \centering
 \includegraphics[width=\textwidth]{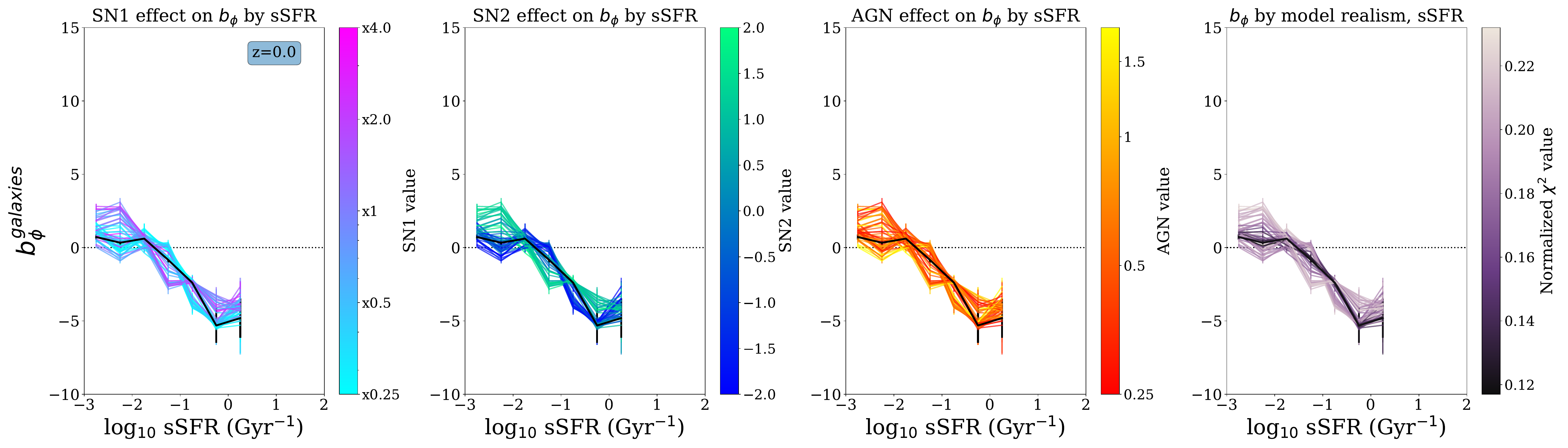}
 \includegraphics[width=\textwidth]{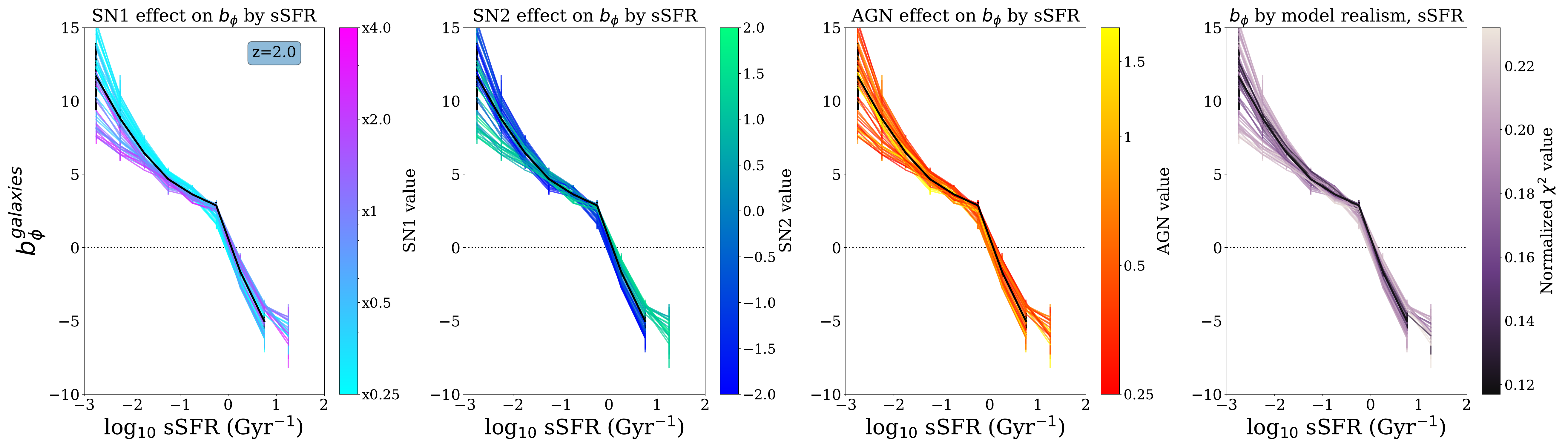}
 \caption{Like Figures \ref{fig:bphi_stemass} and \ref{fig:bphi_sfr}, but for specific SFR selections at $z=0$ (top) and $z=2$ (bottom) (where sSFR = M$_{\mathrm{star}}\ \div$ SFR, Gyr$^{-1}$).}
 \label{fig:bphi_ssfr_0n2}
\end{figure*}

\textbf{Specific SFR}, Figure \ref{fig:bphi_ssfr_0n2}: 

The collapsing of SAM models into a single rough relationship is clear across redshift for sSFR selections. At $z=0$, all models begin near $b_{\phi} \sim 0$ for the smallest sSFR selections, and then dip to strongly negative $b_{\phi} \sim -5$. 
% before rising to $b_{\phi} \sim -2$ at the largest sSFR bin. 

By $z=2$, the smallest sSFR bins are strongly positive (depending on the weakness of the \textsf{SN1/2} parameters); $b_{\phi}$ then crosses from positive to negative at around sSFR=0.5 Gyr$^{-1}$ until bottoming out at about $b_{\phi} \sim -5$ for the largest sSFR bins. The sSFR values where the relationships cross into negative $b_{\phi}$ appear to correspond to the central sSFR of the flat `main sequence' in sSFR vs.\ M$_{\text{halo}}$, which is densest at log$_{10}$sSFR $\sim$ [-1.5, -0.75, -0.25] Gyr$^{-1}$ for $z=[0,1,2]$. The agreement across models regardless of realism indicates sSFR might be a galaxy selection robust to different parameterizations.

%---------------------------%
%---------------------------%
%---------------------------%

\section{Redshift evolution of $\lowercase{b}_{\phi}$ vs.\ $\lowercase{b}_1$ across galaxy selections} \label{app:b1bphi_zevol}

\begin{figure*}
 \centering
 \includegraphics[width=\textwidth]{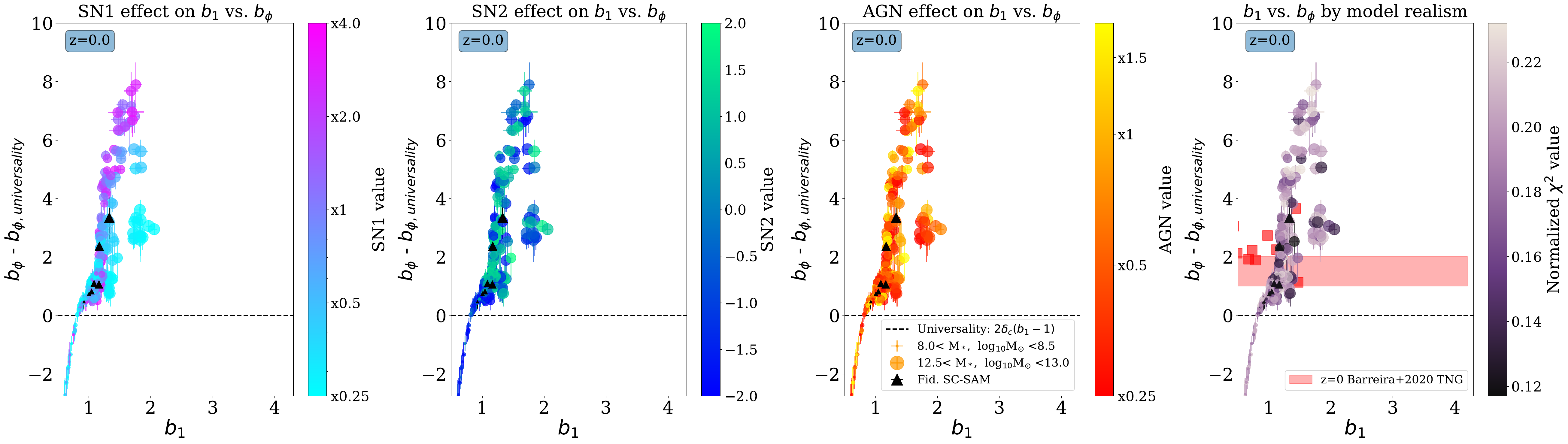}
 \includegraphics[width=\textwidth]{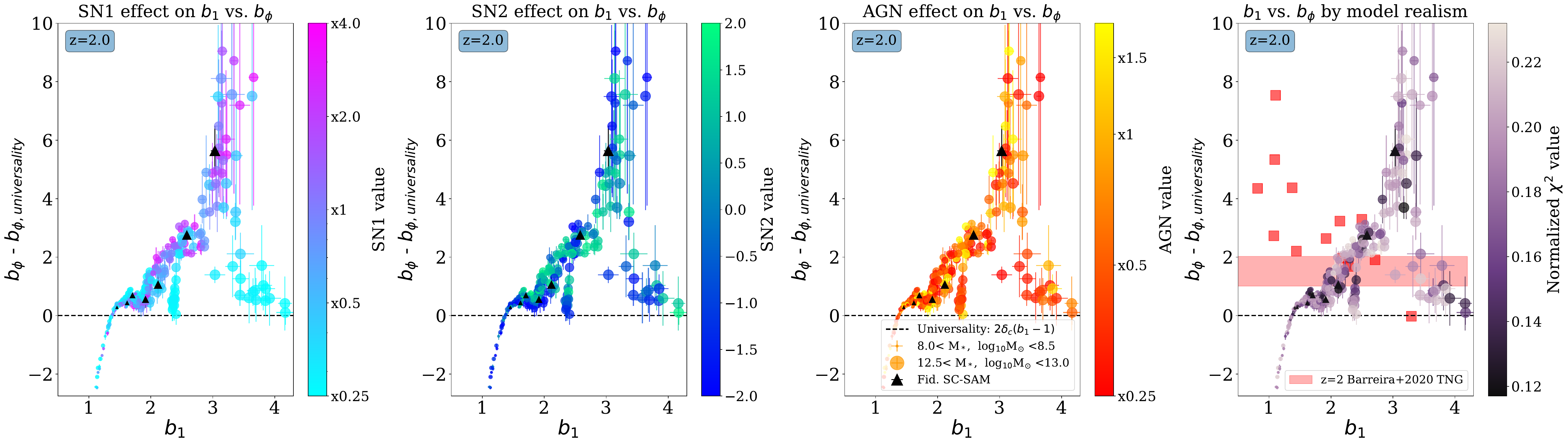}
 \caption{$b_{\phi} -b_{\phi, \textrm{univ.}}$ vs.\ $b_1$ relationship at $z= 0$ (top) and $z=2$ (bottom) across all \texttt{SC-SAM} models for stellar mass selections. The $y$-axis is normalized to the predicted $b_{\phi}$ from the universality relation (Eq. \ref{eq:universality}), so that deviations from zero indicate $b_{\phi}$ far from the universality relation prediction. Colored as earlier figures, with the fiducial \texttt{SC-SAM} model as black triangles.
 The size of markers corresponds qualitatively to the selection, where the largest points indicate the largest stellar mass bins. The red points are replotted from \citet{Barreira+2020_TNG} Fig. 3, and the red shaded area indicates the resulting proposed variant relationship for stellar mass selections they propose, $b_{\phi}=2 \delta_c(b_1 -p),\ p \in [0.4, 0.7]$.
 \texttt{SC-SAM} models show a large positive deviation from the universality relation throughout, where models with weak \textsf{SN1} drop close to universality and into agreement with the \citet{Barreira+2020_TNG} relationship. % at $z>1$.
 }

 \label{fig:b1Vbphi_stemass_0n2}
\end{figure*}

\begin{figure}
 \centering
  
 \includegraphics[width=0.98\textwidth]{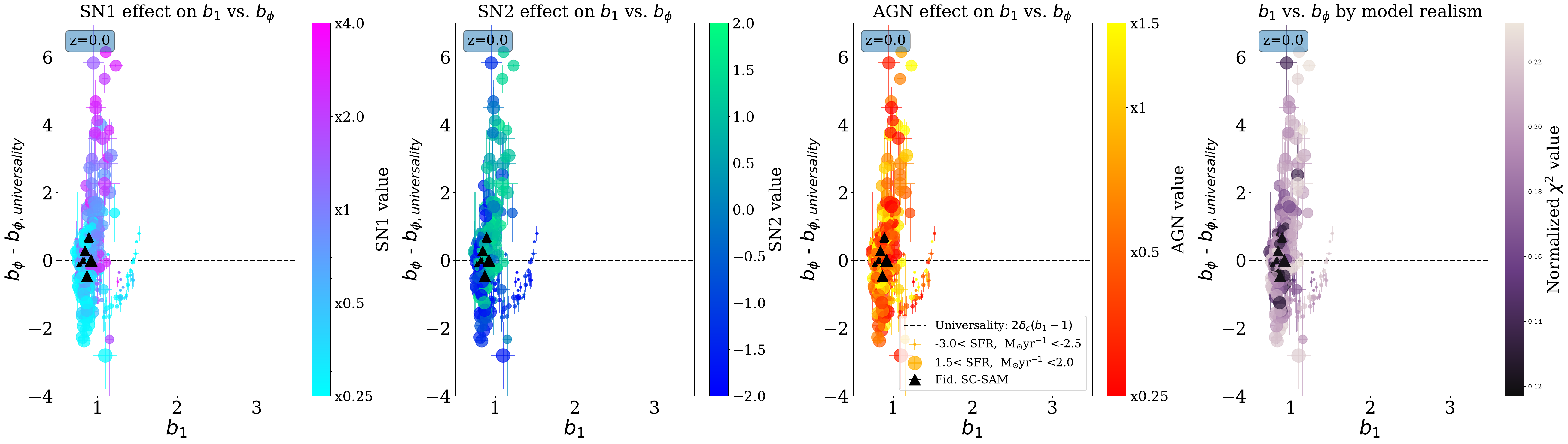}
 \includegraphics[width=0.98\textwidth]{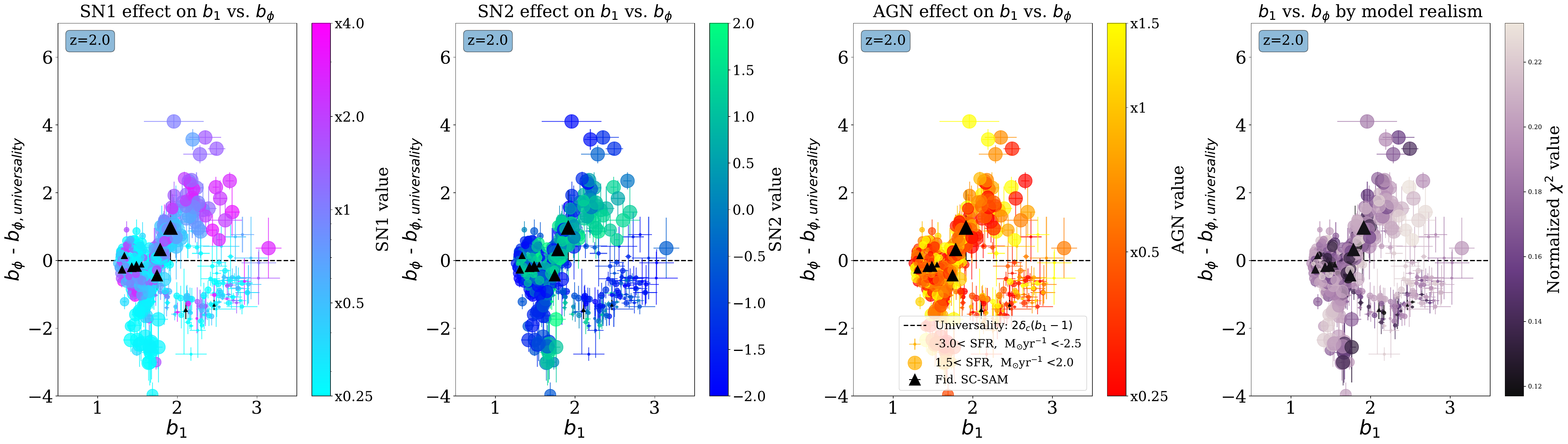}
 \caption{Like Figure \ref{fig:b1Vbphi_stemass}, but for SFR selections at $z=0$ (top) and $z=2$ (bottom). The \texttt{SC-SAM} models span a wide and mostly featureless range of $b_{\phi}$ in a narrow range of $b_1$.  
 The highest- and lowest- value SFR selections show uniquely diverging behavior to mostly positive and negative $b_{\phi}$ respectively, apparently driven by \textsf{SN1} and/or \textsf{SN2}.}
 \label{fig:b1Vbphi_sfr_0n2}
\end{figure}

\begin{figure}
 \centering
 \includegraphics[width=0.98\textwidth]{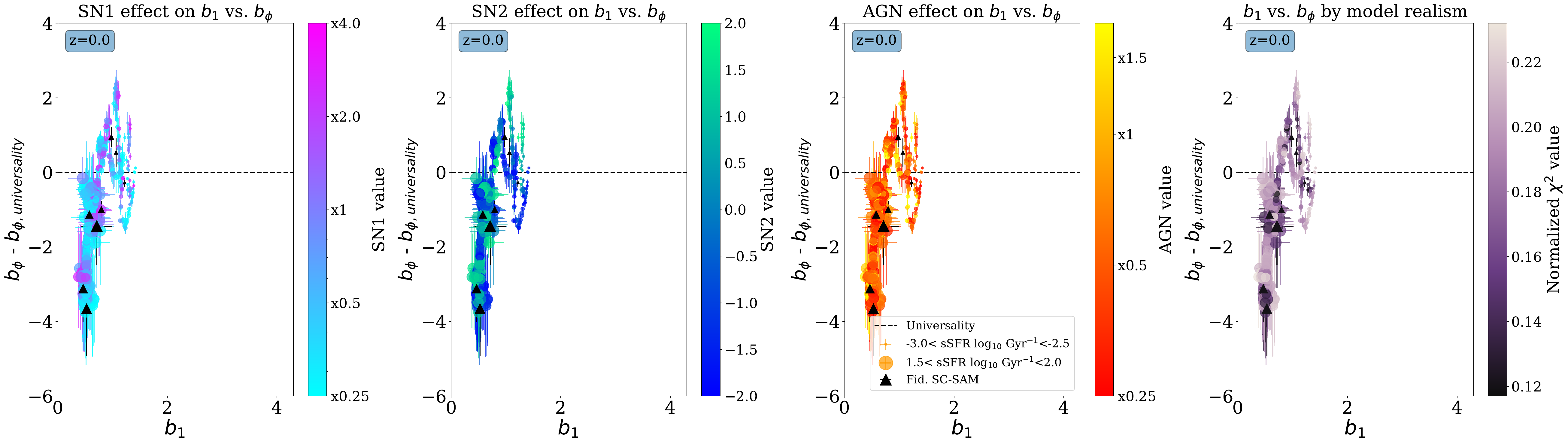}
 \includegraphics[width=0.98\textwidth]{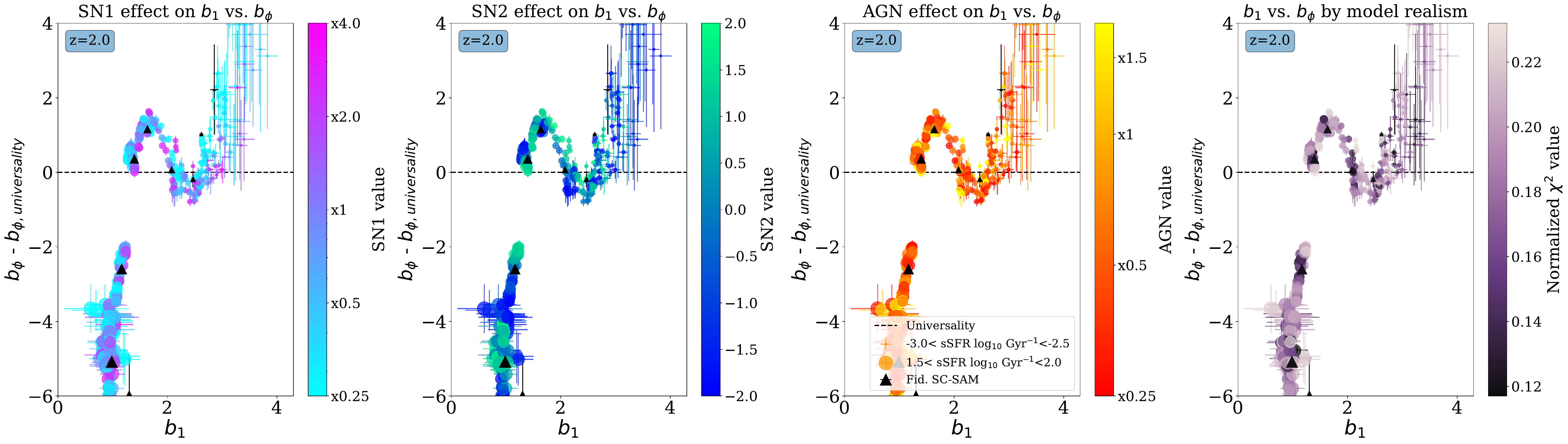}
 \caption{Like Figures \ref{fig:b1Vbphi_stemass} and \ref{fig:b1Vbphi_sfr}, but for specific SFR selections at $z=0$ (top) and $z=2$ (bottom). As seen in Figures \ref{fig:bphi_1Psets} and \ref{fig:bphi_ssfr}, sSFR selections follow a tight $b_{\phi}$ vs.\ $b_1$ relationship whose $b_{\phi}$ spread is narrow appears agnostic to \texttt{SC-SAM} model variations.}
 \label{fig:b1Vbphi_ssfr_0n2}
\end{figure}

In this section, we discuss the redshift evolution of $b_{\phi}$($b_1$) to help inform $f_{\text{NL}}$ studies across redshift. 

\textbf{Stellar mass}, Figure \ref{fig:b1Vbphi_stemass_0n2}: 

For $z=0$, the \texttt{SC-SAM} models disagree with \citep[red points and band]{Barreira+2020_TNG} for most galaxy selections, particularly for low-mass / low $b_1$ selections.

The $z=2$ relation, though still far above universality, sees the low-value \textsf{SN1} `island' at high stellar masses drop closer to universality. Given the agreement with \citet{Barreira+2020_TNG} in $b_{\phi}$, this looks to be a mismatch in our measured $b_1$ for stellar mass samples.

\textbf{Star formation rate}, Figure \ref{fig:b1Vbphi_sfr_0n2}: 

At $z=0$, all models clump at $b_1 \sim 1$ with a broad range of $b_{\phi}$ values. Models with high \textsf{SN1/2} dominate the highest $b_{\phi}$ points, and the most realistic models tend to lie close to $b_{\phi} \sim b_{\phi, \textrm{universality}}$. The smallest SFR selections create an interesting tail, finding slightly higher $1<b_1<2$, though are dominated by models with very low values of \textsf{SN1/2} and of moderately poor realism. %As we identify in Appendix \ref{app:AssemblyBias}, massive quenched galaxies 

Cosmic noon at $z=2$ show a very interesting divergence: as samples reach a wider range of $1.5<b_1<3$, there are two rough relationships that appear depending on SFR selection. The smallest SFR bins still show the negative ($b_{\phi}-b_{\phi, \textrm{univ.}}$ tail) mostly made of low-value \textsf{SN1/2} models (across the breadth of realism). The behavior of high SFR selections appears to depend on \textsf{SN1} the most (with some weaker influence of \textsf{SN2}): the lowest \textsf{SN1} create a clear clump of negative $b{\phi}$ values near $1.5 < b_1 < 2$, and the strongest \textsf{SN1/2} models create another clump of $b_{\phi}$ values above universality  near $2 < b_1 < 2.5$.

\textbf{Specific SFR}, Figure \ref{fig:b1Vbphi_ssfr_0n2}:

Redshift plays a strong effect on sSFR selections, with $z=0$ finding nearly all models and selections clumped around $b_1 <2$ and $b_{\phi}$ below universality, with no visible pattern emerging according to the parameters. 

A small bump in $z=0\ \&\ 1$ becomes more apparent at $z=2$ near $(b_1 \approx 2,\ b_{phi} \approx 4)$. The selections zig-zag up over, down under, and then far above universality. The highest sSFR  selections clump more tightly about $(b_1 \approx 1,\ b_{\phi} - b_{\phi, \textrm{univ.}} \approx -4)$.

%---------------------------%
%---------------------------%
%---------------------------%
%---------------------------%
%---------------------------%
%---------------------------%

\section{Assembly Bias \& the effect of Massive Quenched Galaxies} \label{app:AssemblyBias}

\subsection{Approximating our levels of galaxy assembly bias} \label{app:2ptCFAssemblyBias}

Here, we repeat the \citet{Hadzhiyska2021} measurement of galaxy assembly bias for all our one-at-a-time \texttt{SC-SAM} models. 
The fiducial \texttt{SC-SAM} model is known to recreate a $\sim 10\%$ level of assembly bias for stellar mass selections, like observed in IllustrisTNG (\citealt{Hadzhiyska2021}) -- here we see how this varies for the one-parameter-at-a-time \texttt{SC-SAM} models, and also for SFR and sSFR selections.

We define galaxy assembly bias here as an additional clustering signal beyond what would be expected if galaxy clustering only depended on halo mass. To do this, we first bin a given \texttt{SC-SAM} model galaxy catalog into thin bins of halo mass (here, 0.2 dex-wide). The galaxies are randomly shuffled to another halo in the same halo mass bin, and are placed at the position of whatever galaxy was previously there\footnote{Note that unlike the process of \citealt{Shiferaw2025}, we do maintain the number density of galaxies consistent, we do not distinguish between centrals and satellites, and there will be muddling in the positions of the galaxies.}. This disconnects the galaxy from its original halo's formation history, but maintains the underlying signal of halo assembly bias due to halo mass. We then apply the same galaxy selection to both the unmodified and shuffled catalogs, and calculate both the real-space two-point correlation function and $b_1$ for both as described in \textsection \ref{subsec:b1calcmethod} (N$_{\text{mesh}}$=64, $k_{\textrm{max}}=0.15$ Mpc $h^{-1}$).

For the real-space two-point correlation function $\xi$, we compare the ratio of $\xi_{\text{shuffled}} \div \xi_{\text{original}}$, for a \textit{threshold} galaxy selection, to broadly compare to the $z=0$ results found in \citet{Hadzhiyska2021}. For these threshold selections, we select $z=1$ galaxies with the largest stellar mass, SFR, or sSFR until reaching a density of 0.002 Mpc$^{-3}$. We use the \citet{L-S1993} estimator for the correlation function, and use the same random catalog for both catalogs. We use the jackknife method with 27 total sub-volumes to measure errors. 

\begin{figure}
 \centering
 \includegraphics[width=\textwidth]{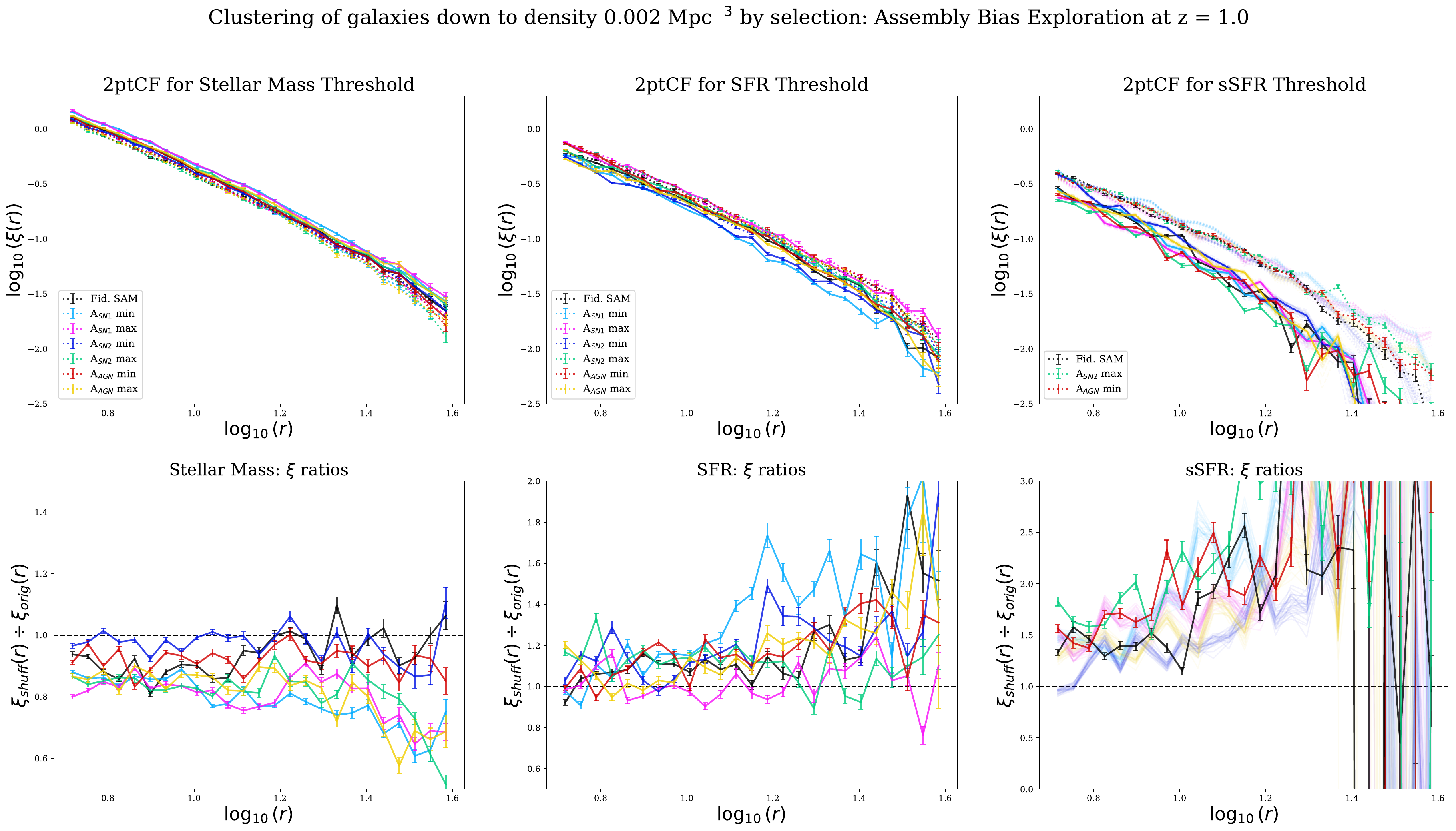}
 \caption{$z=1$ comparison of real-space two-point correlation functions for the unmodified SAM models vs.\ if each galaxy is shuffled to a halo of similar mass before measuring $b_1$. Galaxies are selected by a threshold of stellar mass (left), SFR (center), or sSFR (right) until reaching 0.002 Mpc$^{-3}$.
 }
 \label{fig:2ptCFcombos}
\end{figure}

Figure \ref{fig:2ptCFcombos} shows the real-space 3D two-point correlation functions at $z=1$ (top) for the original catalogs (solid lines) and shuffled catalogs (dotted lines); and the ratio of the correlation functions (shuffled $\div$ original, bottom). We mark a perfect match (i.e.\ no assembly bias) at $\xi_{\textrm{shuffled}}\div\xi_{\textrm{orig}}=1$. The columns of plots indicate which galaxy property was used for the threshold selection for a density of 0.002 Mpc$^{-3}$: stellar mass (left), SFR (center), or sSFR (right). 

For stellar mass, we recreate the result of \citealt{Hadzhiyska2021} at $z=1$, with most of the one-at-a-time models showing that shuffling galaxies to similarly-massive halos decreases their clustering by 5-20\%. What each parameter does is complex and varied -- for example, the minimum \textsf{SN2} model shows nearly no deviation for all distance scales, and oppositely, the two extreme models of \textsf{SN1} show a 20$\%$ or more decrease in clustering after shuffling, depending on the distance scale.

For SFR and sSFR, we find a noisier but inverted assembly bias signal, such that galaxies shuffled in their halo mass bin are \textit{more} correlated than their original distribution. 
We note this anti-correlation does agree qualitatively with the results of \citet{Croton2007}; they noted blue galaxies are often centrals but rarely in the largest halos, and that blue galaxies have a lower formation redshift than red galaxies. 
However, they noted concentration and formation time were not enough to fully account for the assembly bias.
The precise explanation is still not clear, and requires a study into the correlation between (s)SFR and a galaxy's formation history beyond the scope of this work. 

Finally, each one-at-a-time model leads to a different amount of assembly bias, and in a unique way for our three galaxy properties. 
For example, the minimum and maximum \textsf{SN1} parameters both show the most assembly bias in stellar mass, but strongly diverge in SFR. 
More deeply understanding how and why parameter variations change the \texttt{SC-SAM} galaxy assembly bias is intriguing but also beyond our work here.

%---------------------------%
%---------------------------%
%---------------------------%

\subsection{Effect of Assembly Bias on b$_1$} \label{app:b1AssemblyBias}

We now extend this test into the $b_1$ space to answer: how large is the effect of galaxy assembly bias on our measured $b_{\phi}(b_1)$? 
We compare $b_1$ for samples where galaxies live in their original halos vs.\ after they have been shuffled to a halo of a similar mass. We measure $b_1$ for the shuffled catalogs as we did in \textsection \ref{subsec:b1calcmethod}, applying the exact same galaxy selections to the original and shuffled catalogs. 
However, we find that the massive quenched galaxies (i.e.\ those which lie beyond either the SFR vs.\ M$_{\textrm{halo}}$ or sSFR vs.\ M$_{\textrm{halo}}$ `main sequences') strongly affect our $b_1$ experiment for assembly bias. For example, when attempting the comparison of $b_1$ measured on the original SFR- or sSFR-selected sample vs.\ after all galaxies have been shuffled into another halo of similar mass, various selections give \textit{no} information. In particular, very low SFR or sSFR selections will catch a decent spread of halo masses, decreasing the measured $b_1$ for the shuffled galaxies. 
Phrased differently, the strong clustering of the massive quenched objects is easily diluted with shuffling, and the complexity of the (s)SFR-M$_{\textrm{halo}}$ relationships can make their $b_{1}$ challenging to predict. 

\begin{figure}
 \centering
 \includegraphics[width=\textwidth]{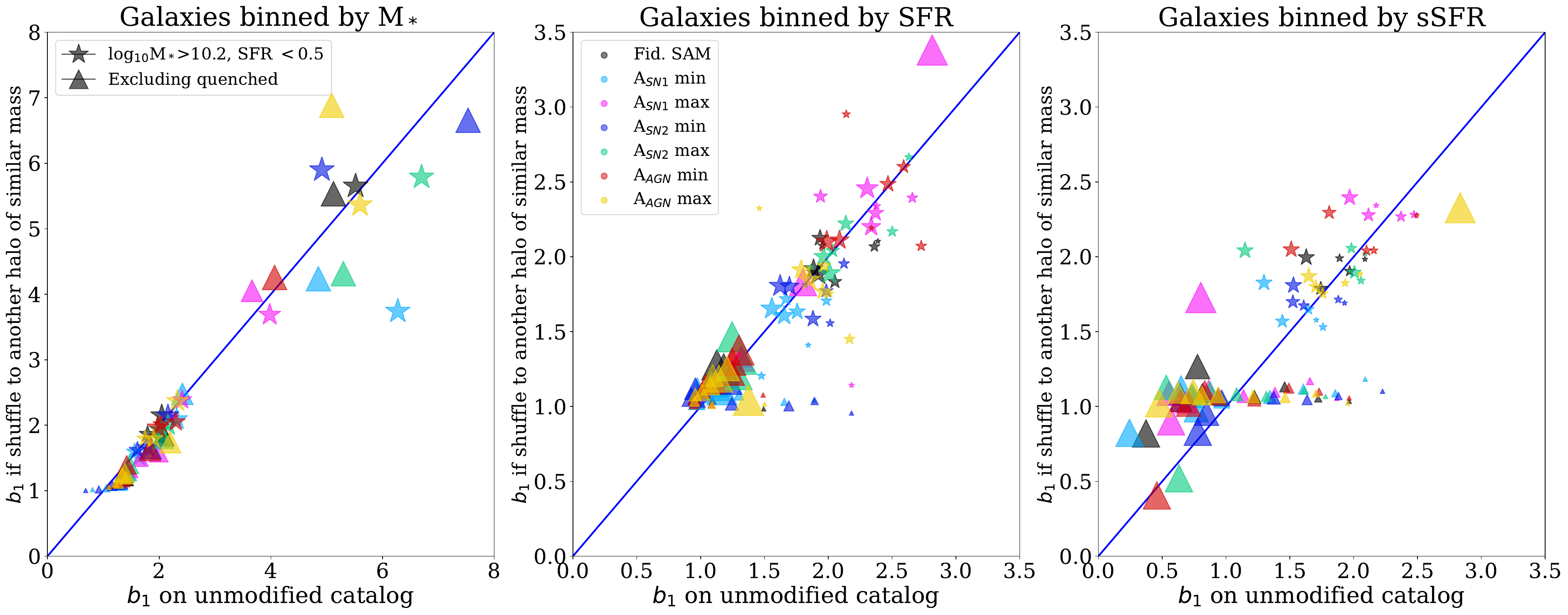}
 \caption{$z=1$ comparison of $b_1$ for the unmodified one-at-a-time \texttt{SC-SAM} models vs.\ if each galaxy is shuffled to a halo of similar mass before measuring $b_1$. Galaxies are binned by stellar mass (left), SFR (center), or sSFR (right), with the size of the marker indicating the relative size of the selection. The stars indicate massive quenched galaxies pulled out to examine assembly bias, and triangles are all other galaxies in a catalog.
 }
 \label{fig:b1_AssemBias_wquenched}
\end{figure}

However, if we repeat this experiment only with massive quenched objects, their true measured and shuffled $b_1$ values are consistent (indicating they are not particularly special galaxies, just within massive halos). 
We separate out massive quenched galaxies with log$_{10}$M$_*>10.2$ M$_{\odot}$ and SFR$<0.5$ M$_{\odot}$ yr$^{-1}$ for $z=1$, and plot their compared $b_1$ values as stars in Figure \ref{fig:b1_AssemBias_wquenched} for the `one-at-a-time' \texttt{SC-SAM} models. We plot the compared $b_1$ for all the remaining galaxies as triangles (`excluding quenched'). On the whole, the massive quenched galaxies behave as expected when separated out, showing the rough level of galaxy assembly bias we saw before. We see still some bins of small SFR or sSFR where the $b_1$ of the shuffled catalog is much smaller than the unmodified catalog, due to a still-wide range of halo masses averaging out to a lower $b_1$ value. 

All together, this confirms that galaxy assembly bias, on the whole, has a limited effect on our measured $b_{\phi}(b_1)$, with most selections for all one-at-a-time models seeing a $<10\%$ effect. 
As a final reminder, the way $b_{\phi}$ is studied in separate-universes is on the effect on galaxy numbers; shuffling the galaxies does not change their galaxy properties and therefore neither their measured $b_{\phi}$. 

%---------------------------%
%---------------------------%
%---------------------------%
%---------------------------%
%---------------------------%
%---------------------------%

\section{Detailed comparisons to studies of non-Gaussian assembly bias with SAMs} \label{app:PNGwSAMsdetails}

Non-Gaussian assembly bias was first derived with the EPS formalism using a peak-background split of the Gaussian field \citep{Slosar2008}, given how the emergent clustering of rare peaks in such a field would respond to non-Gaussian initial conditions \citep{Dalal2008a,Dalal2008b}. As \citet{Slosar2008} show, knowing how mass functions of some tracer responds to changing $\sigma_8$ under Gaussian initial conditions allows one to predict the non-Gaussian scale-dependent halo bias for some $f_{\textrm{NL}}$ for a given halo mass at some redshift. However, given that we observe galaxies and other biased tracers of halos, several works have attempted to derive and study how galaxies' non-Gaussian bias will deviate from this base non-Gaussian scale-dependent halo bias.

In \citet{Marinucci2023} they extend the non-Gaussian halo assembly bias to galaxies using a Halo Occupation Distribution (HOD, i.e.\ assigning how many centrals or satellites inhabit halos according to the halos' properties, expanding the work of \citealt{Voivodic2021}). They express $\bar{b}_{\phi}$ (the see Eq.\ \ref{eq:allNGassemblybiasdefs}) as the weighted average over halo masses incorporating the response of the halo mass function in the context of halo occupation (as do \citealt{Voivodic2021} in their Eq.\ 3.2). Where $b^h_{\phi}$ is the response of the overall halo mass function to changing $\sigma_8$,% ($b^h_{\phi}$, well-approximated by universality for halos to first order), 
$\bar{b}_{\phi}$ functionally scales this by the HOD $\langle N_g(\textbf{S}|M_h,z) \rangle$ of galaxies for some selection \textbf{S} within halos of $M_h$ at redshift $z$:

\begin{equation}
\begin{split}
    \bar{b}_{\phi} (\textbf{S}, z) = \frac{1}{n_g} \int dM_h\ b_{\phi}^h(M_h, z)\ \bar{n}_h(M_h,z)\ \langle N_g(\textbf{S}|M_h,z) \rangle,\\ 
    \textrm{where}\ \langle N_g(\textbf{S}| M_h,z) \rangle \equiv N_c(\textbf{S}| M_h,z) + N_s(\textbf{S}| M_h,z)\ \textrm{and } b^h_{\phi} =2\ \frac{\partial \textrm{ln}\ \bar{n}_h(M_h,z)}{\partial \textrm{ln}\ \sigma_8}
\end{split}
\end{equation}

\citet{Marinucci2023} then consider the response of an HOD model itself to some long-wavelength perturbation \citep[\textsection 3 and Eq.\ 3.2]{Voivodic2021}. As a reminder, they populate galaxies with the \texttt{GALACTICUS} SAM, but work within the lens of the HOD framework for this analysis.
The non-Gaussian assembly bias they call $\Delta b_{\phi}(\textbf{S},z)$ is split into central (\textit{c}) and satellite (\textit{s}) terms, 
as often parameterized in the HOD framework, and a weighted average over halo mass:

\begin{equation}
\Delta b_{\phi}(\textbf{S}, z) = \frac{1}{\bar{n}_g} \int dM_h \left[ f_c \Delta b_\phi^c(\textbf{S}|M_h, z) + f_s \Delta b_{\phi}^s(\textbf{S}|M_h, z) \right] \times \bar{n}_H(M_h, z) \left[ N_c(\textbf{S}|M_h, z) + N_s(\textbf{S}|M_h, z) \right]
\end{equation}

This aligns with \citet{Voivodic2021} Eq.\ 2.11, where the response of the halo occupation to local PNG perturbations  becomes\footnote{Eq.\ \ref{eq:HODresponse} is a correction of Eq.\ 10 in \citet{Marinucci2023} via Marinucci, private communication.}:

\begin{equation}
    R_{\phi}^g (\textbf{S}| M_h, z) = \Big( f_c \Delta b_\phi^c(\textbf{S}|M_h, z) + f_s \Delta b_{\phi}^s(\textbf{S}|M_h, z) \Big)
    \label{eq:HODresponse}
\end{equation}

They compute both $\Delta b_\phi^{c,s}$ components with the peak background split: 

\begin{equation}
    \Delta b_\phi^{c,s} (\textbf{S}|M_h, z) =
    \frac{1}{|\delta_{\sigma_8}|} \left[ \frac{N_{c,s}^{\text{high}}(\textbf{S}|M_h, z) - N_{c,s}^{\text{low}}(\textbf{S}|M_h, z)}{N_{c,s}^{\text{fid}}(\textbf{S}|M_h, z)} \right]
\label{eq:Deltabphi_cs2}
\end{equation}

\begin{figure}
    \centering
    \includegraphics[width=\linewidth]{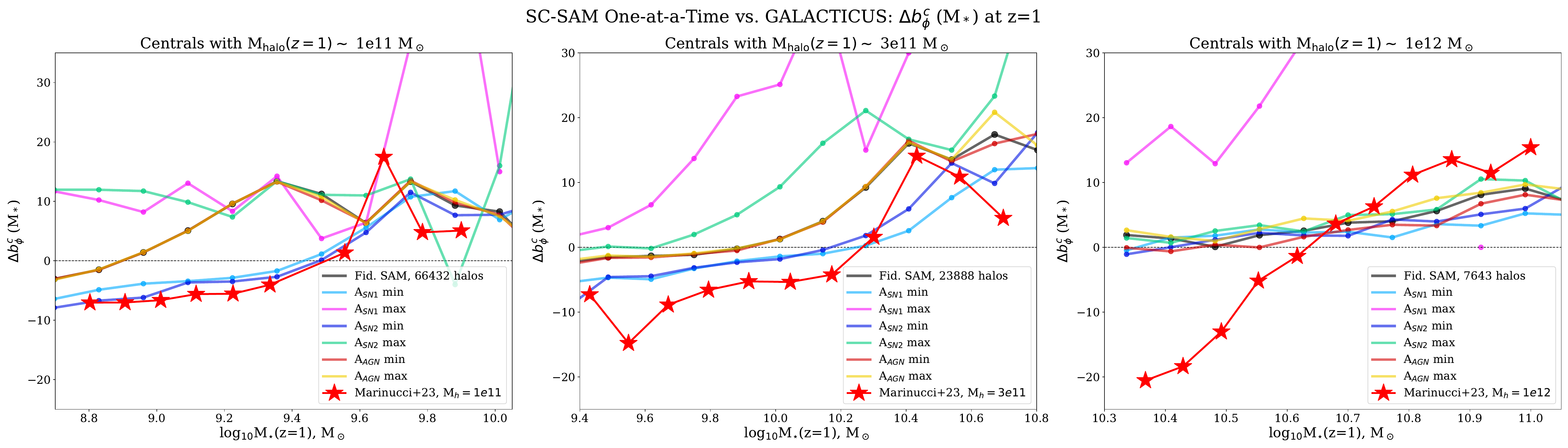}
    \caption{Recreating the non-Gaussian assembly bias for z=1 central galaxies $\Delta b_\phi^{c}(M_*|M_{\textrm{halo}}, z)$ measurement from \citet[Fig. A1, without their errors]{Marinucci2023} with the \texttt{SC-SAM} `one-at-a-time' models (for 30,000 halos). We note the \texttt{SC-SAM} and \texttt{GALACTICUS} were tuned to $z=0$ SMFs with different high-mass ends, and the SAMs were run on halos from simulation-measured merger trees vs.\ halos from EPS-generated merger trees (respectively). }
    \label{fig:Deltabphi}
\end{figure}

\citet{Marinucci2023} run \texttt{Galacticus} over $N_h\sim10^{4-5}$ EPS merger trees with $z=1$ root halos with masses between $[3\times 10^{10}, 10^{12}]$ M$_{\odot}$.
For our one-at-a-time parameter variation models from \textsection \ref{subsec:SCSAM1P}, we directly sample $N_h\sim30,000$ halos close to the target mass at $z=1$ across the three separate-universe cosmologies, bin their central galaxies by stellar mass, and compute Eq.\ \ref{eq:Deltabphi_cs2}. Figure \ref{fig:Deltabphi} compares our $\Delta b_{\phi}^{c}($M$_*|$M$_h, z=1)$ against those calculated in \citet{Marinucci2023} on merger trees of halos with M$_h(z=1)\ =\ [10^{11}, 3\times 10^ {11}, 10^{12}$] M$_{\odot}$. We omit their M$_h(z=1)\ =\ 3\times10^{10}$ to guarantee we avoid resolution effects for M$_*<10^9$ M$_{\odot}$ in our catalogs, and do not consider satellites for the same reason. 

The \texttt{SC-SAM} one-at-a-time models find the best agreement with \texttt{GALACTICUS} $\Delta b_{\phi}^{c}($M$_*|$M$_h=10^{11}, z=1)$ for the minimum A$_{\textrm{SN}}$ \texttt{SC-SAM} models (which have the highest SMFs and SHMRs at these halo masses). The maximum \textsf{SN1} model is wildly distinct from both \texttt{Galacticus} and the other \texttt{SC-SAM} models, especially for the largest target halo mass M$_h\sim 10^{12}\ M_{\odot}$ where all other one-at-time \texttt{SC-SAM} models strongly converge. 
The best agreement overall is at the highest $z=1$ stellar mass bins for all probed halo masses. We attribute the differences between our and \citet{Marinucci2023}'s $\Delta b_{\phi}^{c}($M$_*|$M$_h=10^{11}, z=1)$ to the inherent halo and galaxy assembly biases that our \texttt{SC-SAM} catalogs contain that EPS merger trees do not contain or create.
Additionally, the \texttt{SC-SAM} and \texttt{GALACTICUS} were tuned to $z=0$ SMFs with different high-mass ends, with \texttt{Galacticus} showing a lower SHMR in the low-mass regime (slightly smaller M$_*$ than the \texttt{SC-SAM} sees in M$_h<10^{12}\ M_{\odot}$; comparing against \citealt{RobertsonBenson2025}).

\begin{figure}
    \centering
    \includegraphics[width=0.9\linewidth]{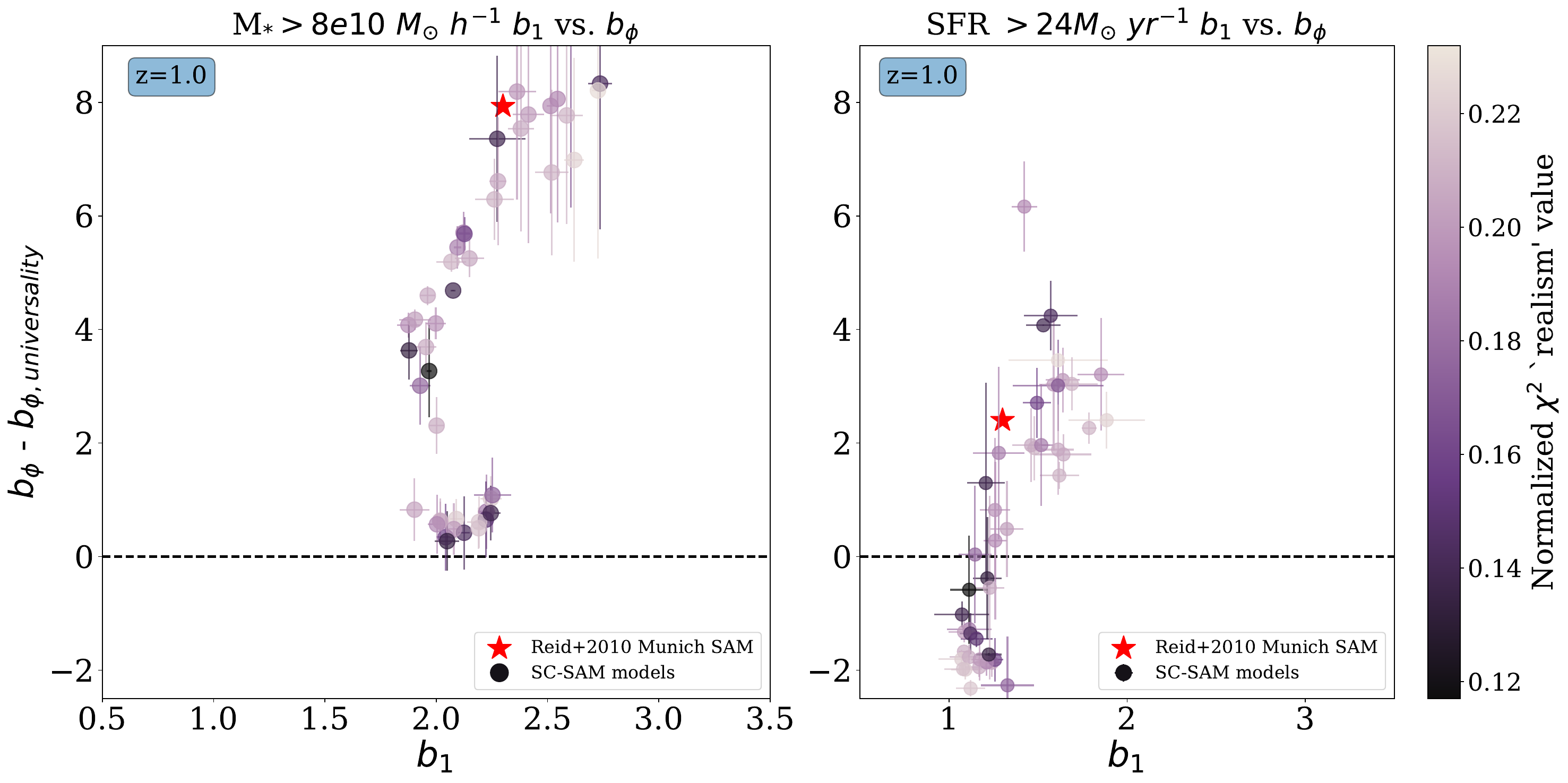}
    \caption{$b_{\phi} -b_{\phi, \textrm{univ.}}$ vs.\ $b_1$ relationship at $z= 1$ for the threshold galaxy selections \citet{Reid2010} studied with the Munich SAM: M$_* >8\times 10^{10}\ M_{\odot} h^{-1}$ (left), and SFR$>24\ M_{\odot}\ yr^{-1}$. Red stars s are the converted values from Table 1 of \citet{Reid2010}. The \texttt{SC-SAM} measurements are circles colored by their `realism' as described throughout this work.}
    \label{fig:Reid2010}
\end{figure}

\citet{Reid2010} leverage a run of the Munich SAM atop the Millenium simulation's merger trees \citep{Bertone2007} to measure $\Delta A^{\text{gal}}_{\text{NG}}$ for $z=1$ galaxies above a selection threshold of stellar mass or SFR, in a complex method involving the formation history of the host halos for a sample. They report in their Table 1, for galaxies sampled to a number density of $4.5\times 10^{-4}\ (h^{-1}$ Mpc$)^{-3}$ by either stellar mass or SFR, their measured Gaussian Eularian bias $b_G$ (functionally equivalent to our $b_1$), $\Delta A^{\text{gal}}_{\text{NG}}$, and $A_{\textrm{NG}}^{\textrm{all}}$ (from $b_G$ assuming universality). To compare to our Figs.\ \ref{fig:b1Vbphi_stemass} and \ref{fig:b1Vbphi_sfr}, we sum the reported values for $A_{\textrm{NG, sample}} = A_{\textrm{NG}}^{\textrm{all}}\ +\ \Delta A_{\textrm{NG}}^{\textrm{sample}}$; convert this into a $b_{\phi}= 2A_{\textrm{NG}}/D(z=1)$; and determine what $b_{\phi, \textrm{ univ.}}$ would be given the reported $b_G$. We note that we match their threshold selection and \textit{not} their targeted number density, given that each \texttt{SC-SAM} will yield a different number density for the same galaxy selection. We find that our varied models agree well with the converted values of \citet{Reid2010} for both selections in $b_{\phi}$ and $b_1$ values, as we report in Section \ref{subsec:TNGnSAMs}.

\bibliography{references}{}
\bibliographystyle{aasjournal}

\end{document}